%% file: arxiv-longer.tex
\documentclass[a4paper,11pt]{article}

\usepackage{placeins}
\usepackage{jheppub}
\makeatletter
\gdef\@fpheader{\phantom{Prepared for submission to JHEP}}
\makeatother
\usepackage[T1]{fontenc}
\usepackage[capitalize,nameinlink]{cleveref}
\usepackage{tikz}
\usepackage{xcolor}
\usepackage{amsmath}
\usepackage{amssymb}
\usepackage[bb=dsserif]{mathalpha}
\usepackage{pgfplots}
\definecolor{darkorange}{RGB}{240,50,0}
\crefname{section}{Sec.}{Secs.}
\crefname{subsection}{Sec.}{Secs.}

\usepgfplotslibrary{groupplots}
\usetikzlibrary{arrows.meta}
\pgfplotsset{compat=1.18}

\usepackage[table]{xcolor}
\usepackage{array}
\usepackage{graphicx}

\definecolor{ClassicalSilver}{HTML}{E8EAED}
\definecolor{HartreeOrchid}{HTML}{E9DDF2}
\definecolor{ScatteringMutedRose}{HTML}{F3DADB}

\definecolor{myblue}{RGB}{49,134,255}      
\definecolor{regiongray}{RGB}{105,105,105}
\definecolor{CouplingLightOrange}{RGB}{253,234,180}
\definecolor{regiongray}{RGB}{110,110,120}
\definecolor{orchidpink}{RGB}{220,70,200}
\definecolor{regiongray}{RGB}{145,145,155}
\usepackage{pifont}
\usepackage{etoolbox}
\usepackage{extarrows}
\usepackage{amsmath}
\usepackage{graphicx}
\usepackage{dcolumn}
\usepackage{bm}
\usepackage{mathrsfs}
\usepackage{wasysym}
\usepackage{verbatim}
\usepackage{mathtools,slashed}
\usepackage{braket}
\newcommand{\bl}{\boldsymbol{\ell}}
\usepackage{float}
\usepackage{multirow}
\usepackage{rotating}
\usepackage{booktabs}

\crefname{appendix}{App.}{Apps.}
\Crefname{appendix}{App.}{Apps.}
\definecolor{orchidfill}{RGB}{248,224,238}
\definecolor{orchidline}{RGB}{163,68,201}
\definecolor{silverfill}{RGB}{233,236,240}
\definecolor{darkgraybullet}{RGB}{85,85,85}
\definecolor{axisblack}{RGB}{20,20,20}

\newcommand{\tr}{\text{tr}}
\newcommand{\Tr}{\text{Tr}}

\newcommand{\baD}{\bar D}
\newcommand{\baF}{\bar F}
\newcommand{\eps}{\epsilon}
\newcommand{\bn}{\hat{\mathbf n}}

\newcommand{\be}{\begin{equation}}
\newcommand{\ee}{\end{equation}}

\newcommand{\mA}{\mathcal{A}}

\newcommand{\bp}{\boldsymbol{p}}
\newcommand{\p}{\partial}

\newcommand{\bx}{{\bf{x}}}
\newcommand{\bq}{{\bf{q}}}
\newcommand{\bT}{{\bf{T}}}

\newcommand{\bea}{\begin{eqnarray}}
\newcommand{\eea}{\end{eqnarray}}

\newcommand{\bk}{{\bf{k}}}

\usepackage{hyperref}
\renewcommand{\thefootnote}{\fnsymbol{footnote}}

\usepackage{tikz}
\usetikzlibrary{arrows.meta,positioning,calc}

\usepackage[T1]{fontenc}

\input{tensor_diagram_preamble.tex}
\input{su2_quadratic_preamble.tex}
\input{born_selfenergy_diagrams_qa.tex}

\input{linear_W_vertex_equation_macros.tex}

\title{\fontfamily{ppl}\bfseries\selectfont
\shortstack[l]{
Stability of Axion--SU(2) Chromomagnetic Condensates \\ with Stochastic Spin-2 Dressing
}}

\abstract{
We investigate whether a stochastic spin-2 gauge-field background can stabilize the chromomagnetic condensate of axion--SU(2) Yang--Mills theory at fixed axion Noether charge in flat spacetime. By coarse-graining the gauge field, we uncover a new class of statistically homogeneous and isotropic spin-2-dressed chromomagnetic condensates satisfying the classical background equations. We then study their fluctuations in two steps. First, at the mean-field (Hartree) level, we determine the fluctuation spectrum and the region where all modes remain positively gapped. Second, we include the leading scattering corrections from fluctuations propagating through the stochastic background. We find that gap survival is controlled by non-Gaussian correlations of the dressing. Statistical positivity constrains the otherwise undetermined higher-point contributions through model-independent consistency conditions, without fixing their detailed momentum dependence. These bounds separate the Hartree-gapped region into a part where a positive Born-level gap is ruled out and a part where stability remains possible. The results are relevant to non-Abelian gauge condensates in early-universe cosmology, including inflation, dark energy, and primordial gravitational waves, and to QCD-like gauge sectors above confinement where thermal effects are subleading.
}

\author[1,2]{Sha Azyzy,}
\emailAdd{sazyzy@mpa-garching.mpg.de}
\author[3]{Azadeh Maleknejad}
\emailAdd{azadeh.maleknejad@swansea.ac.uk}
\author[4,5]{and Vahid Kamali}
\emailAdd{vahid.kamali2@mcgill.ca}
\affiliation[1]{Max Planck Institute for Astrophysics, Karl-Schwarzschild-Str. 1, 85748 Garching, Germany}
\affiliation[2]{Ludwig-Maximilians-Universität München, Schellingstr. 4, D-80799 München, Germany}
\affiliation[3]{Centre for Quantum Fields and Gravity, Swansea University,
Swansea SA2 8PP, United Kingdom}
\affiliation[4]{Department of Physics, McGill University, Montr\'{e}al, QC H3A 2T8, Canada}
\affiliation[5]{Department of Physics, Bu-Ali Sina (Avicenna) University, Hamedan 65178-016016, Iran}

\date{\today}

\begin{document}
\maketitle
\renewcommand{\thefootnote}{\arabic{footnote}}

\section{Introduction}

The stability of chromomagnetic condensates is a longstanding problem
in non-Abelian gauge theory, dating back to the Savvidy vacuum
\cite{Savvidy1977} and Nielsen--Olesen instability
\cite{NielsenOlesen1978}.
\footnote{The latter motivated the Copenhagen or spaghetti vacuum, in which
chromomagnetic flux tubes form an inhomogeneous, disordered condensate
\cite{NielsenOlesen1979,AmbjornOlesen1980}, as discussed in recent
reviews \cite{Savvidy2020,Cea2024}.}
At its core lies the question of whether gauge-field fluctuations
can reorganize an unstable chromomagnetic background into a
self-consistent, stable configuration.
Axion--Yang--Mills theory offers a natural setting to revisit this
question: a rolling axion acts as an effective chemical potential for
Chern--Simons number, modifying the dynamics of non-Abelian gauge fields.
This framework is motivated both by the approximately shift-symmetric
dynamics of the QCD axion above confinement
scales~\cite{PecceiQuinn1977,Borsanyi2016} and by the nonthermal
conditions of the early Universe
~\cite{KofmanLindeStarobinsky1997,Kolb:1990vq,Gorbunov:2011zzc}.
Early investigations of the QCD vacuum also considered homogeneous,
genuinely non-Abelian chromomagnetic backgrounds, whose quantum
fluctuations were found to be unstable~\cite{Takahashi1980}.
A particularly symmetric configuration is the color--spin--locked
ansatz $A_i^a=Q\delta_i^a$, which generates a homogeneous
chromomagnetic field $B_i^a=gQ^2\delta_i^a$ while preserving isotropy
under combined spatial and gauge rotations \cite{SaitoShigemoto1980}.
Despite this symmetry, its static realization possesses a tachyonic
spin-$2$ fluctuation \cite{Takahashi1980}.
Earlier and subsequent studies of homogeneous Yang--Mills
dynamics also revealed nonlinear color oscillations and
stochastic behavior in the enlarged classical phase space
\cite{Baseyan1979,Matinyan1981,Savvidy1983}.
A related phenomenon occurs in chiral plasmas, where chiral
imbalance drives helical gauge instabilities
\cite{AkamatsuYamamoto2014,AkamatsuRothkopfYamamoto2016,
SchlichtingSharma2023}, leading to inverse magnetic-helicity
cascades in Abelian systems
\cite{BoyarskyFrohlichRuchayskiy2012,HironoKharzeevYin2015},
while non-Abelian dynamics involve distinct topological transitions.

A dynamical realization of this non-Abelian ansatz emerged in
inflationary cosmology with gauge-flation
\cite{MaleknejadSheikhJabbari2011}, opening a rich phenomenology
rooted in particle-physics gauge dynamics.
The construction subsequently became central to
axion--$SU(2)$ inflationary models
\cite{AdsheadWyman2012,Maleknejad:2012fw,Maleknejad2016,
Adshead:2016omu,Adshead:2017hnc},
where these gauge tensors provide a distinctive mechanism
for sourcing chiral gravitational waves
\cite{Maleknejad2016,DimastrogiovanniEtAl2017,CampetiEtAl2021}
and potentially large tensor non-Gaussianity
\cite{AgrawalFujitaKomatsu2018}.
The production and backreaction of these spin-$2$ particles
were further quantified, constraining the amplitude of the
sourced gravitational waves \cite{MaleknejadKomatsu2019}.
Such primordial signatures offer targets across more than
twenty decades in frequency, from CMB $B$-mode polarization
and pulsar timing arrays to space- and ground-based laser
interferometers, including LISA \cite{Thorne:2017jft,Komatsu2022,Auclair2023LISA}.

Beyond gravitational waves, spontaneous CP violation in the
axion--gauge background enables baryogenesis through the
$SU(2)$ chiral anomaly
\cite{MaleknejadSU2R2021,Maleknejad2021ChiralAnomaly}
and gravitational leptogenesis
\cite{MaleknejadGravLep2014,MaleknejadGravLep2016,
CaldwellDevulder2018,AlexanderMcDonoughSpergel2018},
without requiring CP violation in the neutrino sector.
The same background has also been explored in Schwinger
pair production during inflation
\cite{LozanovMaleknejadKomatsu2019,Maleknejad2020DarkFermions},
primordial magnetogenesis
\cite{BrandenburgIaryginaSfakianakisSharma2024},
coherent vector dark matter
\cite{Fujita2024VectorDM,Khan2026SU2DM,Daniel2026SU2Resonance},
and late-time cosmic acceleration
\cite{MehrabiMaleknejadKamali2017,
AlexanderBrandenbergerFrohlich2016,RodriguezNavarro2018,
CaldwellDevulder2019,GuarnizoEtAl2020,Papageorgiou2020}.

Despite these developments, the underlying question of
non-Abelian condensate stability remains unresolved:
can stochastic spin-$2$ fluctuations dress the chromomagnetic
background into a self-consistent, perturbatively controlled
stable configuration, once both their backreaction and
scattering effects are taken into account?

\paragraph{Aim and main results.}
We address this question in flat-space $SU(2)$ axion--Yang--Mills
theory at fixed conserved axion shift charge. Our analysis proceeds
through three successive requirements:
(i)~\emph{classical self-consistency} of the coarse-grained stochastic
background;
(ii)~\emph{a positive Hartree gap} for quantum fluctuations; and
(iii)~\emph{survival of the gap under scattering}, with perturbative
control. Our results are summarized in \cref{tab:chromomagnetic-solutions}.
We identify statistically homogeneous and isotropic spin-$2$-dressed
solutions of the classical moment equations and determine the region in which
their mean-field (Hartree) spectrum is positive and gapped. We then include
scattering from the stochastic dressing at Born order. The part of the Born
self-energy fixed by the power spectrum closes the Hartree gap throughout the
backgrounds examined, so a positive gap can survive only through compensating
non-Gaussian scattering contributions. Using exact statistical positivity
inequalities, we bound these otherwise undetermined higher-point
contributions and constrain the region in which a positive Born gap may
survive. These bounds exclude a positive Born gap
in part of the Hartree-gapped region and delimit where it may survive.
The remaining region thus satisfies necessary, but not sufficient,
conditions for Born-level stability; establishing a positive gap and
perturbative control there remains an open question.

\providecolor{ClassicalSilver}{RGB}{232,234,236}
\providecolor{HartreeOrchid}{RGB}{233,222,240}
\providecolor{CouplingLightOrange}{RGB}{251,231,204}
\providecolor{ScatteringMutedRose}{RGB}{243,219,223}

\providecommand{\cmark}{\textcolor[RGB]{0,150,230}{\ding{51}}}
\providecommand{\xmark}{\textcolor{red!80!black}{\ding{55}}}

\begin{table}[htbp]
\centering
\footnotesize
\renewcommand{\arraystretch}{1.15}
\setlength{\tabcolsep}{3pt}

\begin{tabular}{
  | >{\raggedright\arraybackslash}m{\dimexpr0.24\linewidth+3mm\relax}
    >{\centering\arraybackslash}m{0.25\linewidth}
    >{\centering\arraybackslash}m{0.35\linewidth}
  | >{\centering\arraybackslash}m{\dimexpr0.08\linewidth-3mm\relax} |
}
\hline
\textbf{Features}
& \textbf{Color-spin-locked}
& \textbf{Spin-2-dressed}
& \textbf{Sec.}
\\
\hline

\multicolumn{3}{|@{}c@{}|}{%
\begin{tabular}[c]{
    >{\raggedright\arraybackslash}m{\dimexpr0.24\linewidth+3mm\relax}
    >{\centering\arraybackslash}m{0.25\linewidth}
    >{\centering\arraybackslash}m{0.35\linewidth}
}
\multicolumn{3}{>{\columncolor{ClassicalSilver}}c}{%
  \rule[-1.0ex]{0pt}{4.0ex}%
  \textbf{Classical background $\bar A_i$}%
}
\\
\hline
$SU(2)$ ansatz
& $\bar A_i^a=Q\delta_i^a$
& $\bar A_i^a=Q\delta_i^a+W_i^a$
\\
\hline
spin-2 dressing
& Absent
& $W_i^a$: frozen and stochastic
\\
\hline
Spin content
& Spin-$0$
& Spin-$0\oplus2$
\\
\hline
Homogeneity \& isotropy
& Exact
& Statistical at the ensemble level
\\
\end{tabular}%
}
& \rotatebox[origin=c]{90}{%
    \cref{sec:stochastic-spin2-background}%
  }
\\
\hline

\multicolumn{3}{|@{}c@{}|}{%
\begin{tabular}[c]{
    >{\raggedright\arraybackslash}m{\dimexpr0.24\linewidth+3mm\relax}
    >{\centering\arraybackslash}m{0.25\linewidth}
    >{\centering\arraybackslash}m{0.35\linewidth}
}
\multicolumn{3}{>{\columncolor{HartreeOrchid}}c}{%
  \rule[-1.0ex]{0pt}{4.0ex}%
  \textbf{Quantum fluctuations: Hartree sector}%
  \quad
  $\bigl(\bar\Delta,\Gamma_{\mathrm H}^{(1)}\bigr)$%
}
\\
\hline
Mean-field spectrum $\bar\Delta$
& $\bar\Delta=\Delta_Q$
  \newline
  Negative physical mode
& $\bar\Delta=\langle\Delta\rangle_{\rm st}$
  \newline
  $\bar m^2>0$ in the identified region
\\
\hline
One-loop action $\Gamma_{\mathrm H}^{(1)}$
& $\operatorname{Im}\Gamma_{\mathrm H}^{(1)}\ne0$\ \xmark
& $\operatorname{Im}\Gamma_{\mathrm H}^{(1)}=0$\ \cmark
  \newline
  Within the Hartree-gapped region
\\
\end{tabular}%
}
& \rotatebox[origin=c]{90}{%
    \cref{sec:quantum-structure-Hartree}%
  }
\\
\hline

\multicolumn{3}{|@{}c@{}|}{%
\begin{tabular}[c]{
    >{\raggedright\arraybackslash}m{\dimexpr0.24\linewidth+3mm\relax}
    >{\centering\arraybackslash}m{0.25\linewidth}
    >{\centering\arraybackslash}m{0.35\linewidth}
}
\multicolumn{3}{>{\columncolor{CouplingLightOrange}}c}{%
  \rule[-1.0ex]{0pt}{4.0ex}%
  \textbf{Gauge coupling: weak coupling and confinement}%
}
\\
\hline
Background scale $M_{\rm bg}$
& $|gQ|$
& $\max\{|gQ|,g\sqrt{{\rm tr}[W^2]},k_W\}$
\\
\hline
Weak coupling
& \multicolumn{2}{
    >{\centering\arraybackslash}
    m{\dimexpr0.60\linewidth+2\tabcolsep\relax}
  }{%
  $g^2(M_{\rm bg})/4\pi\ll1
   \;\Longrightarrow\;
   \Lambda_{\rm conf}
   \sim M_{\rm bg}\,e^{-12\pi^2/(11g^2)}
   \ll M_{\rm bg}$%
}
\\
\hline
Constant-$g$ window
\newline
(Hartree level)
& Not applicable
  \newline
  (no positive gap)
& $\ln\frac{M_{\rm bg}}{\mu_{\rm IR}}
   \ll\frac{8\pi^2}{b_0g^2}$
  \newline
  $\Longrightarrow\ M_{\rm bg} \gg\Lambda_{\rm conf}$\ \cmark
\\
\end{tabular}%
}
& \rotatebox[origin=c]{90}{%
    \cref{sec:confinement-validity}%
  }
\\
\hline

\multicolumn{3}{|@{}c@{}|}{%
\begin{tabular}[c]{
    >{\raggedright\arraybackslash}m{\dimexpr0.24\linewidth+3mm\relax}
    >{\centering\arraybackslash}m{0.25\linewidth}
    >{\centering\arraybackslash}m{0.35\linewidth}
}
\multicolumn{3}{>{\columncolor{ScatteringMutedRose}}c}{%
  \rule[-1.0ex]{0pt}{4.0ex}%
  \textbf{Quantum fluctuations: Scattering sector}%
  \quad $\Sigma$%
}
\\
\hline
Scattering self-energy $\Sigma$
& $\Sigma=0$
& Born self-energy $\Sigma^{\rm B}$
\\
\hline
Positive Born gap $m^2_{\rm gap} = \bar{m}^2 - \Sigma^{\rm B}$
& No positive Hartree gap to protect\ \xmark
& Candidate Born-gapped region from statistical positivity bounds
\\
\end{tabular}%
}
& \rotatebox[origin=c]{90}{%
    \cref{sec:scattering}%
  }
\\
\hline
\end{tabular}
\caption{\small
Static chromomagnetic backgrounds in axion--Yang--Mills theory at fixed
conserved axion shift charge, with $\bar A_0^a=0$ and $E_i^a=0$.
Stochastic backgrounds satisfy the coarse-grained classical moment
equations. The Hartree operator and effective action are
$\bar\Delta=\langle\Delta\rangle_{\rm st}$ and
$\Gamma_{\mathrm H}^{(1)}=\Gamma^{(1)}[\bar\Delta]$.
Positive-gap entries refer to the identified non-Gaussian
Hartree-gapped region. Weak coupling requires
$g^2(M_{\rm bg})/4\pi\ll1$ and $\bar m\gg\Lambda_{\rm conf}$, so that
the running of $g$ is negligible over the scales relevant to the gap.
Born scattering shifts the gap to
$m_{\rm gap}^2=\min_k[-\bar\Delta(0,k)+\Sigma^{\rm B}(k)]$.
Statistical positivity bounds further restrict the Hartree-gapped
region and identify a candidate Born-gapped domain, without by
themselves establishing a positive gap throughout that region.
}
\label{tab:chromomagnetic-solutions}
\end{table}

The paper is structured as follows. We begin in \cref{sec:2} with
axion--Yang--Mills theory and its homogeneous, color--spin--locked
chromomagnetic background. In \cref{sec:stochastic-spin2-background},
we introduce stochastic spin-2 dressing and derive the self-consistency
conditions. We then turn to
quantum stability in \cref{sec:quantum-structure-Hartree}, contrasting
the instability of the homogeneous background with the possibility of
a positive Hartree mass gap induced by the dressing. The weak-coupling
regime and its separation from confinement are discussed in
\cref{sec:confinement-validity}. In \cref{sec:scattering}, we examine
whether the Hartree gap survives scattering off the stochastic dressing,
combining analytical bounds, numerical evaluation of the Born
self-energies, and an assessment of higher-order corrections.
We conclude in \cref{sec:conclusion}. Mathematical details, higher-point correlators, one-loop quantization and renormalization, the running coupling, and the stochastic scattering self-energy are collected in \crefrange{app:Math}{app:self-energy}.

\section{Yang--Mills Theory with an Axion}
\label{sec:2}

We consider an $SU(N)$ Yang--Mills field coupled to a massless axion
$\phi$ through the action
\begin{equation}
 S[A,\phi]=\int d^4x\,\sqrt{-g}\left[
 -\frac12\operatorname{tr}(\boldsymbol F_{\mu\nu}\boldsymbol F^{\mu\nu})
 -\frac12\partial_\mu\phi\,\partial^\mu\phi
 +\frac{\kappa g^2}{2f}\phi\,
 \operatorname{tr}(\boldsymbol F_{\mu\nu}
 \widetilde{\boldsymbol F}^{\mu\nu})
 \right],
 \label{eq:YM-CS}
\end{equation}
where $g$ is the gauge coupling, $f$ the axion decay constant, and
$\kappa$ a dimensionless coupling. We use the metric signature $(-,+,+,+)$ and
Hermitian generators satisfying
$[\boldsymbol T^a,\boldsymbol T^b]=if^{abc}\boldsymbol T^c$ and
$\operatorname{tr}(\boldsymbol T^a\boldsymbol T^b)=\delta^{ab}/2$.
Our field-strength conventions are
\begin{equation}
 \begin{aligned}
 \boldsymbol A_\mu=A_\mu^a\boldsymbol T^a,\qquad
 \boldsymbol F_{\mu\nu}=\partial_\mu\boldsymbol A_\nu
 -\partial_\nu\boldsymbol A_\mu
 -ig[\boldsymbol A_\mu,\boldsymbol A_\nu],
 \end{aligned}
 \label{eq:YM-definitions}
\end{equation}
where $\epsilon^{\mu\nu\rho\sigma}$ is the totally antisymmetric symbol, and $
 \widetilde{\boldsymbol F}^{\mu\nu}
=\frac{\epsilon^{\mu\nu\rho\sigma}}{2\sqrt{-g}}
 \boldsymbol F_{\rho\sigma}$. The Pontryagin density is a total divergence,
$\operatorname{tr}(\boldsymbol F_{\mu\nu}
\widetilde{\boldsymbol F}^{\mu\nu})=\nabla_\mu K^\mu$,
where the Chern--Simons current, in our conventions, is
\begin{equation}
K^\mu
=\frac{\epsilon^{\mu\nu\rho\sigma}}{\sqrt{-g}}
\operatorname{tr}\!\left(
\boldsymbol A_\nu\boldsymbol F_{\rho\sigma}
+\frac{ig}{3}\boldsymbol A_\nu
\boldsymbol A_\rho\boldsymbol A_\sigma
\right).
\label{eq:CS-current}
\end{equation}

 The axion equation of motion is
\begin{equation}
 \Box\phi+\frac{\kappa g^2}{2f}
 \operatorname{tr}(\boldsymbol F_{\mu\nu}
 \widetilde{\boldsymbol F}^{\mu\nu})=0.
 \label{eq:axion-eom}
\end{equation}
Consequently, the classical Noether current associated with
the axion shift symmetry,
\begin{equation}
 J^\mu_{\rm shift}
 =-\nabla^\mu\phi-\frac{\kappa g^2}{2f}K^\mu,
 \qquad \nabla_\mu J^\mu_{\rm shift}=0,
 \label{eq:shift-current}
\end{equation}
is conserved on shell. The overall sign of the current is conventional;
we choose the sign for which its temporal component contains
$+\dot\phi$ in flat spacetime. The associated Noether charge on a spatial hypersurface $\Sigma$ is
\begin{equation}
 Q_{\rm shift}
 =\int_\Sigma d\Sigma_\mu\,J^\mu_{\rm shift},
 \qquad
 \frac{dQ_{\rm shift}}{dt}=0,
 \label{eq:shift-charge}
\end{equation}
where conservation holds for suitable boundary conditions.
Under a large gauge transformation of winding number $N\in\mathbb Z$,
the integrated Chern--Simons current shifts as
\begin{equation}
 Q_{\rm shift}\longrightarrow
 Q_{\rm shift}-\frac{8\pi^2\kappa}{f}\,N.
 \label{eq:shift-charge-large-gauge}
\end{equation}
We formulate the fixed-charge problem within a specified
gauge sector, with these discrete identifications understood.

\subsection{Chromomagnetic backgrounds with axion constraints}
\label{sec:constrained-axion-backgrounds}

We now investigate chromomagnetic backgrounds supported by an axion coupled
to the non-Abelian gauge sector. Rather than solving the unrestricted
axion--gauge system, we focus on two complementary constrained realizations
of the axion dynamics. These provide controlled settings in which the
existence and properties of the chromomagnetic configurations can be studied
analytically.  We focus on static gauge configurations,
\begin{equation}
 A_0^a=0,\qquad \partial_0A_i^a=0,
 \label{eq:static-gauge-sector}
\end{equation}
for which $E_i^a=F_{i0}^a=0$ and $K^i=0$.
We define $B_i^a=\frac12\epsilon_{ijk}F_{jk}^a$, with
$\epsilon_{ijk}=\epsilon^{0ijk}$. 
The tangential gauge potential is fixed at the boundary,
$ \left.\hat{\mathbf n}\times\delta\mathbf A^a
 \right|_{\partial\Omega}=0$.
We define
the Chern--Simons density and functional by
\begin{equation}
 \begin{aligned}
 n_{\rm CS}
 \equiv\frac12K^0
 =\frac12\epsilon^{ijk}\left(
 A_i^a\partial_jA_k^a
 +\frac{g}{3}f^{abc}A_i^aA_j^bA_k^c\right),
 \end{aligned}
 \label{eq:NCS-def}
\end{equation}
and the corresponding Chern--Simons functional is
\be
 N_{\rm CS}[A]\equiv\int_\Omega d^3x\,n_{\rm CS}.
 \ee
The two realizations correspond, respectively,
to fixing the axion-induced chemical potential  and to fixing the total shift charge:
\begin{equation}
\begin{aligned}
\dot{\phi}=\mathrm{const}
&:\quad \text{fixed chemical potential},\\[1mm]
Q_{\mathrm{shift}}=\mathrm{const}
&:\quad \text{fixed total charge}.
\end{aligned}
\end{equation}

\subsubsection*{Fixed chemical potential}
\label{sec:fixed_axion_chemical_potential}

For a homogeneous axion, a prescribed velocity fixes the effective
chemical potential conjugate to $N_{\rm CS}$,
\begin{equation}
 \alpha\equiv-\frac{\kappa g^2}{f}\dot\phi
 =\mathrm{const}.
 \label{eq:mu5-alpha}
\end{equation}
After integrating the Chern--Simons interaction by parts,
$\mathcal L_{\rm CS}=\alpha n_{\rm CS}$. The reduced gauge Hamiltonian is therefore
\begin{equation}
 H_{\rm red}^{(\alpha)}[A]
 =E_B[A]-\alpha N_{\rm CS}[A],
 \label{eq:Hred_fixed_mu5}
\end{equation}
where
$E_B[A]\equiv\frac12\int_\Omega d^3x\,B_i^aB_i^a$
is the chromomagnetic energy. Its stationary points obey the non-Abelian Beltrami equation,
\begin{equation}
(D\times B)_i^a=\alpha B_i^a,
\label{eq:B-eq}
\end{equation}
with $(D\times B)_i^a\equiv\epsilon_{ijk}(D_jB_k)^a$
and $(D_iX)^a\equiv\partial_iX^a+gf^{abc}A_i^bX^c$.

\subsubsection*{Fixed shift charge}
\label{sec:fixed_axion_shift_charge}

For a dynamical axion, the momentum conjugate to $\phi$ in the derivative
form of the action is
\begin{equation}
 \pi_\phi=J^0_{\rm shift}
 =\dot\phi-\frac{\kappa g^2}{f}n_{\rm CS}.
 \label{eq:pi-phi-shift}
\end{equation}
The total shift charge is
\begin{equation}
 Q_{\rm shift}=\int_\Omega d^3x\,\pi_\phi,
 \label{eq:Q-shift}
\end{equation}
and is conserved under the no-flux boundary condition
$\left.\hat n_iJ^i_{\rm shift}\right|_{\partial\Omega}=0$.
The fixed-charge problem is understood on a chosen Chern--Simons
branch.

In the static gauge sector, the Hamiltonian is
\begin{equation}
 H=E_B[A]+\frac12\int_\Omega d^3x\left[
 \left(\pi_\phi+\frac{\kappa g^2}{f}n_{\rm CS}\right)^2
 +(\boldsymbol\nabla\phi)^2\right].
 \label{eq:H-fixed-Qshift}
\end{equation}
Minimization over $\pi_\phi$ at fixed $Q_{\rm shift}$ makes
$\dot\phi$ spatially uniform. Since $K^i=0$, the no-flux condition
reduces to $\hat{\mathbf n}\cdot\boldsymbol\nabla\phi=0$. Writing $$V\equiv\mathrm{Vol}(\Omega)=\int_\Omega d^3x,$$
for the spatial volume, the conserved shift charge gives
\begin{equation}
 \dot\phi[A,Q_{\rm shift}]
 =\frac{Q_{\rm shift}}{V}
 +\frac{\kappa g^2}{f}\bar n_{\rm CS}[A],
 \qquad
 \bar n_{\rm CS}[A]\equiv\frac{N_{\rm CS}[A]}{V}.
 \label{eq:dotphi-fixed-Q}
\end{equation}
Eliminating the axion variables gives the reduced energy
\begin{equation}
 E_{Q_{\rm shift}}[A]
 =E_B[A]+\frac{1}{2V}
 \left(Q_{\rm shift}+\frac{\kappa g^2}{f}N_{\rm CS}[A]\right)^2.
 \label{eq:fixed_charge_energy}
\end{equation}
Its variation yields \cref{eq:B-eq}, now with the self-consistent
coefficient
\begin{equation}
 \alpha[A]
 =-\frac{\kappa g^2}{f}\dot\phi[A,Q_{\rm shift}]
 =-\frac{\kappa g^2}{fV}
 \left(Q_{\rm shift}+\frac{\kappa g^2}{f}N_{\rm CS}[A]\right).
 \label{eq:B-eq-alpha}
\end{equation}
Only the total shift charge is fixed; $N_{\rm CS}$ remains free to vary.

A useful analogue is provided by the chiral plasma instability, where a chiral imbalance induces an anomalous current \(j_{\chi,i}^a=\sigma_\chi B_i^a\) and allows axial charge to be exchanged with gauge-field Chern--Simons number~\cite{KharzeevMcLerranWarringa2008,FukushimaKharzeevWarringa2008,AkamatsuYamamoto2013,AkamatsuYamamoto2014}. In the magnetostatic limit, the equation \((D\times B)_i^a=\sigma_\chi B_i^a\) coincides with \cref{eq:B-eq} upon identifying \(\alpha=\sigma_\chi\), providing a direct correspondence between fixed axion shift charge and fixed total chirality. More generally, helicity constrains magnetic energetics through Arnold-type bounds and selects Beltrami configurations as stationary points at fixed helicity~\cite{Arnold1974,Moffatt1969,BuniyKephart2014}.

\subsection{Homogeneous and isotropic chromomagnetic background}

Up to this point, the discussion applies to a general $SU(N)$ gauge theory.  A homogeneous and isotropic chromomagnetic configuration can be embedded in any $SU(N)$ through an $SU(2)$ subgroup.  Choosing generators $T_a$, $a=1,2,3$, of such a subalgebra, we consider for constant $\alpha$
\begin{equation}
 \bar A_i=Q\,\delta_i^a T_a ,
 \label{eq:A-Q}
\end{equation}
with all components outside the embedded $SU(2)$ set to zero.
Since the configuration closes within this subalgebra, its classical
dynamics is that of the corresponding $SU(2)$ sector.
For simplicity, we therefore specialize to $SU(2)$ from here onward.
Homogeneous non-Abelian configurations of the form
$\bar A_i^a\propto\delta_i^a$ were first studied in flat-space
Yang--Mills theory~\cite{Baseyan1979,SaitoShigemoto1980,Takahashi1980}.
Their cosmological realization as homogeneous and isotropic gauge-field
backgrounds was later introduced in
gauge-flation~\cite{MaleknejadSheikhJabbari2011,Maleknejad:2011sq}
and subsequently employed in chromo-natural
inflation~\cite{AdsheadWyman2012}.
In temporal gauge, the static configuration gives
\begin{equation}
 \bar B_i^a=gQ^2\delta_i^a,
 \qquad
 \bar n_{\rm CS}=gQ^3,
 \qquad
 (\bar D\times\bar B)_i^a=\alpha\bar B_i^a.
 \label{eq:homogeneous-SU2}
\end{equation}
The classical field equation \cref{eq:B-eq} therefore reduces to
\begin{equation}
 Q^2(2gQ-\alpha)=0,
 \qquad\Longrightarrow\qquad
 Q=0
 \quad\text{or}\quad
 Q=\frac{\alpha}{2g}.
 \label{eq:homogeneous-SU2-branches}
\end{equation}
For $\alpha\neq0$, the latter gives the nontrivial chromomagnetic solution
\begin{equation}
 \bar B_i^a=\frac{\alpha^2}{4g}\delta_i^a .
\end{equation}
At fixed shift charge, $\alpha$ must additionally satisfy
\cref{eq:B-eq-alpha}.

\section{Spin-2-Dressed Chromomagnetic Background}
\label{sec:stochastic-spin2-background}

Chromomagnetic backgrounds in Yang--Mills theory are unstable at the quantum
level \cite{Savvidy1977,NielsenOlesen1978}. The color-locked configuration
\cref{eq:A-Q} is no exception: already in pure ${\rm SU}(2)$ theory it
supports unstable gauge-field fluctuations~\cite{Takahashi1980}, and in the
axion--Yang--Mills system the axion coupling sustains the stationary
background without removing the instability (\cref{subsec:locked-instability}).
The tachyonic mode lies in the spin-$2$ channel of the diagonal rotation group. This section constructs the background on which the question posed in the
Introduction can be asked: can a statistically homogeneous and isotropic
spin-2 dressing lift this instability? The construction is deliberately
restricted. We do not look for the general solution of the classical field
equations. We look for the simplest state that (i)~has the symmetries of the
color-locked configuration \emph{on average}, (ii)~carries spin-$2$ structure
in the channel where the instability lives, and (iii)~can be treated
self-consistently. Four assumptions implement this, and each term dropped
below is dropped because of one of them.

\paragraph{Assumptions and what they remove.}
\begin{itemize}
\item[\textbf{(A1)}] \emph{Static.} We seek a static state, so the
electric field $E^a_i=\partial_tW^a{}_i$ is set to zero at leading order and
the background equation reduces to its chromomagnetic part.

\item[\textbf{(A2)}] \emph{Transverse spin-$2$ dressing.} The deformation of
$Q\,\delta^a_i$ is restricted to its symmetric, traceless, transverse
(helicity-$\pm2$) part. This is the sector that couples to the tachyonic
channel, and it is the only one whose stochastic superposition is compatible
with statistical isotropy of the diagonal ${\rm SO}(3)$ without introducing a
preferred direction on average. Vector and scalar deformations are not
included.

\item[\textbf{(A3)}] \emph{Classical-statistical regime.} The populated tensor
modes are highly occupied, $n_\lambda(k)\gg1$, so the dressing is a
stochastic classical field and vacuum contributions to its correlators are
dropped.

\item[\textbf{(A4)}] \emph{Coarse-grained closure.} The field equations are
imposed only on self-averaged, rotationally invariant projections: the scalar
trace and the helicity-$2$ two-point projection. Components whose ensemble
mean vanishes by statistical homogeneity and isotropy (the vector part and the
helicity-$0,\pm1$ parts of the tensor equation) are not imposed realization by
realization.
\end{itemize}

Together, (A1)--(A4) define a \emph{closure}, not an exact solution: the
stochastic configuration does not solve the classical equations pointwise. The
closure is matched to the question. 
\paragraph{Ansatz.}
We take an ${\rm SU}(2)$ configuration consisting of the color-locked
condensate and a stochastic deformation,
\be
\bar A^a{}_i(t,\bx)
=Q\,\delta^a_i+W^a{}_i(t,\bx),
\qquad
\bar A^a_0=0.
\label{eq:BG-gauge-field}
\ee
The symmetric-traceless part of $W^a{}_i$ transforms in the spin-$2$
representation of the diagonal subgroup of spatial and global gauge
rotations. Following (A2), we keep only its transverse helicity-$\pm2$
sector,
\be
W_{ai}=W_{ia},
\qquad
W^a{}_a=0,
\qquad
\partial^i W_i{}^a=0.
\label{eq:W-TT-conditions}
\ee
The gauge-covariant transversality condition, $D^iW^a{}_i=0$ with $D_i$ built
from the full background \cref{eq:BG-gauge-field}, contains the non-Abelian
terms $gQ\,\epsilon^{abc}W_{cb}+g\,\epsilon^{abc}(W^2)_{bc}$. Both vanish
identically for symmetric $W$, so covariant and ordinary transversality
coincide,
\be
D^iW_i{}^a=0
\qquad \xLeftrightarrow{\;W_{ia}=W_{ai}\;} \qquad
\partial^i W_i{}^a=0.
\ee

\paragraph{Field equation.}
With the matrix powers
\be
(W^2)_{ij}=W_{i}{}^a W_{aj}, \qquad (W^3)_{ij}=W_{i}{}^a W_{a}{}^l W_{lj},
\ee
and using \cref{eq:CHI}, the chromomagnetic field is
\be
B^a{}_i
=
gQ^2\delta^a_i
+\epsilon_{ijk}\partial^j W^{ak}
-gQ\,W^a{}_i
+g\left(
(W^2)^a{}_i
-\frac12\delta^a_i \tr[W^2]
\right).
\ee
Using \cref{eq:CH-1,eq:CHI}, its covariant curl is
\be
\begin{aligned}
(D\times B)^a{}_i
& =\;
2g^2Q^3\delta^a_i
-\nabla^2W^a{}_i
-2gQ\,\epsilon_{ijk}\partial^jW^{ak}
+g^2Q
\left[
\delta^a_i\,\tr[W^2]
-3(W^2)^a{}_i
\right] \\
&+g^2
\left[
\frac12 \tr[W^2]\,W^a{}_i
 - \frac13 \tr[W^3]\, \delta^a_i
\right]  +g\,\epsilon^{abc}
\left(
W_{bj}\partial_iW^{cj}
-2W_{bj} \partial^j W_{ci}
\right).
\end{aligned}
\ee
By (A1), the background equation \cref{eq:B-eq,eq:B-eq-alpha} reduces to
\be
X^a{}_i\equiv (D\times B)^a{}_i - \alpha B^a{}_i =0.
\label{eq:Bar---}
\ee
Up to this point no approximation beyond (A1) and (A2) has been made.

\subsection{Statistical description of the dressing}
\label{sec:spin2-ordering}

This subsection makes (A1), (A3) and (A4) precise.

\paragraph{Quasi-static.}
In temporal gauge, $\bar A_0^a=0$, assumption (A1) means
\be
|\partial_t W^a{}_i|\ll |\partial_j W^a{}_i| ,
\label{eq:quasistatic-condition}
\ee
so that $E^a_i=\partial_tW^a{}_i$ is subleading and the leading background
equation is \cref{eq:Bar---}. If subleading time dependence is retained, the
Gauss constraint must be imposed consistently at the same order.

\paragraph{Mode decomposition and ensemble.}
We expand $W^a{}_i$ in transverse-traceless helicity-$2$ modes,
\be
W^a{}_{i}(\bx) =
\sum_{\lambda=\pm}
\int d^3 k \, \delta^{aj} \,
e^\lambda_{ij}(\hat \bk) \,
W_\lambda(\bk) \,
e^{i\bk\cdot\bx},
\label{eq:W-decompose}
\ee
where $e^\lambda_{ij}(\hat{\bk})$ are helicity-$2$ polarization tensors normalized as
$e^\lambda_{ij}(\hat{\bk})\,e^{\lambda'}_{ij}(-\hat{\bk})=\delta_{\lambda\lambda'}$;
further identities are collected in \cref{app:Math}. The amplitudes
$W_\lambda(\bk)$ are stochastic variables. Statistical homogeneity and
isotropy are encoded in
\be
\langle W_\lambda(\bk)\rangle_{\rm st} =0, \qquad
\langle W_\lambda(\bk) W^{*}_{\lambda'}(\bk')\rangle_{\rm st}
= \delta_{\lambda\lambda'} \, \delta^{(3)}(\bk-\bk') \, P_{\lambda}(k),
\label{eq:W-power}
\ee
where $\langle \dots \rangle_{\rm st}$ is an average over classical
realizations, not a quantum expectation value, and $P_\lambda(k)$ is the power
in helicity $\lambda$. Hence
\be
\langle \tr[W^2]\rangle_{\rm st}
=
\sum_{\lambda=\pm}\int d^3k\,P_\lambda(k).
\ee
Beyond homogeneity and isotropy, the radial spectrum and the helicity content
are left unrestricted; they are fixed by the field equations below.

\paragraph{Classicality.}
A mode with occupation $n_\lambda(k)$ carries
$P_\lambda(k)=[n_\lambda(k)+\tfrac12]/[(2\pi)^3\omega_\lambda(k)]$. Assumption
(A3), $n_\lambda(k)\gg1$, is therefore the statement
\be
P_{\lambda}(k)
\gg
P^{\rm vac}_{\lambda}(k)=\frac{1}{(2\pi)^3\,2\omega_\lambda(k)},
\ee
under which the vacuum contribution to the correlators of $W$ is negligible
and $W$ may be treated as a classical stochastic field.

\begin{figure}[t]
\centering
\includegraphics[width=0.38\textwidth]{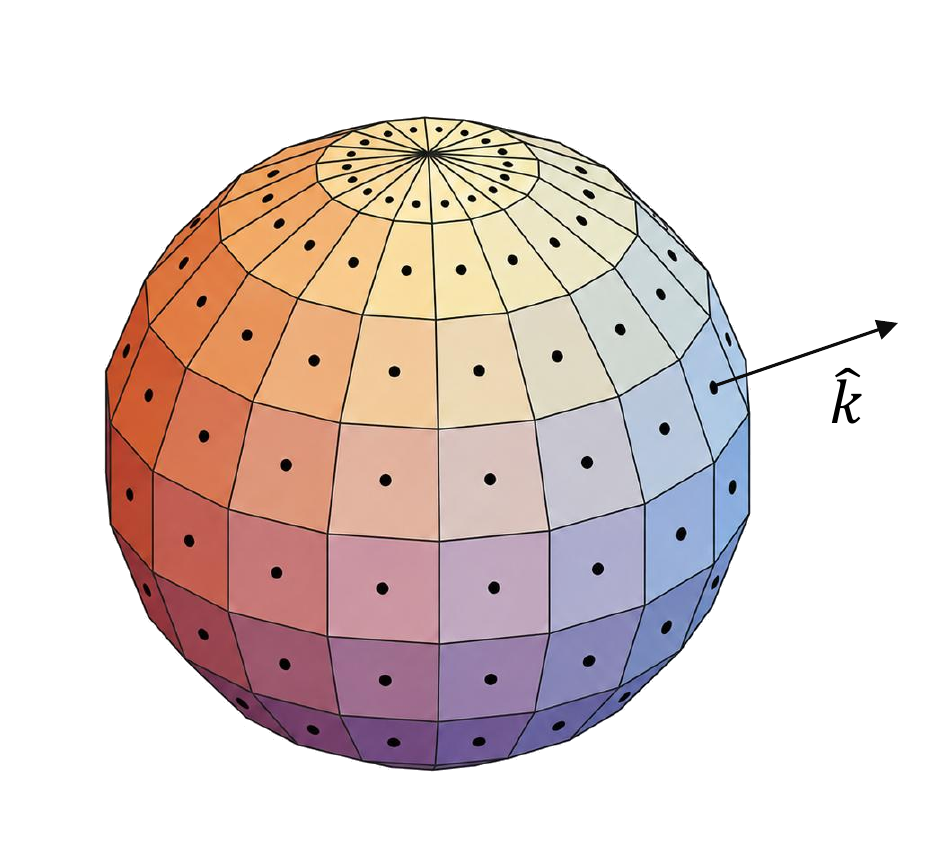}
\caption{Discretization of the momentum-space sphere into angular cells
$\Delta\Omega_I$, centred on directions $\hat{\bk}_I$.}
\label{fig:angular-cells}
\end{figure}

\paragraph{Angular self-averaging.}
Statistical isotropy is an ensemble property, whereas only one realization is
physically available. Assumption (A4) replaces angular averages within that
realization by ensemble averages when sufficiently many independent directions
are populated. Discretizing the momentum-space sphere into cells $\Delta\Omega_I$
(\cref{fig:angular-cells}), we define
\be
p_{\lambda,I}(k)
\equiv
\frac{(2\pi)^3}{\mathrm{Vol}}\,k^2
\int_{\Delta\Omega_I}d\Omega_{\hat\bk}\,
|W_\lambda(k\hat\bk)|^2 ,
\qquad
\langle p_{\lambda,I}(k)\rangle_{\rm st}
=
k^2\Delta\Omega_I\,P_\lambda(k).
\label{eq:helicity-angular-cell-power-density}
\ee

With
$p_I\equiv\sum_{\lambda}\int_{\mathcal B}dk\,p_{\lambda,I}(k)$,
the effective number of populated directions is
\be
N_{\Omega}^{\rm eff}
\equiv
\frac{\big(\sum_Ip_I\big)^2}{\sum_Ip_I^2},
\qquad
1\le N_{\Omega}^{\rm eff}\le N_\Omega .
\label{eq:angular-Neff}
\ee
For mode-resolved quantities we similarly define
\be
N_{\Omega}^{\rm eff}(k)
\equiv
\frac{\big[\sum_Ip_I(k)\big]^2}{\sum_Ip_I^2(k)},
\qquad
p_I(k)\equiv\sum_\lambda p_{\lambda,I}(k),
\label{eq:angular-Neff-k}
\ee
and require
$N_{\Omega}^{\rm eff}(k)\gg1$ for
$k\in\mathcal B_{\rm inst}$. Provided the populated directions also sample the sphere approximately
uniformly, angular averages are self-averaging:
\be
\overline{\Phi}_{\mathsf A,\Omega}
=
\langle\Phi_{\mathsf A}\rangle_{\rm st}
+
\mathcal O_{\rm rms}\!\left(
\frac{\sigma_{\Phi_{\mathsf A}}}
{\sqrt{N_\Omega^{\rm eff}}}
\right).
\ee
Thus (A4) holds up to corrections suppressed by
$(N_\Omega^{\rm eff})^{-1/2}$. We further assume that correlations between distinct angular cells
are sufficiently weak for the variance of their weighted average
to scale as $1/N_\Omega^{\rm eff}$. This mixing assumption,
together with $N_\Omega^{\rm eff}\gg 1$, ensures self-averaging.

\subsection{Projected field equations}

We now impose \cref{eq:Bar---} in the form dictated by (A4). Because color
and spatial indices are locked by the background, the natural decomposition
is into irreducible representations of the diagonal ${\rm SO}(3)$,
${\bf 3}\otimes{\bf 3} = {\bf 1}\oplus{\bf 3}\oplus{\bf 5}$:
\be
X^a{}_i = X_{\rm S}\,\delta^a_i + X^a{}_{i,{\rm V}} + X^a{}_{i,{\rm T}},
\ee
with
\be
X_{\rm S} \equiv \frac13 X^k{}_k, \qquad
X^a{}_{i,{\rm V}} \equiv \frac12\left(X^a{}_i-X^i{}_a\right), \qquad
X^a{}_{i,{\rm T}} \equiv \frac12\left(X^a{}_i+X^i{}_a\right) -\frac13\delta^a_i X^k{}_k .
\ee
Each sector is treated in turn. Only the scalar sector and the helicity-$2$
part of the tensor sector have a nonvanishing ensemble mean; these are the
equations we impose.

\paragraph{Scalar sector.}
The trace of the field equation, $X_{\rm S}/g=0$, is exact:
\be
(2gQ - \alpha) Q^2 + \frac{\alpha}{6} \tr[W^2] -\frac{g}{3}\tr[W^3] +\frac{1}{3}\epsilon_{abc} W^b{}_j \partial^a W^{c\,j}  =0. \label{eq:scalar-trace-exact}
\ee
Pointwise, its $W$-dependent terms fluctuate from place to place. By (A4) we
impose its angular average, which for $N_\Omega^{\rm eff}\gg1$ equals the
ensemble average. Since $Q$ and $\alpha$ are direction-independent, only the
$W$-dependent terms are affected. They are captured by the invariant moments
\be
\mathcal{W}_2 \equiv \langle \tr[W^2] \rangle_{\rm st}, \qquad
\mathcal{W}_3 \equiv \langle \tr[W^3] \rangle_{\rm st},
\ee
and the parity-odd helicity moment
\be
H_W \equiv \langle \epsilon_{abc} W^b{}_j\partial^a W^{c\,j} \rangle_{\rm st}
=-\big\langle \tr[W\,(\nabla\times W)]\big\rangle_{\rm st},
\qquad
(\nabla\times W)^a{}_i\equiv\epsilon_{ijk}\partial^jW^{ak}.
\ee
The cubic shape parameter and helicity wavenumber can be written as
\be
\begin{aligned}
b_W
&=
\frac{1}{\mathcal W_2^{3/2}}
\sum_{\lambda,\lambda',\lambda''}
\int_{\mathbf k,\mathbf p}
\mathcal D_{\lambda\lambda'\lambda''}
(\mathbf k,\mathbf p,\mathbf k-\mathbf p)\,
B_{\lambda\lambda'\lambda''}
(\mathbf k,\mathbf p,\mathbf k-\mathbf p),
\\
h_W
&=
-\frac{1}{\mathcal W_2}
\sum_\lambda
\int_{\mathbf k}\lambda k\,P_\lambda(k),
\end{aligned}
\ee
where $\mathcal D_{\lambda\lambda'\lambda''}$ is the
helicity-dependent polarization overlap,
\begin{equation}
\mathcal D_{\lambda\lambda'\lambda''}(\mathbf k,\mathbf p,\mathbf q)
=
\operatorname{tr}\left[
e^{\lambda *}(\hat{\mathbf k})
e^{\lambda'}(\hat{\mathbf p})
e^{\lambda''}(\hat{\mathbf q})
\right],
\end{equation}
and $B_{\lambda\lambda'\lambda''}$ denotes the tensor bispectrum,
\begin{equation}
B_{\lambda\lambda'\lambda''}
(\mathbf k,\mathbf p,\mathbf q)
\equiv
\left\langle
W_\lambda^*(\mathbf k)
W_{\lambda'}(\mathbf p)
W_{\lambda''}(\mathbf q)
\right\rangle'.
\label{eq:b-W----}
\end{equation}
Using the above, the scalar condition reads
\be
(2gQ - \alpha) Q^2 + \left(\frac{\alpha}{6}  + \frac13 h_W\right) \mathcal{W}_2 -\frac{g}{3}\,b_W \mathcal{W}_2^{3/2}  =0.
\label{eq:scalar-trace}
\ee
It ties the condensate $Q$ to the amplitude, cubic structure and helicity
content of the dressing.

\paragraph{Vector sector.}
The antisymmetric part is conveniently written through its dual,
\begin{equation}
 X_{{\rm V},\ell}\equiv
\frac12\epsilon_{\ell ai}X^a{}_i
=-\frac{g}{2}\partial_j(W^2)_{j\ell}.
\label{eq:vector-source-corrected}
\end{equation}
Being a total derivative, its ensemble mean vanishes by statistical
homogeneity, and its average over regions large compared with the correlation
length $\ell_W$ of $W$ is suppressed by surface terms. Pointwise, however, it
would require $\partial_j(W^2)_{j\ell}=0$, which a generic superposition of
transverse-traceless modes does not satisfy. In accordance with (A4), the
vector equation is imposed only after coarse-graining. This is the first
place where the closure departs from an exact solution.

\paragraph{Tensor sector.}
Identifying color and spatial indices through the diagonal background, the
symmetric-traceless part is exactly
\be
\begin{aligned}
X^a{}_{i,{\rm T}}
=\;&\Big(-\nabla^2+\frac12g^2\mathcal{W}_2+\alpha g Q\Big)\, W^a{}_i
-(2gQ +\alpha) \,\epsilon_{ijk}\partial^j W^{a\,k}\\
&+g\, U^a{}_i-(3g^2Q+\alpha g)\,  Z^a{}_i + \frac12 g^2 V^a{}_i
=0,
\end{aligned}
\label{eq:tensor-field}
\ee
with the traceless combinations\footnote{The combination
$\tr[W^2]-\mathcal W_2$ in $V$ is the centred fluctuation of the local
spin-$2$ intensity. The mean $\langle V^a{}_i\rangle_{\rm st}=\langle
\tr[W^2]\,W^a{}_i\rangle_{\rm st}$ vanishes by statistical isotropy, since
there is no invariant symmetric-traceless rank-$2$ tensor.}
\begin{align}
U^a{}_i & = Y^a{}_i + Y_i{}^a
-\frac{2}{3} \delta^a_i\tr[Y] ,\\
V^a{}_i & = \big(\tr[W^2] - \mathcal{W}_2\big) W^a{}_i, \label{eq:def-V}\\
Z^a{}_i &=
(W^2)^a{}_i
-\frac13\delta^a_i \tr[W^2] ,
\end{align}
and
\be
Y^a{}_i \equiv
\frac12 \epsilon^{abc}
\left(
W^b{}_j\partial_iW^c{}_j
-
2W^b{}_j\partial_jW^c{}_i
\right).
\ee
The nonlinear terms $U$, $V$ and $Z$ are symmetric and traceless but not
transverse, so \cref{eq:tensor-field} has helicity-$0$, $\pm1$ and $\pm2$
components. The helicity-$0,\pm1$ components have vanishing one-point mean by
statistical isotropy but are not zero realization by realization; as with the
vector sector, (A4) imposes them only on average. The helicity-$2$ component
is the one that carries the spectrum, and we now impose it.

\paragraph{Helicity-$2$ projection.}
With the helicity components of a symmetric tensor $O^a{}_i$ defined by
$O_\lambda(\bk)\equiv [e^\lambda_{ij}(\hat\bk)]^*\,\delta_{aj}O^a{}_i(\bk)$,
in the Fourier convention of \cref{eq:W-decompose}, the exact mode equation is
\be
\begin{aligned}
\Big[
k^2
&+\frac12g^2\mathcal W_2
+\alpha gQ
-\lambda(2gQ+\alpha)k
\Big]
W_\lambda(\bk)
+gU_\lambda(\bk)-(3g^2Q+\alpha g)Z_\lambda(\bk)
\\
&+\frac12g^2V_\lambda(\bk)
=0.
\end{aligned}
\label{eq:tensor-helicity-exact}
\ee
Following (A4), we project onto the two-point level: multiply by
$W^*_\lambda(\bk)$ and average. Every term then carries the factor
$\delta^{(3)}(\mathbf 0)=\mathrm{Vol}/(2\pi)^3$; writing
\be
\langle W^*_\lambda(\bk)O_\lambda(\bk)\rangle_{\rm st}\equiv\delta^{(3)}(\mathbf 0)\,\langle W^*_\lambda O_\lambda\rangle'_{\rm st}(k)
\ee
and removing it, we find
\be
\begin{aligned}
\Big[
k^2
&+\frac12g^2\mathcal W_2
+\alpha gQ
-\lambda(2gQ+\alpha)k
\Big]
P_{\lambda}(k)
+g \langle W^*_\lambda U_\lambda\rangle'_{\rm st}
-(3g^2Q+\alpha g) \langle W^*_\lambda Z_\lambda\rangle'_{\rm st}
\\
&+\frac12g^2 \langle W^*_\lambda V_\lambda\rangle'_{\rm st}
=0.
\end{aligned}
\label{eq:tensor-helicity-averaged}
\ee
The nonlinear terms enter only through three cross-spectra as
\begin{equation}
\label{eq:UVZ}
\begin{aligned}
g\,U_\lambda W_\lambda^*
&= \DiagramUW
\\[1em]
\frac{g^2}{2}\,V_\lambda W_\lambda^*
&= \DiagramVW - \DiagramVTadpole
\\[1em]
\bigl(3g^2Q + \alpha g\bigr)\,Z_\lambda W_\lambda^*
&= \DiagramZQ + \DiagramZphi,
\end{aligned}
\end{equation}
in which the lines and vertices
are represented as follows:
\begin{equation*}
\begin{aligned}
\LegendW \quad &: \quad
\text{stochastic tensor field } W_i^{\,a},
\\[0.5em]
\LegendQ \quad &: \quad
\text{homogeneous gauge background } Q\delta_i^{\,a},
\\[0.5em]
\LegendAxion \quad &: \quad
\text{rolling axion background } \dot\phi,
\\[0.5em]
\LegendYM\,,\ \LegendQuartic \quad &: \quad
\text{Yang--Mills vertices},
\\[0.5em]
\LegendCS \quad &: \quad
\text{axion Chern--Simons vertex},
\end{aligned}
\end{equation*}
and a small open circle marks a derivative coupling.
The nonlinear cross-spectra can be expressed in terms of the
tensor bispectrum and four-point function. The quadratic
interactions $U$ and $Z$ give
\begin{align}
\langle W_\lambda^*(\mathbf k)U_\lambda(\mathbf k)\rangle'
&=
\sum_{\lambda',\lambda''}
\int_{\mathbf p}
(\lambda k+\lambda'p+\lambda''q)\,
\mathcal D_{\lambda\lambda'\lambda''}
(\mathbf k,\mathbf p,\mathbf q)\,
B_{\lambda\lambda'\lambda''}
(\mathbf k,\mathbf p,\mathbf q),
\\
\langle W_\lambda^*(\mathbf k)Z_\lambda(\mathbf k)\rangle'
&=
\sum_{\lambda',\lambda''}
\int_{\mathbf p}
\mathcal D_{\lambda\lambda'\lambda''}
(\mathbf k,\mathbf p,\mathbf q)\,
B_{\lambda\lambda'\lambda''}
(\mathbf k,\mathbf p,\mathbf q),
\qquad \mathbf q=\mathbf k-\mathbf p.
\end{align}
The quartic Yang--Mills cross-spectrum reads
\begin{align}
\langle W_\lambda^*(\mathbf k)V_\lambda(\mathbf k)\rangle'
&=
\frac{2}{5}\mathcal W_2 P_\lambda(k)
+
\sum_{\lambda',\lambda'',\lambda'''}
\int_{\mathbf p,\mathbf q}
\mathcal K_{\lambda\lambda'\lambda''\lambda'''}
(\mathbf k,\mathbf p,\mathbf q,\mathbf s)\,
T^c_{\lambda\lambda'\lambda''\lambda'''}
(\mathbf k,\mathbf p,\mathbf q,\mathbf s),
\label{eq:response-V--}
\end{align}
where $\mathbf s=\mathbf k-\mathbf p-\mathbf q$.
The two terms are the Gaussian (\cref{app:trispectrum-gaussian})  and connected non-Gaussian contributions, respectively. The quartic Yang--Mills polarization kernel is
\begin{equation}
\mathcal K_{\lambda\lambda'\lambda''\lambda'''}
=
\operatorname{tr}\left[
e^{\lambda'}(\hat{\mathbf p})
e^{\lambda''}(\hat{\mathbf q})
\right]
\operatorname{tr}\left[
e^{\lambda *}(\hat{\mathbf k})
e^{\lambda'''}(\hat{\mathbf s})
\right],
\label{eq:math K}
\end{equation}
and the connected tensor trispectrum is defined by
\begin{equation}
T^c_{\lambda\lambda'\lambda''\lambda'''}
(\mathbf k,\mathbf p,\mathbf q,\mathbf s)
\equiv
\left\langle
W_\lambda^*(\mathbf k)
W_{\lambda'}(\mathbf p)
W_{\lambda''}(\mathbf q)
W_{\lambda'''}(\mathbf s)
\right\rangle'_c.
\label{eq:T--} 
\end{equation}
All helicity combinations are retained. We can now parametrize them by response functions,
\be
\begin{aligned}
\langle W^*_\lambda Z_\lambda\rangle'_{\rm st}
&=\mathcal R^{(3)}_\lambda(k)\,P_\lambda(k),\\
\langle W^*_\lambda U_\lambda\rangle'_{\rm st}
&=\Big[\lambda k\,\mathcal R^{(3)}_\lambda(k)
      +\tilde{\mathcal R}^{(3)}_\lambda(k)\Big]P_\lambda(k),\\
\langle W^*_\lambda V_\lambda\rangle'_{\rm st}
&=\Big[\frac25\,\mathcal W_2+\mathcal R^{(4)}_\lambda(k)\Big]P_\lambda(k).
\end{aligned}
\label{eq:response-exact}
\ee
Using the above, the cubic shape
parameter in \cref{eq:scalar-trace} reads
\be
b_W
=
\frac{1}{\mathcal W_2^{3/2}}
\sum_\lambda\int_{\mathbf k}
P_\lambda(k)\,\mathcal R^{(3)}_\lambda(k).
\ee

\paragraph{Spectral condition.}
Substituting \cref{eq:response-exact} into \cref{eq:tensor-helicity-averaged}
and dividing by $P_\lambda(k)$, every $k$ in the support of $P_\lambda$ must
satisfy
\begin{equation}
k^2
-\lambda\big(2gQ+\alpha-g\mathcal R^{(3)}_\lambda(k)\big)k
+\tfrac{7}{10}\,g^2\mathcal W_2
+\alpha gQ
-g(3gQ+\alpha)\mathcal R^{(3)}_\lambda(k)
+g\tilde{\mathcal R}^{(3)}_\lambda(k)
+\tfrac12\,g^2\mathcal R^{(4)}_\lambda(k)
=0,
\label{eq:tensor-dispersion-responses}
\end{equation}
with the responses evaluated at $k$. Besides, the scalar condition
\cref{eq:scalar-trace} can be written as
\be
(2gQ - \alpha) Q^2 + \frac{\alpha}{6}  \mathcal{W}_2  - \frac13 \sum_\lambda
\int_{\mathbf k}\, \big(\lambda k + g \, \mathcal R^{(3)}_\lambda(k)\big)\,P_\lambda(k) =0.
\label{eq:scalar-trace---}
\ee
They make a set of coarse-grained background
equations. Their solutions are the candidate states of the first requirement,
classical self-consistency. The remainder of this section solves them, first
for the simplest spectrum and then for broad spectra.

Throughout the rest of this section we use the dimensionless variables
\be
\chi\equiv\frac{2gQ}{\alpha},
\qquad
\chi_W\equiv\frac{g\mathcal W_2^{1/2}}{\alpha},
\qquad
\chi_k\equiv\frac{\lambda k}{\alpha}.
\label{eq:ss-dimless}
\ee
The background equations are invariant under
$(\alpha,Q,b_W,\lambda)\to(-\alpha,-Q,-b_W,-\lambda)$ at fixed $k$,
so we take $\alpha>0$ without loss of generality. Then
$k=\alpha|\chi_k|$ and $\lambda=\operatorname{sgn}\chi_k$.

\subsection{Thin-shell solution}
\label{sec:thin-shell}

The simplest spectrum places all stochastic power on a single thin shell of
radius $k_\star>0$ and helicity $\lambda_\star=\pm1$,
\be
P_\lambda(k)=
\delta_{\lambda\lambda_\star}\,
\frac{\mathcal W_2}{4\pi k_\star^2}\,
\delta(k-k_\star),
\qquad
\sum_\lambda\int d^3k\,P_\lambda(k)=\mathcal W_2 ,
\label{eq:ss-spectrum}
\ee
where $k_\star$ and $\lambda_\star$ are determined by the field equations.

Since all populated Fourier modes have the same curl eigenvalue
$\lambda_\star k_\star$,
\be
H_W
=
-\big\langle\tr[W(\nabla\times W)]\big\rangle_{\rm st}
=
-\lambda_\star k_\star\,\mathcal W_2,
\qquad
h_W=-\lambda_\star k_\star .
\label{eq:ss-hW}
\ee
The response functions therefore reduce to constants
(App.~\ref{app:bispectrum-responses}),
\be
\mathcal R^{(3)}_{\lambda_\star}(k_\star)
=b_W\mathcal W_2^{1/2},
\qquad
\widetilde{\mathcal R}^{(3)}_{\lambda_\star}(k_\star)
=2\lambda_\star k_\star b_W\mathcal W_2^{1/2},
\qquad
\mathcal R^{(4)}_{\lambda_\star}(k_\star)
=\tau_W\mathcal W_2 .
\label{eq:ss-responses}
\ee

\paragraph{Tensor condition.}
Inserting \cref{eq:ss-responses} into
\cref{eq:tensor-dispersion-responses} at
$(k,\lambda)=(k_\star,\lambda_\star)$ gives
\be
k_\star^2
-\lambda_\star
\big(2gQ+\alpha-3g b_W\mathcal W_2^{1/2}\big)k_\star
+\alpha gQ
+\frac12g^2\left(\frac75+\tau_W\right)\mathcal W_2
-g b_W\mathcal W_2^{1/2}(3gQ+\alpha)
=0 .
\label{eq:ss-tensor}
\ee

\paragraph{Scalar condition.}
Using \cref{eq:ss-hW}, the scalar equation becomes
\be
k_\star
=
\lambda_\star
\left[
\frac{\alpha}{2}
-gb_W\mathcal W_2^{1/2}
+\frac{3Q^2(2gQ-\alpha)}{\mathcal W_2}
\right].
\label{eq:ss-scalar-k}
\ee
In dimensionless form this fixes the signed shell momentum,
\be
\chi_{k_\star}
\equiv
\frac{\lambda_\star k_\star}{\alpha}
=
\frac12-b_W\chi_W
+\frac{3\chi^2(\chi-1)}{4\chi_W^2}.
\label{eq:ss-sigma}
\ee

The tensor condition can be written as
\be
F(\chi_{k_\star})=0,
\label{eq:ss-F-shell}
\ee
where
\be
F(\chi_k)
=
\chi_k^2
-\big(1+\chi-3b_W\chi_W\big)\chi_k
+\frac{\chi}{2}
+\frac12\left(\frac75+\tau_W\right)\chi_W^2
-b_W\chi_W\left(\frac{3\chi}{2}+1\right).
\label{eq:ss-F}
\ee
Thus \cref{eq:ss-sigma,eq:ss-F-shell} determine the thin-shell solutions.
A physical shell requires $\chi_{k_\star}\neq0$, with
\be
k_\star=\alpha|\chi_{k_\star}|,
\qquad
\lambda_\star=\operatorname{sgn}\chi_{k_\star} .
\ee

\subsection{Broad spectra}
\label{sec:broad-spectrum-W}

For a general spectrum it is useful to introduce the normalized power density
in signed wavenumber,
\be
\rho(\chi_k)
\equiv
\frac{1}{\mathcal W_2}
\sum_{\lambda=\pm}
\int d^3k\,
P_\lambda(k)\,
\delta\left(\chi_k-\frac{\lambda k}{\alpha}\right),
\qquad
\int d\chi_k\,\rho(\chi_k)=1 .
\label{eq:broad-rho-def}
\ee
Thus $\rho(\chi_k)d\chi_k$ is the fraction of the total stochastic power
carried by modes with signed wavenumber in
$[\chi_k,\chi_k+d\chi_k]$. Power-weighted averages are then denoted by an
overbar,
\be
\bar\chi_k
=
\int d\chi_k\,\rho(\chi_k)\chi_k,
\qquad
\overline{\chi_k^2}
=
\int d\chi_k\,\rho(\chi_k)\chi_k^2 ,
\label{eq:broad-sigma-moments}
\ee
and the spectral variance is
\be
\rm Var(\chi_k)
\equiv
\overline{\chi_k^2}-\bar\chi_k^2
\ge0 .
\label{eq:broad-sigma-variance}
\ee

Since $h_W=-\alpha\bar\chi_k$, the scalar condition fixes the mean signed
wavenumber,
\be
\bar\chi_k
=
\frac12-b_W\chi_W
+\frac{3\chi^2(\chi-1)}{4\chi_W^2}.
\label{eq:broad-sigma}
\ee

Averaging the tensor equation over the spectrum gives
\be
0
=
\overline{\chi_k^2}
-\big(1+\chi-3b_W\chi_W\big)\bar\chi_k
+\frac{\chi}{2}
+\frac12\left(\frac75+\tau_W\right)\chi_W^2
-b_W\chi_W\left(\frac{3\chi}{2}+1\right),
\ee
where we have made the assumption that the cubic response retains its thin-shell momentum dependence mode by mode across the
support of the spectrum, so that the derivative-weighted cubic terms are fixed
by $b_W$ rather than only their power-weighted averages. Using $\overline{\chi_k^2}=\bar\chi_k^2+\rm Var(\chi_k)$, this becomes
\be
\rm Var(\chi_k)=-F(\bar\chi_k).
\label{eq:broad-width-F}
\ee
Thus $\rm Var(\chi_k)=0$ reproduces the thin-shell condition, while
$\rm Var(\chi_k)>0$ requires a spectrum with nonzero width.

\paragraph{Gaussian closure.}
For a Gaussian dressing all connected responses vanish, so
$b_W=\tau_W=0$ and every populated mode obeys
\be
\chi_k^2-(1+\chi)\chi_k
+\frac{\chi}{2}
+\frac{7}{10}\chi_W^2=0 .
\label{eq:broad-gaussian-quadratic}
\ee
The spectrum can therefore contain at most the two shells
\be
\chi_{k_\pm}
=
\frac{
1+\chi
\pm
\sqrt{1+\chi^2-\frac{14}{5}\chi_W^2}
}{2},
\label{eq:broad-two-roots}
\ee
which exist provided
\be
1+\chi^2-\frac{14}{5}\chi_W^2\ge0 .
\ee
If both are populated, their fractions are fixed by the mean,
\be
f_+
=
\frac{\bar\chi_k-\chi_{k_-}}{\chi_{k_+}-\chi_{k_-}},
\qquad
f_-=1-f_+,
\label{eq:broad-two-shell-weights}
\ee
with $0\le f_\pm\le1$. A single shell is recovered when one of the fractions
vanishes.

\paragraph{Continuous spectra.}
A continuous spectrum requires a momentum-dependent nonlinear response.
We retain the cubic response parameterized by $b_W$ and allow the connected
quartic response to depend on $\chi_k$. For this purpose, define the quadratic
with the spectrum-averaged quartic contribution removed,
\be
F_0(\chi_k)
\equiv
F(\chi_k)\big|_{\tau_W=0}
=
F(\chi_k)-\frac12\tau_W\chi_W^2 ,
\label{eq:broad-F0}
\ee
or explicitly,
\be
F_0(\chi_k)
=
\chi_k^2
-\big(1+\chi-3b_W\chi_W\big)\chi_k
+\frac{\chi}{2}
+\frac{7}{10}\chi_W^2
-b_W\chi_W\left(\frac{3\chi}{2}+1\right).
\ee
The tensor equation is then satisfied at every populated wavenumber by
choosing
\be
\frac{g^2}{\alpha^2}\,
\mathcal R^{(4)}(\chi_k)
=
-2F_0(\chi_k).
\label{eq:broad-R4-profile}
\ee
Its spectrum average must reproduce the connected fourth moment,
\be
\tau_W
=
\frac{1}{\chi_W^2}\,
\overline{
\frac{g^2}{\alpha^2}\mathcal R^{(4)}(\chi_k)
}.
\label{eq:broad-tau-average}
\ee
Since $F_0$ is a monic quadratic,
\be
\overline{F_0(\chi_k)}
=
F_0(\bar\chi_k)+\rm Var(\chi_k) ,
\ee
so averaging \cref{eq:broad-R4-profile} reproduces
\cref{eq:broad-width-F}.

An explicit continuous realization is obtained by choosing the signed-wavenumber
power density to be uniform over a finite interval,
\be
\rho(\chi_k)
=
\frac{1}{2h},
\qquad
\chi_k\in[\bar\chi_k-h,\bar\chi_k+h],
\qquad
h=\sqrt{3\rm Var(\chi_k)},
\label{eq:broad-uniform-band}
\ee
with $\rho(\chi_k)=0$ outside the band. Its mean and variance are
$\bar\chi_k$ and $\rm Var(\chi_k)$ by construction. The band endpoints need not
coincide with the roots of $F_0$, so
$\mathcal R^{(4)}(\chi_k)$ may have either sign on the populated region,
allowing negative, zero, or positive $\tau_W$.

This construction fixes the projection of the connected four-point function
required by the background equations, but not the full connected trispectrum
entering the scattering corrections considered below.

\subsection{Consistency bounds}
\label{sec:consistency-bounds}
The coarse-grained background equations require $\rm Var(\chi_k)\ge0$. We therefore
define the background region by
\be
\mathcal{R}_{\rm bg}(b_W,\tau_W)
=
\left\{
(\chi,\chi_W):
\chi_W>0,\;
\rm Var(\chi_k)\ge0
\right\}, \quad \text{(background region),}
\label{eq:allowed-region}
\ee
where $\bar\chi_k$ and $\rm Var(\chi_k)$ are fixed by
\cref{eq:broad-sigma,eq:broad-width-F}. The boundary $\rm Var(\chi_k)=0$ gives the
thin-shell solutions, while $\rm Var(\chi_k)>0$ admits spectra with finite width,
such as the continuous band \cref{eq:broad-uniform-band}.

The stochastic moments must in addition be realizable. Defining
\be
S\equiv\tr W^2,
\qquad
T\equiv\tr W^3 ,
\ee
their relevant moments are
\be
\langle S\rangle=\mathcal W_2,
\qquad
\langle S^2\rangle
=
\left(\frac75+\tau_W\right)\mathcal W_2^2,
\qquad
\langle T\rangle
=
b_W\mathcal W_2^{3/2}.
\ee
Positivity of the variance of $S$ gives
\be
\operatorname{Var}(S)
=
\left(\frac25+\tau_W\right)\mathcal W_2^2
\ge0,
\ee
and therefore
\be
\tau_W\ge-\frac25 .
\label{eq:tau-bound}
\ee
The pointwise inequality $6T^2\le S^3$, together with Cauchy--Schwarz, gives
the second bound
\be
6b_W^2\le\frac75+\tau_W .
\label{eq:ss-moment-bound-corrected}
\ee
These conditions depend only on the local stochastic moments and are
independent of the radial shape of the spectrum.

\section{One-Loop Effective Action and Hartree Stability}
\label{sec:quantum-structure-Hartree}

In this section and the next we ask whether the spin-2-dressed backgrounds
$\bar A^a_\mu$ of \cref{sec:stochastic-spin2-background} are stable against
quantum fluctuations. Solving the (coarse-grained) field equations makes the
background a stationary point of the action, not necessarily a minimum.
Stability is decided by the spectrum of fluctuations, and the quantity that
records it is the imaginary part of the effective action. We first write this
quantity in full for a static stochastic background. We then split it into a
mean-field (Hartree) part and a scattering part, which are treated in this
section and in \cref{sec:scattering} respectively.

We decompose the gauge field into a stationary background and a quantum fluctuation,
\begin{equation}
A_\mu^a=\bar A_\mu^a+\mathcal A_\mu^a,
\label{eq:A-dec}
\end{equation}
with $\bar A_\mu^a$ satisfying the background field equations, so that the action contains no term linear in $\mathcal A_\mu^a$,
and
\be
S[\bar A+\mathcal A]
=S_{\rm cl}[\bar A]
+\frac12\int d^4x\,\mathcal A^a_\mu\,\Delta^{\mu\nu}_{ab}[\bar A]\,\mathcal A^b_\nu
+\mathcal O(\mathcal A^3),
\label{eq:quadratic-expansion}
\ee
where $\Delta[\bar A]$ is the quadratic fluctuation operator in background-field
gauge (\cref{subsec:one-loop-formalism})\footnote{The background is taken to be on-shell with respect to the coarse-grained effective action defined by the closure of \cref{sec:stochastic-spin2-background}; accordingly, the linear term in the fluctuation expansion vanishes at the retained order.}. For simplicity, we have neglected quantum fluctuations of the axion and restricted the one-loop analysis to the gauge sector. For the static, purely magnetic backgrounds considered here, axion--spatial-gauge mixing vanishes at zero frequency; we therefore do not expect this omission to alter our static gauge-sector gap bounds. Integrating out $\mathcal A$ and the
ghosts gives
\be
\Gamma_{\rm eff}[\bar A]=S_{\rm cl}[\bar A]+\Gamma_{\rm 1\text{-}loop}[\bar A]+\mathcal O(\hbar^2),
\qquad
\Gamma_{\rm 1\text{-}loop}[\bar A]=\frac{i}{2}{\rm Tr}\ln\Delta[\bar A]-i\,{\rm Tr}\ln\Delta_{\rm gh}[\bar A],
\label{eq:one-loop-effective-action}
\ee
with counterterms understood (\cref{app:1-loop-action}). The real part of
$\Gamma_{\rm 1\text{-}loop}$ describes the vacuum's dispersive response: it
renormalizes the couplings and shifts the saddle. The imaginary part measures
the loss of vacuum persistence,
\be
P_{\rm vac}[\bar A]=\big|\langle0_{\rm out}|0_{\rm in}\rangle_{\bar A}\big|^2
=\exp\!\big[-2\,{\rm Im}\,\Gamma_{\rm 1\text{-}loop}[\bar A]\big].
\label{eq:vac-persistence}
\ee

\paragraph{Hartree and scattering contributions.}
The dressing enters $\Delta$ in two ways. Its local
intensity appears through terms quadratic in $W$ without derivatives; its
Fourier components appear through a term linear in $W$, which transfers
momentum. We separate the ensemble-averaged operator from its zero-mean stochastic fluctuation,
\begin{equation}
\Delta[\bar A]
=
\langle\Delta[\bar A]\rangle_{\rm st}
+\delta\Delta[\bar A],
\qquad
\langle\delta\Delta[\bar A]\rangle_{\rm st}=0.
\label{eq:deltaK}
\end{equation}
Although $\delta\Delta$ has vanishing ensemble mean, its stochastic fluctuations are nontrivial, i.e.
\begin{equation}
\left\langle
\delta\Delta[\bar A]\,
\delta\Delta[\bar A]
\right\rangle_{\rm st}
\neq0,
\end{equation}
and this two-point correlator controls the leading scattering corrections beyond the ensemble-averaged (Hartree) operator, to which we return in \cref{sec:scattering}.
The ensemble-averaged one-loop effective action naturally separates into
\begin{equation}
\langle\Gamma_{\rm 1\text{-}loop}\rangle_{\rm st}
=
\Gamma_{\rm H}
+
\Gamma_{\rm scat},
\label{eq:Gamma-decomposition}
\end{equation}
where $\Gamma_{\rm H}$ and $\Gamma_{\rm scat}$ are, respectively, the Hartree contribution from the ensemble-averaged quadratic operator and the scattering contribution generated by its stochastic fluctuations, i.e.
\begin{equation}
\Gamma_{\rm H}
=
\frac{i}{2}{\rm Tr}\ln\langle\Delta\rangle_{\rm st}
-i\,{\rm Tr}\ln\langle\Delta_{\rm gh}\rangle_{\rm st},
\qquad
\Gamma_{\rm scat}
=
\frac{i}{2}
\Big\langle
{\rm Tr}\ln\!\left(1+\bar G\,\delta\Delta\right)
\Big\rangle_{\rm st}
+(\mathrm{gh}),
\label{eq:Gamma-H-scat}
\end{equation}
In the following we treat the Hartree part and construct the mean-field operator
$\langle\Delta\rangle_{\rm st}$, and determine where
${\rm Im}\,\mathcal L^{(1)}_{\rm H}=0$. 
\subsection{One-loop effective action in the background-field formalism}
\label{subsec:one-loop-formalism}

The decomposition \cref{eq:A-dec} allows the quantum fluctuation to be gauge fixed
while preserving explicit gauge covariance with respect to the
background field. We employ the background-field $R_\xi$ gauge,
\be
\mathcal L_{\rm gf}
=
-\frac{1}{2\xi}
\left(
\bar D^\mu\mathcal A_\mu
\right)^a
\left(
\bar D^\nu\mathcal A_\nu
\right)^a,
\label{eq:gf-BG2}
\ee
where $\bar D_\mu$ is constructed from $\bar A_\mu$. The BRST-invariant gauge-fixing construction and the derivation of the fluctuation operators are given in \cref{app:1-loop-action}; here we only quote the results. At quadratic order in the quantum fields, the ghost Lagrangian is
\be
\mathcal L_{\rm gh}^{(2)}
=
\bar c^a
\big(\Delta_{\rm gh}\big)^{ab}
c^b,
\qquad
\big(\Delta_{\rm gh}\big)^{ab}
=
-\big(\bar D^2\big)^{ab}.
\label{eq:Delta-gh}
\ee
Expanding the gauge-fixed axion--Yang--Mills action to second order in
$\mathcal A_\mu^a$ gives
\be
S^{(2)}
=
\frac12\int d^4x\,
\mathcal A_\mu^a
\big[\Delta_{\rm YM}^{\mu\nu} + \Delta_{\rm CS}^{\mu\nu}\big]^{ab}
\mathcal A_\nu^b,
\label{eq:2nd-action}
\ee
where the Yang--Mills (YM) contribution is
\be
\big[\Delta_{\rm YM}^{\mu\nu}\big]^{ab}
=
\eta^{\mu\nu}
\big(\bar D^2\big)^{ab}
-
\left(1-\frac{1}{\xi}\right)
\big(\bar D^\mu\bar D^\nu\big)^{ab}
-
2g f^{abc}\bar F^{c\,\mu\nu},
\label{eq:Delta-gluon}
\ee
and the axion Chern--Simons (CS) coupling gives
\be
\big[\Delta_{\rm CS}^{\mu\nu}\big]^{ab}
=
\frac{\kappa\, g^2}{f}\,
\epsilon^{\mu\nu\lambda\sigma}
(\partial_\lambda\phi)\,
\big(\bar D_\sigma\big)^{ab}.
\label{eq:Delta-cs}
\ee
The complete gauge-field fluctuation operator is therefore
\be
\Delta^{\mu\nu}
=
\Delta_{\rm YM}^{\mu\nu}
+
\Delta_{\rm CS}^{\mu\nu}.
\label{eq:Delta-total}
\ee
For the static chromomagnetic backgrounds considered below, $\bar F_{0i}=0$, and for a homogeneous axion velocity, $\partial_\mu\phi=(\dot\phi,\boldsymbol 0)$, the mixed components vanish in Feynman background gauge, $\Delta^{0i}=0$. The temporal and spatial sectors thus decouple, and once the temporal and ghost contributions are included, the stability problem reduces to the physical spatial fluctuations. In the $R_\xi$ family, we work in the 't Hooft--Feynman gauge, $\xi=1$. For $SU(2)$, $f^{abc}=\epsilon^{abc}$, and the nonvanishing blocks are
\begin{align}
\Delta^{00}_{ab} &= - \big(\bar D^2\big)_{ab},\\
\Delta^{ij}_{ab} &= \delta^{ij} \, \big(\bar D^2\big)_{ab} -2 g\, \epsilon_{abc}\, \bar F^{c\,ij} + \frac{\kappa \, g^2\, \dot\phi}{f} \, \epsilon^{ijk} \, \big(\bar D_k\big)_{ab}.
\end{align}

\subsection{Quantum instability of the color--spin-locked background}
\label{subsec:locked-instability}

We first determine whether the conventional color--spin-locked background
\be
\bar A_i^a=Q\delta_i^a
\ee
is quantum-mechanically stable. A physical mode with $\Omega^2<0$ corresponds to an exponentially growing real-time fluctuation and, equivalently, to a negative eigenvalue of the Euclidean Hessian, which gives rise to an imaginary part of the one-loop effective action. Since the background is homogeneous, the quadratic operator is diagonal
in momentum space. Using the background relation between $Q$ and
$\dot\phi$, \cref{eq:homogeneous-SU2-branches}, the physical branches are
\be
\lambda_\ell(\omega,k)
=
\omega^2-\Omega_\ell^2(k),
\qquad
\Omega_\ell^2(k)
=
k^2-2\ell k\,gQ+2g^2Q^2,
\qquad
\ell=-2,\ldots,2 ,
\label{eq:locked-spectrum}
\ee
where $\ell$ is the total (color plus spin) helicity and the $\ell=0$ branch is doubly degenerate. These six branches remain after the temporal and ghost determinants cancel one copy of the $\ell=0,\pm1$ branches among the nine spatial ones. The cancelled branches, $\Omega_\ell^2=(k-\ell gQ)^2+(2-\ell^2)g^2Q^2>0$, are positive, so the ghosts contribute only to the real, UV-sensitive part of the effective action and do not cancel the negative mode.

The instability occurs in the helicity sector aligned with the sign of
$Q$ (for $g>0$): the $\ell=+2$ branch for $Q>0$ and the $\ell=-2$ branch for $Q<0$. Defining $\ell_\star=2\,\operatorname{sgn}(Q)$, the unstable branch reads
\be
\Omega_{\ell_\star}^2(k)
=
\bigl(k-2|gQ|\bigr)^2-2g^2Q^2 ,
\ee
which is negative over the finite band
\be
(2-\sqrt2)|gQ|
<
k
<
(2+\sqrt2)|gQ|,
\label{eq:unstable-band}
\ee
with minimum $\Omega_{\ell_\star}^2=-2g^2Q^2$ at $k=2|gQ|$.
Hence the conventional color--spin-locked background has a single
unstable branch, of helicity $\ell_\star$ \cite{Takahashi1980}, analogous to the Nielsen--Olesen instability
\cite{Savvidy1977,NielsenOlesen1978}. 
Writing ${\rm Im}\,\Gamma_{\rm 1\text{-}loop}=VT\,{\rm Im}\,\mathcal L_Q^{(1)}$, with $VT$ the spacetime volume, this band gives
\be
\operatorname{Im}\mathcal L_Q^{(1)}
=
\frac12
\int_{\Omega_{\ell_\star}^2<0}
\frac{d^3\bk}{(2\pi)^3}
\sqrt{-\Omega_{\ell_\star}^2(k)}
=
\frac{9}{8\pi}g^4Q^4 .
\label{eq:locked-imaginary-part}
\ee
The instability is infrared in origin: it arises from a finite band of
physical modes at $k\sim|gQ|$, not from ultraviolet renormalization.

\subsection{Quantum stability of the spin-2-dressed background: mean-field operator}\label{sec:Hartree--Delta}

We now turn to the spin-2-dressed background, \cref{eq:BG-gauge-field}, shown to be a classical stationary solution in \cref{sec:stochastic-spin2-background}. As above, its quantum stability is controlled by the spectrum of \cref{eq:Delta-total}. For the spin-2-dressed configuration, the quadratic operator takes the
schematic form
\be
\Delta[Q,W]
=
\Delta_Q+\Delta_W+\Delta_{W^2},
\label{eq:Delta-expansion-W}
\ee
where $\Delta_W$ and $\Delta_{W^2}$ are, respectively, linear and
quadratic in $W_i^a$. Explicitly,
\begin{equation}
\begin{aligned}
(\Delta_Q)^{ij}_{ab}={}&
\delta^{ij}\left[
\left(\partial^\mu\partial_\mu-2g^2Q^2\right)\delta_{ab}
-2gQ\eps_{abn}\partial_n
\right]
\\
&-\alpha\eps^{ijk}\partial_k\delta_{ab}
+gQ(\alpha-2gQ)
\left(\delta_a^i\delta_b^j-\delta_a^j\delta_b^i\right),
\end{aligned}
\end{equation}
while
\begin{equation}
\begin{aligned}
(\Delta_W)^{ij}_{ab}={}&
2\delta^{ij}
\left[g^2QW_{ab}-g\eps_{abc}W_{cn}\partial_n\right]
-2g\eps_{abc}
\left(\partial^iW^c_j-\partial^jW^c_i\right)
\\
&-2g^2Q\eps_{abc}
\left(\eps^c{}_{ie}W^e_j+\eps^c{}_{dj}W^d_i\right)
+\alpha g\eps^{ijk}\eps_{abc}W^c_k ,
\end{aligned}
\label{eq:Delta-W}
\end{equation}
and
\begin{equation}
(\Delta_{W^2})^{ij}_{ab}
=
g^2\delta^{ij}
\left[(W^2)_{ab}-\tr(W^2)\delta_{ab}\right]
-2g^2\eps_{abc}\eps^c{}_{de}W^{di}W^{ej}.
\label{eq:Delta-W2-}
\end{equation}
We write the quadratic action diagrammatically, using the notation of
\cref{eq:UVZ} and representing quantum gauge fluctuations by
\begin{equation}
\label{eq:quantum-line:qa}
\mathcal A_i^a
\quad :\quad
\QALegendQuantum .
\end{equation}
Defining
\(S_X^{(2)}\equiv\frac12\int d^4x\,
\mathcal A_i^a(\Delta_X)^{ij}_{ab}\mathcal A_j^b\),
we obtain
\begin{equation}
\label{eq:quantum-action-diagrams:qa}
\begin{gathered}
S_{\mathcal A}^{(2)}[Q,W]
= S_Q^{(2)}+S_W^{(2)}+S_{W^2}^{(2)},
\\[.5em]
\begin{aligned}
S_Q^{(2)}
&= \QADiagramKinetic
 + \QADiagramQDerivative
 + \QADiagramQQ
\\[-.1em]
&\quad
 + \QADiagramPhiDerivative
 + \QADiagramQPhi
\\[.65em]
S_W^{(2)}
&= \QADiagramWDerivative
 + \QADiagramQW
 + \QADiagramPhiW
\\[.65em]
S_{W^2}^{(2)}
&= \QADiagramWW .
\end{aligned}
\end{gathered}
\end{equation}
In the above, a small open circle marks a derivative coupling.
Each diagram includes the full differential kernel, with numerical
coefficients and tensor contractions implicit.

For a fixed realization of $W$, integrating out $\mathcal A_\mu^a$ gives
\be
\Gamma_{\rm 1\text{-}loop}[Q,W]
=
\frac{i}{2}{\rm Tr}\ln\Delta[Q,W],
\label{eq:one-loop-fixed-W}
\ee
with the ($W$-dependent) ghost contribution understood.
This expression is one-loop in $\mathcal A_\mu^a$ but retains the full
dependence of the quadratic operator on the classical background $W$. Since the ensemble average of $\Delta_{W^2}$ is parametrically comparable to $\Delta_Q$, it cannot be
treated as a perturbative insertion. We therefore include it in
the leading fluctuation operator (see \cref{eq:deltaK}),
\be
\langle\Delta \rangle_{\rm st}
\equiv
\Delta_Q+\langle\Delta_{W^2}\rangle_{\rm st},
\qquad
\bar G_0\equiv\bar\Delta^{-1},
\label{eq:Delta-bar}
\ee
so that $\bar G_0$ resums arbitrary insertions of the isotropic
quadratic tensor background. The remaining stochastic part is
\be
\delta\Delta
=
\Delta_W+\delta\Delta_{W^2},
\qquad
\delta\Delta_{W^2}
\equiv
\Delta_{W^2}
-\langle\Delta_{W^2}\rangle_{\rm st},
\ee
and for a centered ensemble, $\langle W\rangle_{\rm st}=0$, one has $\langle\delta\Delta\rangle_{\rm st}=0$.

\subsubsection*{Mean-field (Hartree) operator of the spin-2-dressed state}\label{sec:Delta-bar}

We first determine the isotropically spin-2-dressed fluctuation
operator $\bar{\Delta}$ defined in \cref{eq:Delta-bar}, and then analyze
its spectrum and contribution to the one-loop effective action. Since $W_{ai}=W_{ia}$ is symmetric and traceless and its stochastic
distribution is isotropic under the diagonal rotation group, its
equal-point two-point function is fixed uniquely by the spin-$2$
projector,
\be
\begin{aligned}
\left\langle W^a_{i}W^b_{j}\right\rangle_{\rm st}
= \frac{\mathcal W_2}{5}
\Pi^{ab}_{ij},
\end{aligned}
\label{eq:W-isotropic-two-point}
\ee
where the symmetric-traceless projector is
\be
\Pi^{ab}_{ij}
=
\frac12
\left(
\delta^{ab}\delta_{ij}
+
\delta^a_{j}\delta^b_{i}
\right)
-
\frac13\delta^a_{i}\delta^b_{j}.
\ee
The normalization in \cref{eq:W-isotropic-two-point} follows from
$\tr \,\Pi^{(2)}=5$. The contractions entering $\Delta_{W^2}$ are consequently
\be
\left\langle (W^2)_{ab}\right\rangle_{\rm st}
=
\frac{\mathcal W_2}{3}\delta_{ab},
\label{eq:W2-contraction-one}
\ee
and
\be
\eps_{abc}\eps^c{}_{de}
\left\langle W^{di}W^{ej}\right\rangle_{\rm st}
=
\frac{\mathcal W_2}{6}
\left(
\delta_a^j\delta_b^i
-
\delta_a^i\delta_b^j
\right).
\label{eq:W2-contraction-two}
\ee
It follows that 
\be
\begin{aligned}
 \left\langle
(\Delta_{W^2})^{ij}_{ab}
\right\rangle_{\rm st}
=
-\frac{2g^2\mathcal W_2}{3}
\delta^{ij}\delta_{ab}+
\frac{g^2\mathcal W_2}{3}
\left(
\delta_a^i\delta_b^j
-
\delta_a^j\delta_b^i
\right).
\end{aligned}
\label{eq:DeltaW2-isotropic-average}
\ee
Substituting this result into \cref{eq:Delta-bar}, the dressed
fluctuation operator becomes
\be
\begin{aligned}
\langle\Delta^{ij}_{ab} \rangle_{\rm st}
&=
\delta^{ij}
\left[
\left(
\partial^\mu\partial_\mu
-
2g^2Q^2
-
\frac{2}{3}g^2\mathcal W_2
\right)\delta_{ab}
-
2gQ\eps_{ab\ell}\partial_\ell
\right]
\\
&-
\alpha
\eps^{ijk}\partial_k\delta_{ab}+
\left[
gQ\left(\alpha-2gQ\right)
+
\frac{g^2\mathcal W_2}{3}
\right]
\left(
\delta_a^i\delta_b^j
-
\delta_a^j\delta_b^i
\right).
\end{aligned}
\label{eq:Deltabar-explicit}
\ee

Since both the color-locked background $Q\delta_i^a$ and the
stochastic average $\langle\Delta_{W^2}\rangle_{\rm st}$ are
homogeneous and isotropic under the diagonal rotation group, a
spatial-color tensor $X_{ia}$ decomposes into irreducible
$SO(3)_{\rm diag}$ representations according to
\be
{\bf 3}\otimes{\bf 3}
=
{\bf 1}\oplus{\bf 3}\oplus{\bf 5},
\ee
corresponding respectively to its trace, antisymmetric, and
symmetric-traceless components. We introduce the algebraic locking operator
\be
(\mathsf L X)_{ia}
\equiv
\delta_{ia}\tr(X)-X_{ai},
\qquad
\mathsf  L^{ij}_{ab}
=
\delta_a^i\delta_b^j-\delta_a^j\delta_b^i,
\ee
which acts diagonally on these sectors,
\be
\mathsf L X^{(J)}=\ell_J X^{(J)},
\qquad
\ell_0=2,\quad \ell_1=1,\quad \ell_2=-1.
\ee
At zero momentum, these $J=0,1,2$ sectors determine the corresponding
homogeneous mass gaps. At finite momentum, the derivative terms preserve
helicity with respect to $\hat\bk$ rather than $J$ separately; we therefore
organize the full dispersion relations by helicity below.
The momentum-independent part of the dressed
frequency-squared operator in \cref{eq:Deltabar-explicit} can be written as
\be
\mathsf M^2
=
\left(
2g^2Q^2+\frac{2}{3}g^2\mathcal W_2
\right)\mathbb 1
-
\left[
gQ(\alpha-2gQ)
+\frac{1}{3}g^2\mathcal W_2
\right]\mathsf L .
\ee
Using the eigenvalues of $\mathsf L$ given above, the corresponding
$J=1$ and $J=0$ eigenvalues are
\be
m_1^2
=
4g^2Q^2-\alpha gQ+\frac{1}{3}g^2\mathcal W_2,
\label{eq:J1-gap}
\ee
and
\be
m_0^2
=
2gQ(3gQ-\alpha),
\label{eq:J0-gap}
\ee
while the $J=2$ eigenvalue is
\be
m_2^2
=
\alpha gQ+g^2\mathcal W_2.
\label{eq:J2-gap}
\ee
An important feature of \cref{eq:J0-gap} is that the stochastic
tensor contribution cancels identically from the spin-$0$ mode. Indeed,
the contribution proportional to $\mathcal W_2$ is
\be
\frac{g^2\mathcal W_2}{3}
\left(
2\mathbb 1-\mathsf L
\right),
\ee
which annihilates the trace sector since $\ell_0=2$. By contrast, it
shifts the spin-$1$ and spin-$2$ sectors by
$g^2\mathcal W_2/3$ and $g^2\mathcal W_2$, respectively.
Intuitively, the spin-$2$ tensor background primarily dresses the non-scalar fluctuations: after isotropic averaging, its quadratic Yang--Mills contribution cancels exactly on the scalar trace mode, but gives positive mass shifts to the spin-$1$ and spin-$2$ sectors.

\paragraph{Spin-$2$ sector ($J=2$).}

The chromomagnetic instability resides in the transverse-traceless
sector. Consider a plane-wave perturbation
\be
\mathcal A_i^a(t,\bx)
= \int d^4k \, 
\mathcal A_\lambda(k^\mu) \,
e^\lambda_{ia}(\hat\bk) \,
e^{ik_\mu x^\mu},
\ee
where $e^\lambda_{ia}=e^\lambda_i e^\lambda_a$ is a helicity-$2$
polarization tensor. We choose the helicity convention
\be
i\eps^{ijk}\hat k_k e^\lambda_{ja}
=
-\lambda e^\lambda_{ia},
\qquad
i\eps_{ab\ell}\hat k_\ell e^\lambda_{ib}
=
-\lambda e^\lambda_{ia},
\qquad
\lambda=\pm1.
\label{eq:tensor-helicity-identities}
\ee
Applying \cref{eq:Deltabar-explicit} to this mode gives 
\be
 \langle \Delta_\lambda(\omega,k)\rangle_{\rm st}
=
\left[
\omega^2-\omega_\lambda^2(k)+i0
\right],
\label{eq:Deltabar-tensor-eigenvalue}
\ee
where
\be
\omega_\lambda^2(k)
=
k^2
-
\lambda(2gQ+\alpha)k
+
g\alpha Q+g^2\mathcal W_2.\label{eq:Spin-2 dressed-dispersion}
\ee
Thus, the isotropic tensor condensate produces a positive shift
$g^2\mathcal W_2$ in the helicity-$2$ frequency squared.

The potentially dangerous quantum fluctuation is the helicity for which the
term linear in $k$ lowers the frequency,
\begin{equation}
\lambda_{\rm d}
=
{\rm sgn}\!\left(gQ+\frac{\alpha}{2}\right).
\end{equation}
Completing the square gives
\be
\omega_{\lambda_{\rm d}}^2(k)
=
\left(
k-\left|gQ+\frac{\alpha}{2}\right|
\right)^2
+
\bar{m}^2,
\label{eq:dangerous-tensor-square}
\ee
where the minimum tensor gap is
\be
\begin{aligned}
\bar{m}^2
=
g^2\mathcal W_2
-
g^2Q^2
-
\frac{\alpha^2}{4}.
\end{aligned}
\label{eq:Spin-2 dressed-gap}
\ee
The cancellation of the mixed term $g\alpha Q$ in
\cref{eq:Spin-2 dressed-gap} is exact. The opposite-helicity mode has
\be
\omega_{-\lambda_{\rm d}}^2(k)
=
\left(
k+\left|gQ+\frac{\alpha}{2}\right|
\right)^2
+
\bar{m}^2.
\ee
The tensor sector is free of negative-frequency-squared modes provided that
the quadratic operator is positive and gapped, which requires
\be
g^2\mathcal W_2
>
g^2Q^2+\frac{\alpha^2}{4}.
\label{eq:tensor-strict-stability}
\ee
This condition establishes stability of the helicity-$\pm2$ sector at the
level of the ensemble-averaged quadratic operator. Full stability must
additionally be assessed after including the stochastic scattering
self-energy, to which we return in \cref{sec:scattering}.

\paragraph{Helicity-$\pm1$ fluctuation sectors.} At finite momentum, the derivative terms in
\cref{eq:Deltabar-explicit} are rotationally covariant but need not
preserve $J$ separately for a mode of fixed $\bk$. Rather, components
with the same helicity with respect to $\hat\bk$ can mix. This does
not signal any anisotropy of the dressed background: the background
remains homogeneous and isotropic, while $\bk$ simply labels the
quantum fluctuation being propagated through it. For helicity $\lambda=\pm1$, the relevant two-dimensional subspace is
spanned by
\be
\left\{
e_i^\lambda e_a^0,\,
e_i^0 e_a^\lambda
\right\},
\qquad
e_i^0\equiv\hat k_i ,
\ee
and the dressed inverse propagator takes the form 
\be
\bar\Delta_{\lambda}^{(1)}
=
\left[
\omega^2\mathbb 1
-
\mathsf H_{\lambda}^{(1)}(k)
\right],
\ee
with
\be
\mathsf H_{\lambda}^{(1)}(k)
=
\begin{pmatrix}
k^2+2g^2Q^2+\frac{2}{3}g^2\mathcal W_2
-\lambda\alpha k
&
gQ(\alpha-2gQ)+\frac13g^2\mathcal W_2
\\[1mm]
gQ(\alpha-2gQ)+\frac13g^2\mathcal W_2
&
k^2+2g^2Q^2+\frac{2}{3}g^2\mathcal W_2
-2\lambda gQ\,k
\end{pmatrix}.
\label{eq:helicity-one-matrix}   
\ee
The two corresponding dispersion relations are therefore
\be
\begin{aligned}
\omega_{\lambda,\pm}^2(k)
={}&
k^2
+
2g^2Q^2
+
\frac{2}{3}g^2\mathcal W_2
-
\lambda
\left(
gQ+\frac{\alpha}{2}
\right)k
\\
&\pm
\sqrt{
\frac{(\alpha-2gQ)^2}{4}\,k^2
+
\left[
gQ(\alpha-2gQ)
+
\frac13g^2\mathcal W_2
\right]^2
}.
\end{aligned}
\label{eq:helicity-one-dispersion}
\ee
At $k=0$, these two eigenvalues reduce, as expected, to the
spin-$1$ and spin-$2$ eigenvalues \cref{eq:J1-gap,eq:J2-gap}. Notice in particular that the tensor-stability condition
\cref{eq:tensor-strict-stability} already guarantees a positive
spin-$1$ mass, since
\be
\begin{aligned}
m_1^2
=
\frac13 \bar m^2
+
\frac{(\alpha-6gQ)^2+16g^2Q^2}{12}
>0.
\end{aligned}
\label{eq:m1-mgap}
\ee

\paragraph{Helicity-zero fluctuation sector.}  

In the basis
\be
\left\{
e_i^+e_a^-,
\,
e_i^0e_a^0,
\,
e_i^-e_a^+
\right\},
\ee
the helicity-zero frequency-squared matrix takes the simple form 
\be
\mathsf H^{(0)}(k)
=
\begin{pmatrix}
k^2+m_1^2-Dk
&
m_0^2-m_1^2
&
0
\\[1mm]
m_0^2-m_1^2
&
k^2+2g^2Q^2+\dfrac23g^2\mathcal W_2
&
m_0^2-m_1^2
\\[1mm]
0
&
m_0^2-m_1^2
&
k^2+m_1^2+Dk
\end{pmatrix},
\label{eq:helicity-zero-matrix}  
\ee
where
\be
D\equiv\alpha-2gQ.
\label{eq:D-def}
\ee
The three
helicity-zero dispersion relations are determined by
\be
\det
\left[
\omega^2\mathbb 1-\mathsf H^{(0)}(k)
\right]
=0.
\label{eq:helicity-zero-dispersion}
\ee
At $k=0$, the three roots reduce exactly to the spin-$0$, spin-$1$,
and spin-$2$ eigenvalues given above. The spin-$0$ mode therefore provides an additional necessary stability
condition. Since its homogeneous gap is $m_0^2$, positivity requires
\be
m_0^2 =2gQ(3gQ-\alpha)>0.
\label{eq:scalar-stability-condition}
\ee
At finite momentum, however, the derivative terms mix different $J$
sectors carrying the same helicity. Positivity of the homogeneous gaps alone
therefore does not, in general, guarantee positivity of the helicity-$\pm1$
and helicity-0 spectra. Neglecting the stochastic Born self-energy, the spatial quantum
fluctuation operator $\bar\Delta$ is therefore positive and gapped
provided
\begin{align}
g^2\mathcal W_2
>
g^2Q^2+\frac{\alpha^2}{4},
\qquad \text{(positivity of spin-2 sector)}, \label{eq:Deltabar-full-stability-2}\\ 
2gQ(3gQ-\alpha)>0, \qquad \text{(positivity of spin-0 sector)}.
\label{eq:Deltabar-full-stability-1}  
\end{align}
The first condition removes the finite-momentum chromomagnetic
instability in the helicity-$\pm2$ sector, while the second excludes
the instability associated with the spin-$0$ trace mode; for
$gQ/\alpha<0$ the helicity-$0$ condition discussed above must be added. Both
conditions refer to the mean-field operator $\bar\Delta$. For later convenience, we define the minimum Hartree gap as
\begin{equation}
    m_H^2 = \rm{min}_k [-\bar\Delta(0,k)],
    \label{eq:mH}
\end{equation}
where the minimum runs over all helicity sectors of the spatial Hartree
operator.

\subsection{Mass gap around the background solutions}
\label{sec:gap-on-solutions}

We now determine the part of parameter space in which the background equations
of \cref{sec:stochastic-spin2-background} are satisfied and, at the same time,
all quantum fluctuations described by $\bar\Delta$ are gapped. Since
$\bar\Delta$ depends on $W$ only through $\mathcal W_2$, its spectrum is a function
of the dimensionless variables \cref{eq:ss-dimless} alone,
\be
\frac{\bar m^2}{\alpha^2}=\chi_{_{\!W}}^2-\frac{1+\chi^2}{4},
\qquad
\frac{m_0^2}{\alpha^2}=\chi\Big(\frac{3\chi}{2}-1\Big),
\label{eq:gaps-dimless}
\ee
together with the helicity-$0,\pm1$ eigenvalues of
\cref{eq:helicity-one-matrix,eq:helicity-zero-matrix}. 

 The Hartree fluctuation operator is gapped whenever $m_H^2>0$.
We therefore define the corresponding Hartree-gapped region by
\begin{equation}
\mathcal{R}_{\rm Har}
=
\left\{
(\chi,\chi_W):
\chi_W>0,\;
m_H^2>0
\right\}, \quad (\text{Hartree-gapped region)}.
\label{eq:hartree-region}
\end{equation}
The region $\mathcal{R}_{\rm Har}$ depends only on the Hartree operator and is independent
of the detailed spectrum of the dressing. The power spectrum, the bispectrum and the connected trispectrum enter only through the
background equations, which select the allowed points of the
$(\chi,\chi_{_{\!W}})$ plane $\mathcal{R}_{\rm bg}(b_W,\tau_W)$ as introduced in
\cref{sec:consistency-bounds}. 

\paragraph{Gapped background solutions.}
The task is to find, for given
$(b_W,\tau_W)$, the allowed region $\mathcal{R}_{\rm bg}$ and its intersection with $\mathcal{R}_{\rm Har}$,
\be
 \mathcal{R}_{\rm bg}(b_W,\tau_W)\cap \mathcal{R}_{\rm Har} .
\label{eq:A-intersect-G}
\ee
Here $\mathcal{R}_{\rm bg}$ is fixed by the coarse-grained background equations through
$\rm Var(\chi_k)\ge0$. Its boundary $\rm Var(\chi_k)=0$ consists of thin-shell solutions,
while points with $\rm Var(\chi_k)>0$ require a spectrum with nonzero width.
As discussed in \cref{sec:broad-spectrum-W}, the detailed spectrum must still
satisfy the corresponding mode-by-mode tensor condition.

\begin{figure}[h]
\centering
\includegraphics[width=\textwidth]{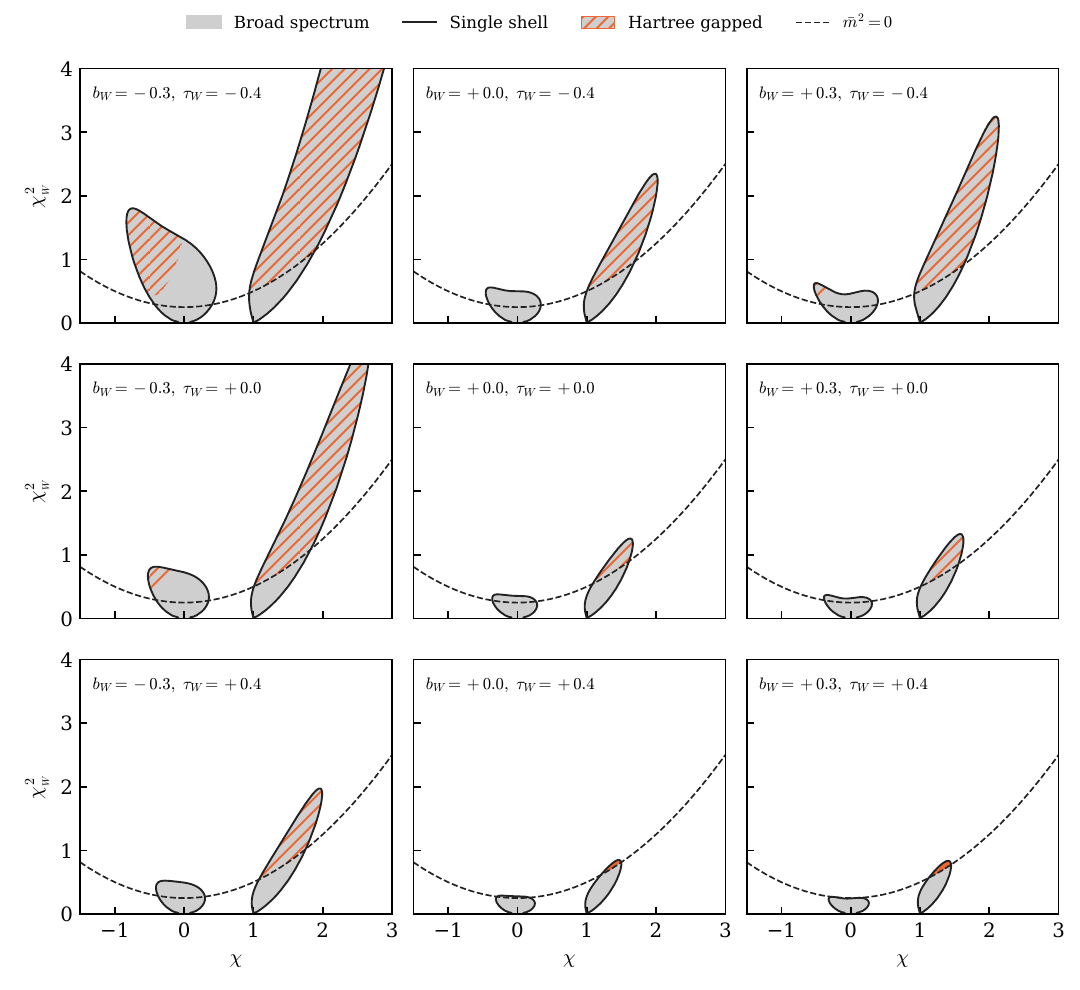}
\caption{
Region allowed by the coarse-grained background equations in the
$(\chi,\chi_W^2)$ plane for representative values of $(b_W,\tau_W)$.
Grey denotes $\rm Var(\chi_k)>0$, while the black boundary $\rm Var(\chi_k)=0$
corresponds to thin-shell solutions. Orange hatching shows the intersection
with the gap region $\mathcal{R}_{\rm Har}$, and the dashed curve marks $\bar m^2=0$.
The parameters $(b_W,\tau_W)$ change $\mathcal{R}_{\rm bg}$, whereas $\mathcal{R}_{\rm Har}$
itself is fixed.
}
\label{fig:param-panels}
\end{figure}

\Cref{fig:param-panels} shows $\mathcal{R}_{\rm bg}$ and its overlap with $\mathcal{R}_{\rm Har}$
for representative values of $(b_W,\tau_W)$. Negative $\tau_W$ generally
increases the overlap, whereas positive $\tau_W$ reduces it. The parameter
$b_W$ changes the shape and position of the allowed region; in particular,
negative $b_W$ extends the gapped region towards larger positive $\chi$ and
can also open a branch at negative $\chi$.

\begin{figure}[h]
    \centering
    \includegraphics[width=0.95\textwidth]{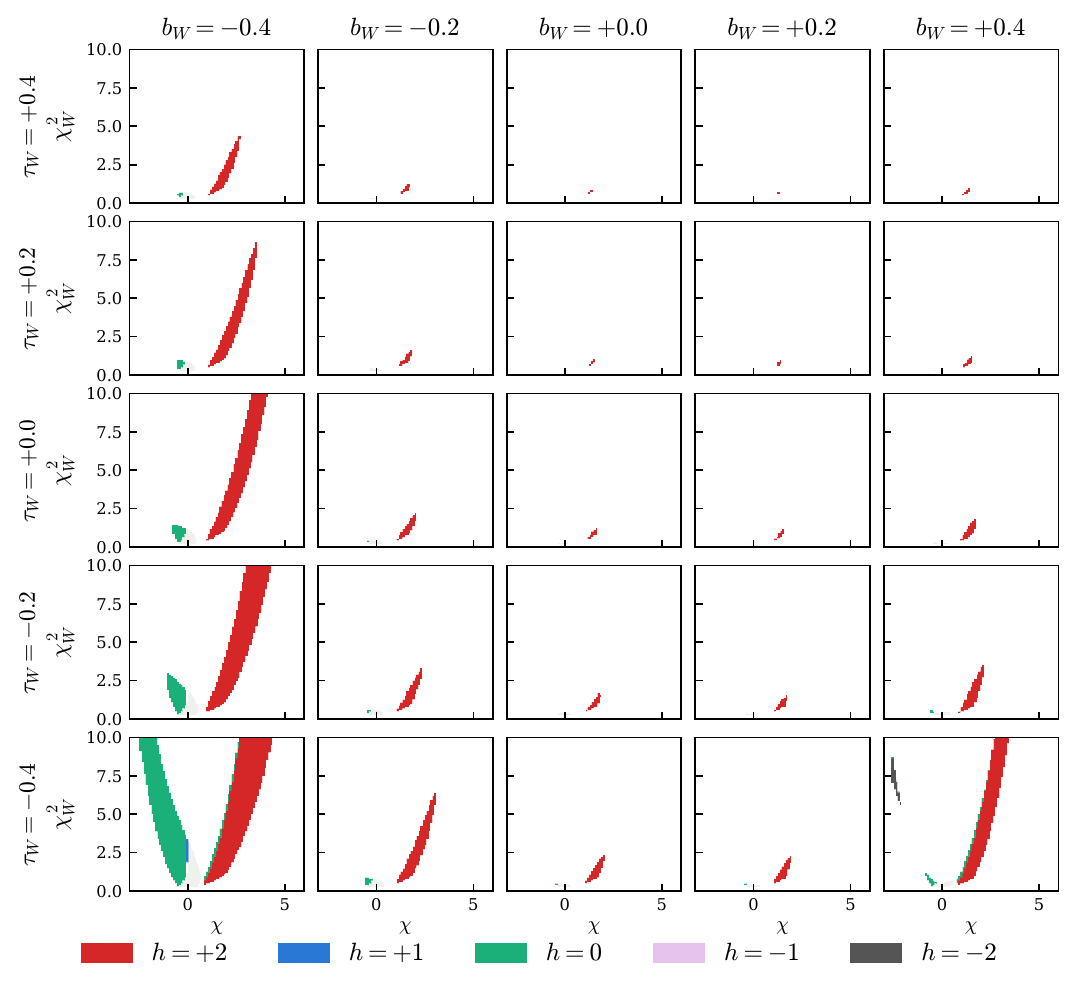}
    \caption{Helicity sector with the smallest Hartree gap.
    At each background point we compare the Hartree gaps
    $m_h^2=\min_k\lambda_{\min}[-\bar\Delta_h]$ of the five helicity sectors
    $h=+2,+1,0,-1,-2$, without scattering corrections.
    Colours indicate the sector with the smallest gap:
    red ($+2$), blue ($+1$), green ($0$),
    light orchid ($-1$), and grey ($-2$).
    The tensor sector is softest over most of the plane (red),
    while the scalar sector is softest in the
    $\chi<0$ lobe, near $\chi=0$, and at the base of the
    corner strip (green), where $m_0^2<\bar m^2$.
    Light grey indicates backgrounds satisfying the background
    equations but with at least one ungapped Hartree channel.}
    \label{fig:smallest-gap-helicity}
\end{figure}

\begin{figure}[h]
\centering
\includegraphics[width=0.75\textwidth]{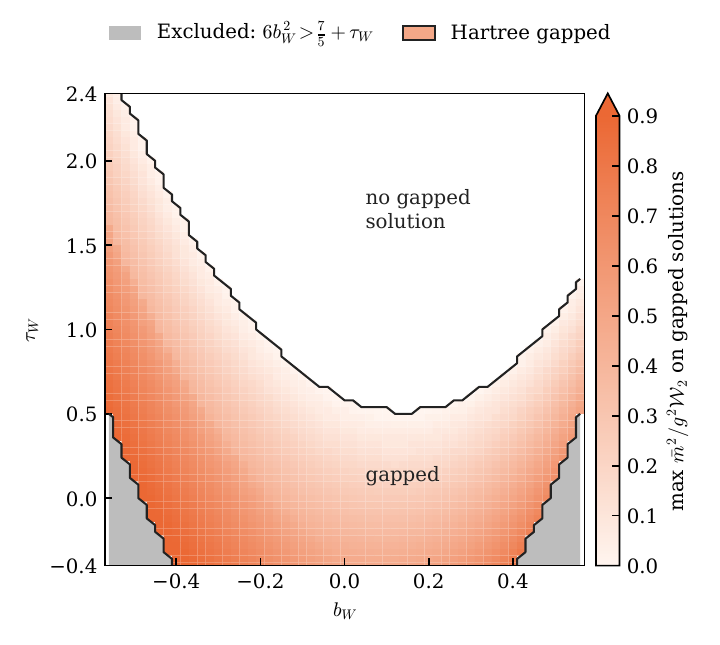}
\caption{
Existence of gapped backgrounds in the $(b_W,\tau_W)$ plane.
The color scale gives the largest
$\bar m^2/(g^2\mathcal W_2)$ in $\mathcal{R}_{\rm bg}\cap\mathcal{R}_{\rm Har}$;
white denotes parameters for which no gapped solution exists.
Grey denotes parameters excluded by the realizability bounds
$\tau_W\ge-2/5$ and
$6b_W^2\le7/5+\tau_W$.
The black curve separates the regions with and without gapped solutions.
}
\label{fig:param-phase}
\end{figure}

The dependence on the stochastic moments is summarized in
\cref{fig:param-phase}. Gapped solutions exist only below a limiting curve
$\tau_W=\tau_c(b_W)$. Negative $\tau_W$, especially together with negative
$b_W$, allows the largest overlap with $\mathcal{R}_{\rm Har}$, while sufficiently
positive $\tau_W$ removes it. Part of the $(b_W,\tau_W)$ plane is excluded
independently by the realizability bounds of
\cref{sec:consistency-bounds}. The helicity dependence of the gap is illustrated in
\cref{fig:smallest-gap-helicity}, which identifies the sector with the
smallest upper bound on the Hartree gap at each background point.
The tensor sector ($h=+2$) dominates most of the parameter space,
while the scalar sector ($h=0$) becomes limiting in localized regions.
These bounds closely track the smallest Hartree mass wherever
the Born bounds are loose.

These results test stability only at the level of the averaged operator
$\bar\Delta$. The remaining terms $\Delta_W$ and
$\Delta_{W^2}-\langle\Delta_{W^2}\rangle_{\rm st}$ transfer momentum between
the fluctuations and the stochastic dressing and generate scattering
corrections. Whether the gap survives these corrections is studied in
\cref{sec:scattering}.

\section{Weak Coupling and Separation from Confinement}
\label{sec:confinement-validity}

Before examining scattering corrections to the Hartree spectrum,
we specify the regime in which the gauge coupling remains weak
and its running can be neglected. The dressed background contains
several characteristic scales. We choose the renormalization scale
$\mu_{\rm bg}\sim M_{\rm bg}$, where
\be
M_{\rm bg}\equiv
\max\left\{
|\alpha|,\,
|gQ|,\,
g\sqrt{\mathcal W_2},\,
\left[
\frac{1}{\mathcal W_2}
\sum_{\lambda=\pm}\int_{\bk}k^2P_\lambda(k)
\right]^{1/2}
\right\}.
\label{eq:M-bg-def}
\ee
The last entry characterizes the typical momentum of the
stochastic dressing and reduces to $k_\star$ for a single thin
shell. It should be distinguished from the infrared edge of
the spectrum,
\be
k_{\min}\equiv
\inf\operatorname{supp}\left[P_+(k)+P_-(k)\right].
\label{eq:kmin-def}
\ee
For broad spectra, $k_{\min}$ need not coincide with the typical
momentum. Moreover, the infrared scale relevant to the fluctuation
propagators is not necessarily $k_{\min}$: scattering can connect
momenta below this edge, while a positive mass gap regulates the
propagator. When all intermediate Hartree channels are gapped,
their lowest eigenvalue defines $m_{\rm H}^2>0$, and we take
$\mu_{\rm IR}\sim m_{\rm H}$ as the characteristic infrared scale,
provided no lighter physical scale is involved.

As shown in \cref{app:heat-kernel-running}, the gauge coupling
obeys the standard one-loop Yang--Mills beta function,
\be
\beta(g)\equiv\mu\frac{dg}{d\mu}
=-\frac{b_0}{16\pi^2}g^3+\mathcal O(g^5),
\qquad
b_0=\frac{11}{3}C_A=\frac{22}{3}
\quad [SU(2)].
\label{eq:beta-main}
\ee
The corresponding RG-invariant strong-coupling scale is
\be
\Lambda_{\rm conf}
=
\mu_{\rm bg}
\exp\left[
-\frac{8\pi^2}{b_0g^2(\mu_{\rm bg})}
\right],
\label{eq:Lambda-main}
\ee
or, for $SU(2)$,
\be
\frac{\Lambda_{\rm conf}}{M_{\rm bg}}
\sim
\exp\left[
-\frac{3\pi}{11\,\alpha_g(M_{\rm bg})}
\right],
\qquad
\alpha_g\equiv\frac{g^2}{4\pi}.
\label{eq:Lambda-SU2-main}
\ee
Here $\Lambda_{\rm conf}$ denotes the perturbative strong-coupling
scale, parametrically associated with confinement in vacuum
Yang--Mills theory, up to scheme-dependent factors of order unity.
Thus $\alpha_g(M_{\rm bg})\ll1$ implies
$M_{\rm bg}\gg\Lambda_{\rm conf}$.

However, weak coupling at the background scale alone does not
guarantee weak coupling throughout the fluctuation spectrum.
At one loop,
\be
g^2(\mu)
=
\frac{g^2(M_{\rm bg})}{
1-\dfrac{b_0g^2(M_{\rm bg})}{8\pi^2}
\ln\dfrac{M_{\rm bg}}{\mu}
}.
\label{eq:constant-g-validity}
\ee
For the coupling to remain approximately constant over
$\mu_{\rm IR}\lesssim\mu\lesssim M_{\rm bg}$, we therefore require
\be
\alpha_g(M_{\rm bg})\ll1,
\qquad
\frac{b_0\,\alpha_g(M_{\rm bg})}{2\pi}
\ln\frac{M_{\rm bg}}{\mu_{\rm IR}}\ll1 .
\label{eq:constant-g-window}
\ee
Since
$\ln(M_{\rm bg}/\Lambda_{\rm conf})
=2\pi/[b_0\alpha_g(M_{\rm bg})]$,
the second condition implies
\be
\mu_{\rm IR}\gg\Lambda_{\rm conf}.
\label{eq:weak-coupling-confinement}
\ee
The entire dynamically relevant range then lies above the
strong-coupling scale, rather than only its ultraviolet end.
For comparable background and infrared scales,
\cref{eq:constant-g-window} follows from weak coupling alone;
for $m_{\rm H}\ll M_{\rm bg}$, it becomes an additional restriction.

Importantly, the classical background equations and the
Hartree--Born kernels depend on $gQ$ and $g^2\mathcal W_2$,
rather than on $g$ separately. At fixed values of these
combinations, $\alpha$, and the normalized stochastic spectrum,
decreasing $g$ leaves the background spectra unchanged while
exponentially suppressing $\Lambda_{\rm conf}/M_{\rm bg}$.
Thus, for a strictly positive infrared gap, a parametrically
weakly coupled regime satisfying \cref{eq:constant-g-window}
is available within the Yang--Mills sector, provided the
corresponding axion background remains within the validity
of the effective theory. The spacetime-dependent axion modifies the fluctuation spectrum
but does not change the universal coefficient $b_0$ governing
the Yang--Mills gauge coupling at the order considered here.
Additional axion-dependent counterterms belong to the
higher-dimensional operator basis of the effective theory
and must be treated separately.\footnote{Including dynamical
axion fluctuations generates further higher-dimensional
counterterms. For a massless axion in dimensional regularization,
these do not modify the dimension-four Yang--Mills beta
function at the order considered
\cite{Villalba-Chavez:2018eql,Chala:2020wvs}.}

Finally, separation from confinement and perturbative control
of the dressed background are distinct requirements.
The absence of an imaginary part of the one-loop effective
action within the Hartree approximation, established in
\cref{sec:quantum-structure-Hartree}, does not by itself
guarantee stability beyond Hartree. Quantum-loop corrections
must remain controlled, and the stochastic scattering
self-energy must be small relative to the Hartree kernel
in the relevant fluctuation channels. In particular, Born
corrections can be enhanced by inverse powers of a small
Hartree gap even when $\alpha_g\ll1$ and
$\mu_{\rm IR}\gg\Lambda_{\rm conf}$.
Such an enhancement signals a breakdown of the expansion
around the dressed background, not the onset of confinement.
We investigate these scattering corrections in the next section.

\section{Scattering off the Dressing: Is the Gap Protected?}
\label{sec:scattering}

In \cref{sec:quantum-structure-Hartree} we studied the mean-field stability of quantum
fluctuations around the spin-2-dressed background and identified
classically self-consistent configurations for which the
ensemble-averaged (Hartree) operator is positive and gapped.  This is
only the second of the three stability tests.  The Hartree contribution
retains the isotropic mean of the stochastic dressing, whereas the
spatially varying remainder scatters quantum fluctuations between
different momenta.  Correspondingly, the one-loop effective action
contains both the Hartree and scattering contributions,
\cref{eq:Gamma-decomposition,eq:Gamma-H-scat}.  We now ask whether this
momentum-changing scattering is sufficiently controlled for the Hartree
gap to survive.

Since stochastic scattering can appreciably modify propagation, we organize the analysis around the ensemble-averaged propagator. Defining $G[W]\equiv\Delta[W]^{-1}$ for each realization,
\be
\bar G\equiv\left\langle G[W]\right\rangle_{\rm st}
=\left\langle\Delta^{-1}\right\rangle_{\rm st}.
\label{eq:averaged-propagator}
\ee
and  
\be
\bar G_0\equiv\langle\Delta\rangle_{\rm st}^{-1},
\ee 
as the \emph{Hartree propagator}, namely the propagator associated with the ensemble-averaged quadratic operator. The difference $\bar G-\bar G_0$ therefore isolates the effect of stochastic scattering. The propagator can be written exactly as
\be
G[W]
=
\left(1+\bar G_0\delta\Delta\right)^{-1}\bar G_0 .
\ee
When the corresponding Neumann series converges,
$\rho(\bar G_0\delta\Delta)<1$, this identity admits the Born expansion~\cite{Born:1926,Sakurai:2020},
\be
G[W]
=
\bar G_0
-\bar G_0\delta\Delta\bar G_0
+\bar G_0\delta\Delta\bar G_0\delta\Delta\bar G_0
-\cdots .
\ee
After stochastic averaging, the term linear in $\delta\Delta$ vanishes,
\be
\bar G
=
\bar G_0
+
\bar G_0
\left\langle
\delta\Delta\,\bar G_0\,\delta\Delta
\right\rangle_{\rm st}
\bar G_0+\cdots .
\ee
This defines the leading Born self-energy,
\be
\Sigma^{\rm B}
\equiv
\left\langle
\delta\Delta\,\bar G_0\,\delta\Delta
\right\rangle_{\rm st}.
\label{eq:SigmaW-scattering}
\ee
The Born approximation consists of truncating the irreducible scattering kernel at this leading non-vanishing order in $\delta\Delta$.

\paragraph{Scattering vertices.}
The stochastic field enters the fluctuation operator through
$S^{(2)}_W$ and $S^{(2)}_{W^2}$ of \cref{eq:quantum-action-diagrams:qa}. We
split it into its mean and its fluctuating part,
\begin{equation}
\label{eq:se-split}
\Gamma_1\equiv\Delta_W,\qquad
\Gamma_2\equiv\Delta_{W^2}-\langle\Delta_{W^2}\rangle_{\rm st}.
\end{equation}
$\Gamma_1$ is linear in $W$ and collects the three vertices of $S^{(2)}_W$,
while $\Gamma_2$ is the single vertex of $S^{(2)}_{W^2}$ with its stochastic
mean removed. We represent them by
\begin{equation}
\label{eq:se-vertices}
\begin{aligned}
\SEVertexGammaOne
&= \QADiagramWDerivative
 + \QADiagramQW
 + \QADiagramPhiW
\\[.65em]
\SEVertexGammaTwo
&= \QADiagramWW
 - \SEDiagramWWMean\;.
\end{aligned}
\end{equation}

\subsection{Born self-energy}\label{sec:Born-Scattering-----}
The Born kernel of \cref{eq:SigmaW-scattering} is the second-order
self-energy of a quantum gluon scattering off the background,
\begin{equation}
\label{eq:se-def}
\Sigma^{\rm B}=\sum_{i,j\in\{1,2\}}\Sigma_{ij},\qquad
\Sigma_{ij}\equiv\big\langle\Gamma_i\,\bar G_0\,\Gamma_j\big\rangle_{\rm st}.
\end{equation}
The stochastic average joins the $W$ legs of the two vertices into the
correlators of the background: a pair into the power spectrum, a triple into
the bispectrum $B$ and a quadruple into the connected trispectrum $T$,
\begin{equation}
\label{eq:se-legend}
\big\langle WWW\big\rangle_{\rm st}
\quad :\quad \SELegendBlob{B_W}\,,
\qquad
\big\langle WWWW\big\rangle_{\rm st,c}
\quad :\quad \SELegendBlob{T_W}\,.
\end{equation}
The Born self-energy decomposes as
\begin{equation}
\label{eq:born-selfenergy-diagrams}
\Sigma^{\rm B}
=\Sigma_{11}+\Sigma_{12}+\Sigma_{21}+\Sigma_{22}.
\end{equation}
The linear--linear scattering contribution is
\begin{equation}
\Sigma_{11}=\SEDiagramEleven\,,
\end{equation}
while the mixed linear--quadratic contributions are
\begin{equation}
\Sigma_{12}+\Sigma_{21}
=\SEDiagramTwelve+\SEDiagramTwentyOne\,.
\end{equation}
Finally, the quadratic--quadratic scattering contribution reads
\begin{equation}
\Sigma_{22}
=2\times\SEDiagramTwentyTwoG
+\SEDiagramTwentyTwoT\,.
\end{equation}

\subsubsection{Linear--linear scattering}

The detailed derivation of the projected linear--linear Born
self-energy is given in \cref{app:Sigma11}. To define the transition amplitudes, let $I=(i,a)$
denote a composite spatial--color index. We use
the helicity basis
\be
u_I^{(\lambda_1\lambda_2)}(\hat\bp)
\equiv e_i^{\lambda_1}(\hat\bp)e_a^{\lambda_2}(\hat\bp),
\qquad
e_I^\sigma(\hat\bk)
\equiv u_I^{(\sigma\sigma)}(\hat\bk),
\ee
where $\lambda_1,\lambda_2\in\{+,0,-\}$, $e_i^\pm$ are normalized
transverse helicity vectors, and $e_i^0=\hat p_i$.
The helicity-projected vertex function $\mathcal M$
is associated with the linear fluctuation operator
$\Delta_W$ in \cref{eq:Delta-W}, whose explicit
vertex $\mathcal V_\lambda$ is given in
\cref{eq:Vplus-exact}. Its matrix elements between
the external tensor state of helicity $\sigma$
and an intermediate helicity state $(\lambda_1,\lambda_2)$ are
\be
\mathcal M_{\sigma;(\lambda_1\lambda_2)}(\bk,\bq)
\equiv
e_I^{\sigma *}(\hat\bk)
[\mathcal V_\lambda(\bq;\bp)]_{IJ}
u_J^{(\lambda_1,\lambda_2)}(\hat\bp).
\label{eq:M-main}
\ee
Following the ordering of the Hartree helicity
blocks, these amplitudes are collected into
\be
\begin{aligned}
\mathbf m_\pm &=
\begin{pmatrix}
\mathcal M_{\sigma;(\pm0)}\\
\mathcal M_{\sigma;(0\pm)}
\end{pmatrix},
\qquad
\mathbf m_0 =
\begin{pmatrix}
\mathcal M_{\sigma;(+-)}\\
\mathcal M_{\sigma;(00)}\\
\mathcal M_{\sigma;(-+)}
\end{pmatrix}.
\end{aligned}
\label{eq:m-main}
\ee
The dependence on $(\bk,\bq)$ and the background
helicity $\lambda$ is implicit. 

Here we report
its static limit, $\omega=0$, separating the contributions
from the intermediate helicity sectors:
\be
\Sigma_{11,\sigma}(0,k)
=
-g^2\sum_{\lambda=\pm}\int_{\bq}P_\lambda(q)
\Bigg[
\sum_{\lambda'=\pm}
\frac{|\mathcal M_{\sigma;(\lambda'\lambda')}|^2}
{\omega_{\lambda'}^2(p)}
+
\sum_{h=\pm,0}
\mathbf m_h^\dagger
[\mathsf H^{(h)}(p)]^{-1}
\mathbf m_h
\Bigg],
\label{eq:Sigma11-static-}
\ee
where $\bp=\bk-\bq$. The first term accounts for
intermediate tensor modes of helicity $\pm2$, with
dispersion relations $\omega_{\lambda'}^2(p)$,
while the second includes the helicity-$\pm1$
and helicity-$0$ sectors, described by the Hartree
blocks $\mathsf H^{(h)}(p)$ of
\cref{sec:Hartree--Delta}. For a positive-definite Hartree operator, every
term in the brackets of \cref{eq:Sigma11-static-}
is non-negative, implying
$\Sigma_{11,\sigma}(0,k)\leq0$.
The static self-energy therefore lowers the dressed
gap, $\omega_\sigma^2(k)+\Sigma_{11,\sigma}(0,k)$.
The most vulnerable mode is the dressed helicity
$\sigma=\lambda_{\rm d}$ near $k=k_{\rm d}$,
where the tree-level restoring scale is $\bar m^2$.
Nevertheless, both external helicities must be
evaluated, since their self-energy magnitudes
need not be ordered.

\paragraph{Role of the vector and scalar channels.}
The intermediate vector and scalar modes make an appreciable
contribution to the Born self-energy and cannot be neglected.
At the minimum of the dangerous branch, they typically account
for an order-one fraction of $\Sigma_{11}$. These contributions arise from helicity-changing scattering
amplitudes, which are generally comparable to the tensor
ones away from collinear configurations.

\paragraph{Infrared and ultraviolet limits.}
The two limiting regimes of \cref{eq:Sigma11-static-}
provide useful physical insight.
For long-wavelength dressing modes ($q\to0$), the
intermediate momentum approaches the external one,
$p\to k$, while the scattering amplitudes remain finite
(see \cref{app:Sigma11}). The dressing then acts locally
as a nearly homogeneous random tensor, whose orientation
mixes the different helicity sectors. In particular,
near the minimum $k=k_{\rm d}$ of the dangerous branch,
the tensor propagator is enhanced by
$1/\omega_{\lambda_{\rm d}}^2(k_{\rm d})=1/\bar m^2$.
Thus, infrared power can produce a large negative
self-energy when the Hartree gap is small. In the opposite limit ($q\gg k$), derivative couplings
dominate: the scattering amplitudes grow as $q$, while
the intermediate propagators scale as $q^{-2}$.
The angular-integrated kernel approaches a universal
constant, yielding
\be
\Sigma_{11,\sigma}(0,k)\big|_{\rm UV}
=-\frac{32}{15}g^2\mathcal W_2^{\rm UV},
\label{eq:Sigma11-UV}
\ee
where $\mathcal W_2^{\rm UV}$ denotes the dressing power
in short-wavelength modes. Hence, the Born self-energy
introduces no additional ultraviolet divergence beyond
that of $\mathcal W_2$. Notably, short-wavelength power
produces a negative correction approximately twice the
positive Hartree shift $+g^2\mathcal W_2$, illustrating
how scattering can undermine mean-field stabilization.

\paragraph{Thin-shell limit.}
For a dressing concentrated on a single momentum shell
$q=k_{\rm s}$ with helicity $\lambda_{\rm s}$ and total
power $\mathcal W_2$, the radial integral can be performed
exactly. The Born self-energy reduces to a single angular
integral,
\be
\Sigma_{11,\sigma}(k)
= -
\frac{g^2\mathcal W_2}{2}
\int_{-1}^{1}d\mu\,
\Bigg[ 
\sum_{\lambda'=\pm}
\frac{|\mathcal M_{\sigma;(\lambda'\lambda')}|^2}
{\omega_{\lambda'}^2(p)}
+
\sum_{h=\pm,0}
\mathbf m_h^\dagger
[\mathsf H^{(h)}(p)]^{-1}
\mathbf m_h
\Bigg]_{q=k_{\rm s},\,\lambda=\lambda_{\rm s}},
\label{eq:Sigma11-thin-shell}
\ee
where $\mu=\hat\bk\cdot\hat\bq$ and
$p=\sqrt{k^2+k_{\rm s}^2-2kk_{\rm s}\mu}$.
The shell radius $k_{\rm s}$ need not coincide with the
minimum $k_{\rm d}$ of the dangerous branch.
Scattering through its softest tensor modes is enhanced
when the intermediate momentum can reach $k_{\rm d}$,
which requires
\be
|k-k_{\rm s}|\leq k_{\rm d}\leq k+k_{\rm s}.
\ee

\subsubsection{Quadratic--quadratic scattering}

The quadratic--quadratic contribution arises from two
insertions of the mean-subtracted $W^2$ operator.
Its detailed derivation is given in \cref{app:Sigma22}. The corresponding helicity-projected vertex function
$\mathcal N$ is associated with the quadratic operator
$\Delta_{W^2}$ in \cref{eq:Delta-W2-}.
It is obtained by projecting the quadratic vertex
$\mathcal U_{\lambda\lambda'}$, defined explicitly
in \cref{eq:U22-definition}, between the external
tensor state and an intermediate helicity state:
\be
\mathcal N_{\sigma;(\lambda_1\lambda_2)}(\bk;\bp,\bp')
\equiv
e_I^{\sigma *}(\hat\bk)
[\mathcal U_{\lambda\lambda'}(\hat\bp,\hat\bp')]_{IJ}
u_J^{(\lambda_1\lambda_2)}(\hat{\boldsymbol{\ell}}).
\label{eq:N22-main}
\ee
Following the same helicity-block ordering as
$\mathbf m_h$ in \cref{eq:m-main}, we define
\be
\begin{aligned}
\mathbf n_\pm &=
\begin{pmatrix}
\mathcal N_{\sigma;(\pm0)}\\
\mathcal N_{\sigma;(0\pm)}
\end{pmatrix},
\qquad
\mathbf n_0 =
\begin{pmatrix}
\mathcal N_{\sigma;(+-)}\\
\mathcal N_{\sigma;(00)}\\
\mathcal N_{\sigma;(-+)}
\end{pmatrix}.
\end{aligned}
\label{eq:n-main}
\ee
The dependence on $(\bk,\bp,\bp')$ and the two
background helicities $(\lambda,\lambda')$ is implicit.
Thus, $\mathbf n_h$ and $\mathbf m_h$ have identical
helicity-block structures and couple to the same
Hartree matrices, but differ in their underlying
quadratic and linear scattering vertices.

We first consider the Gaussian (Wick) contribution,
whose static limit reads
\be
\begin{aligned}
\Sigma_{22,\sigma}^{\rm G}(0,k)
=
-2g^4\sum_{\lambda,\lambda'=\pm}
\int_{\bp,\bp'}
P_\lambda(p)P_{\lambda'}(p')
\Bigg[
\sum_{\lambda''=\pm}
\frac{
|\mathcal N_{\sigma;(\lambda''\lambda'')}|^2
}{
\omega_{\lambda''}^2(\ell)
}
+
\sum_{h=\pm,0}
\mathbf n_h^\dagger
[\mathsf H^{(h)}(\ell)]^{-1}
\mathbf n_h
\Bigg],
\end{aligned}
\label{eq:Sigma22-static}
\ee
where $\boldsymbol{\ell}=\bk-\bp-\bp'$ and $\ell=|\boldsymbol{\ell}|$.
The factor of two accounts for the equivalent
cross-vertex Wick contractions. As in the
linear--linear case, the two terms describe scattering
through the intermediate tensor and lower-helicity
sectors, respectively. For a positive-definite Hartree operator,
$\Sigma_{22,\sigma}^{\rm G}(0,k)\leq0$.
Like the linear--linear contribution, quadratic
scattering therefore lowers the dressed gap.
Unlike $\Sigma_{11}$, however, this contribution
is of order $g^4P_W^2$ and involves two independent
background momenta.

\paragraph{Non-Gaussian contribution.}
For non-Gaussian dressings, the connected four-point
function of $W$ contributes an additional term to the
quadratic--quadratic self-energy,
\be
\Sigma_{22,\sigma}(0,k)
=
\Sigma_{22,\sigma}^{\rm G}(0,k)
+
\Sigma_{22,\sigma}^{\rm NG}(0,k),
\label{eq:Sigma22-static-total}
\ee
whose explicit form is given in \cref{eq:Sigma22-NG}.
This contribution involves a different momentum and
polarization projection of the trispectrum from that
entering the nonlinear background response in
\cref{eq:response-V--}, and is therefore independent
of $\mathcal R_{\lambda}^{(4)}$
in general. To quantify its effect, we introduce
\be
\mathcal S_\sigma^{\rm NG}(k)
\equiv
\frac{\Sigma_{22,\sigma}^{\rm NG}(0,k)}
{\Sigma_{22,\sigma}^{\rm G}(0,k)},
\qquad
\Sigma_{22,\sigma}(0,k)
=
\bigl[1+\mathcal S_\sigma^{\rm NG}(k)\bigr]
\Sigma_{22,\sigma}^{\rm G}(0,k).
\label{eq:Sigma22-with-SNG-}
\ee
Two background modes with nearly opposite momenta
can transfer arbitrarily small total momentum,
$q=|\bp+\bp'|\to0$, so that the intermediate
momentum approaches $\ell\to k$. Near the minimum
of the dangerous branch, this allows the same
$1/\bar m^2$ propagator enhancement as in
linear--linear scattering. At large $q$, the
propagator instead falls as $1/q^2$, without
compensating derivative factors from
$\mathcal U$. Thus, small momentum transfers
can be particularly important, although their
relative weight depends on the trispectrum.

\subsubsection{Mixed scatterings}

The mixed contributions arise from the interference
between the linear and quadratic vertices and are
controlled by the stochastic bispectrum
$B_{\lambda\lambda'\lambda''}$ of \cref{eq:b-W----}.
As shown in \cref{app:Sigma12},  the total static mixed self-energy:
\be
\begin{aligned}
\Sigma_{\rm mix,\sigma}(0,k) \equiv \Sigma_{12,\sigma}(0,k)+\Sigma_{21,\sigma}(0,k)
&=
-2g^3\,\mathrm{Re}
\sum_{\lambda,\lambda',\lambda''}
\int_{\bp,\bq}
B_{\lambda\lambda'\lambda''}
(\bp+\bq,\bp,\bq)
\\
\times\Bigg[
&\sum_{\lambda_1=\pm}
\frac{
\mathcal N_{\sigma;(\lambda_1\lambda_1)}^*
\mathcal M_{\sigma;(\lambda_1\lambda_1)}
}{
\omega_{\lambda_1}^2(\ell)
}
+
\sum_{h=\pm,0}
\mathbf n_h^\dagger
[\mathsf H^{(h)}(\ell)]^{-1}
\mathbf m_h
\Bigg],
\end{aligned}
\label{eq:Sigma12-static}
\ee
where $\ell=|\bk+\bp+\bq|$. The linear amplitudes
$\mathcal M$ and $\mathbf m_h$ are evaluated at
$(\bk,-\bp-\bq)$, while the quadratic amplitudes
$\mathcal N$ and $\mathbf n_h$ are evaluated at
$(\bk;-\bp,-\bq)$, with their background helicities
implicit.

\paragraph{Sign and bound.}
Unlike the linear--linear and Gaussian
quadratic--quadratic contributions, the mixed
self-energy is not sign-definite and can therefore
either enhance or reduce the dressed gap. It vanishes
for a Gaussian or sign-symmetric stochastic background,
for which the bispectrum is zero.

\subsection{Born stability from moment-positivity bounds}
\label{sec:bounds-k}

We now ask whether the Hartree tensor gap can survive the full Born
correction. At a fixed background, the power spectrum determines
$\bar\Delta$, $\Sigma_{11}$ and the Wick contribution
$\Sigma_{22}^{G}$. The remaining terms depend on the bispectrum and
connected trispectrum. Although the local moments do not determine
these functions, exact positivity constraints bound their contribution
and exclude a Born gap in part of the parameter space.

\begin{figure}[!htbp]
\centering
\includegraphics[width=\textwidth,height=0.69\textheight,keepaspectratio]{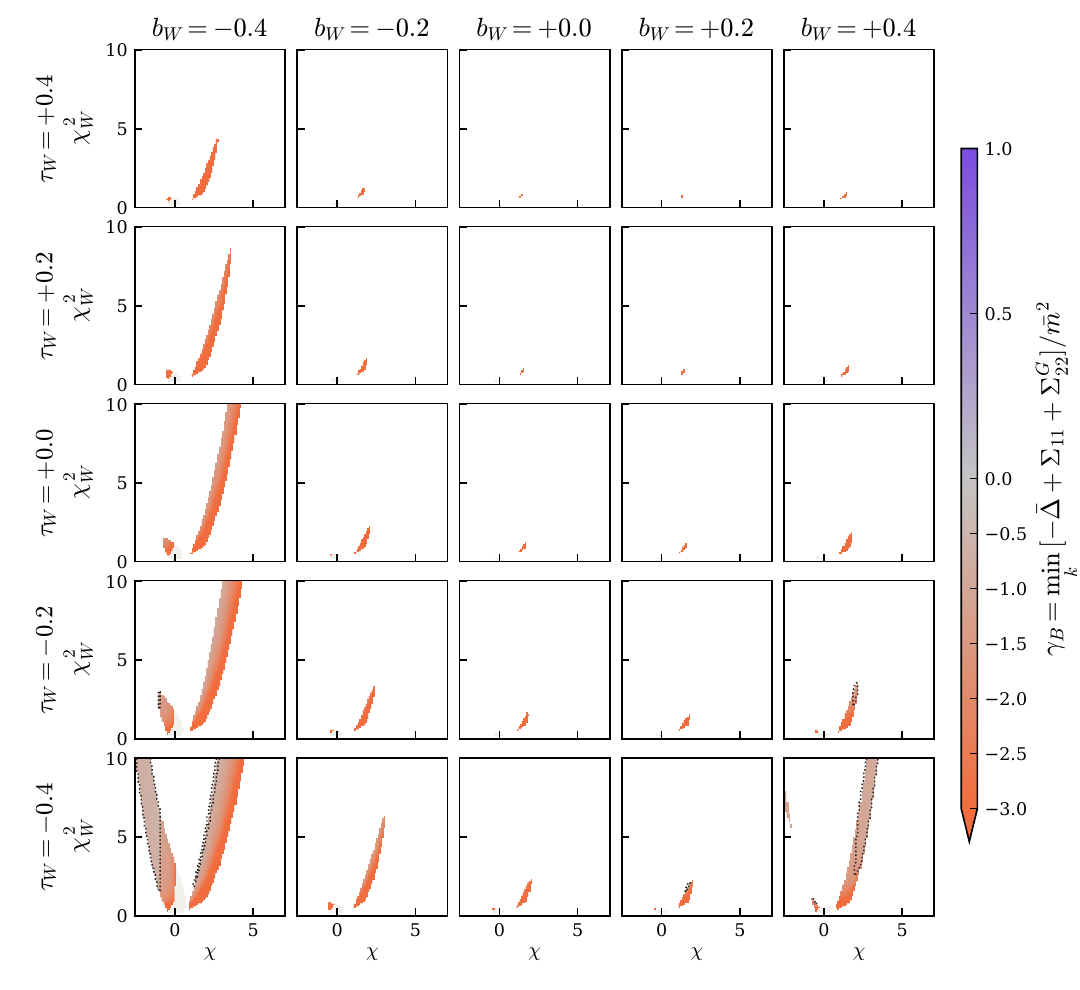}
\caption{Minimum partial Born-corrected gap obtained from the known self-energy
contributions for the soft tensor mode $h={\rm sgn}(1+\chi)\,2$, $\min_k[-\bar\Delta(0,k)+\Sigma_{11}(k)+\Sigma_{22}^{G}(k)]/\bar m^2$,
over backgrounds satisfying the classical moment equations with a positive Hartree gap.
The columns and rows fix $b_W$ and $\tau_W$, respectively.
This does not include the higher-point contributions $\Sigma_{22}^{NG}+\Sigma_{\rm mix}$; the known part alone yields no positive Born gap in the displayed scan.}
\label{fig:gbox-map}
\end{figure}

We suppress the external helicity label and evaluate all self-energies
at zero frequency. For the canonically normalized tensor mode,
\cref{eq:born-selfenergy-diagrams} and the discussion in \cref{sec:Born-Scattering-----}
give the mass gap in the static Born approximation as
\begin{align}
m_{\rm gap}^{2}
=\min_k\!\left[-\bar\Delta(0,k)+\Sigma^{\rm B}(k)\right],
\label{eq:Born-static-diagnostic}
\end{align}
where Born self-energy can be decomposed into the contributions fixed by the two-point statistics and those controlled by higher-point correlations,
\begin{align}
\Sigma^{\rm B}
=\underbrace{\Sigma_{11}+\Sigma_{22}^{G}}_{\text{known}}
+\underbrace{\Sigma_{22}^{NG}+\Sigma_{\rm mix}}_{\text{higher-point correlations}}.
\label{eq:born-split}
\end{align}
Here $\min_k$ selects the lowest point of the Born-corrected spectrum over momentum, and therefore defines the physical mass gap.
The Hartree gap satisfies
$m_H^2>0$ in the region considered.
For this gap to survive Born scattering, the higher-point contributions
must satisfy
\begin{equation}
\Sigma_{22}^{NG}(k)+\Sigma_{\rm mix}(k)
>
\bar\Delta(0,k)-\Sigma_{11}(k)-\Sigma_{22}^{G}(k)
\label{eq:Born-margin}
\end{equation}
at every momentum, with a strictly positive margin uniform in $k$. The known part has a negative minimum at every sampled background
(\cref{fig:gbox-map}). A gap therefore requires a positive contribution
from the unknown terms at the offending momenta. This can come from
suppression of the quadratic scattering, destructive interference,
or both. For a Gaussian ensemble both unknown terms vanish;
$b_W=\tau_W=0$ alone does not imply Gaussian statistics.

\paragraph{Bounds on the unknown contributions.}
Assume that $-\bar\Delta(0)$ is strictly positive on all retained
intermediate states, so that
\begin{equation}
-\bar G_0(0)=[-\bar\Delta(0)]^{-1}>0.
\label{eq:Born-positive-resolvent}
\end{equation}
For a normalized external tensor state $|e_k\rangle$, the self-adjoint
linear vertex $\Delta_W$ and centred quadratic vertex
$\delta\Delta_{W^2}=\Delta_{W^2}
-\langle\Delta_{W^2}\rangle_{\rm st}$ give
\begin{equation}
\Sigma^{\rm B}(k)
=-\left\langle
\left\|[-\bar G_0(0)]^{1/2}
(\Delta_W+\delta\Delta_{W^2})|e_k\rangle\right\|^2
\right\rangle_{\rm st}\leq0.
\label{eq:Born-positivity}
\end{equation}
The norm includes intermediate momenta and spatial--color components.
Thus $\Sigma_{11},\Sigma_{22}\leq0$: scattering lowers the Hartree
eigenvalue. Destructive interference can reduce this downward shift,
and corresponds to $\Sigma_{\rm mix}>0$ in this convention.

The local quartic moment fixes the variance of the quadratic vertex,
\begin{equation}
\left\langle\|\delta\Delta_{W^2}|e_k\rangle\|^2\right\rangle_{\rm st}
=L(\tau_W)
=g^4\mathcal W_2^2
\left[\frac{13}{10}\left(\frac75+\tau_W\right)-1\right].
\label{eq:Born-local-variance}
\end{equation}
As defined in \cref{eq:mH}, $m_H^2>0$ is a lower bound on the Hartree eigenvalues in every
intermediate channel, and $\Lambda_{\max}(k)$ an upper bound on the
states reached by $\delta\Delta_{W^2}|e_k\rangle$. Then
\begin{equation}
-\frac{L}{m_H^2}
\leq\Sigma_{22}(k)
\leq-\frac{L}{\Lambda_{\max}(k)},
\label{eq:S22-bounds-k}
\end{equation}
or, equivalently,
\begin{equation}
-\frac{L}{m_H^2}-\Sigma_{22}^{G}(k)
\leq\Sigma_{22}^{NG}(k)
\leq-\frac{L}{\Lambda_{\max}(k)}-\Sigma_{22}^{G}(k).
\label{eq:Born-connected-band}
\end{equation}
If the reachable Hartree spectrum is unbounded, take
$\Lambda_{\max}=\infty$; a numerical cutoff alone does not justify
a finite upper bound. Also, $m_H^2$ must include every intermediate
channel and need not equal the external tensor gap $\bar m^2$.
Where the lower bound in \cref{eq:Born-connected-band} is positive,
the connected trispectrum must suppress the magnitude of the Wick
contribution.

The interference is constrained by Cauchy--Schwarz. A sharper bound
also uses the spectrum-determined quantity
\begin{equation}
J(k)\equiv
\left\langle\|\bar G_0(0)\Delta_W|e_k\rangle\|^2\right\rangle_{\rm st},
\qquad
J(k)\leq-\frac{\Sigma_{11}(k)}{m_H^2}.
\label{eq:Born-J}
\end{equation}
It is obtained from $-\Sigma_{11}$ by replacing
$-\bar G_0(0)$ with $\bar G_0(0)^2$ in the intermediate propagator.
Applying Cauchy--Schwarz with these two propagator weightings gives
\begin{equation}
|\Sigma_{\rm mix}(k)|
\leq2\min\left\{
\sqrt{\Sigma_{11}(k)\Sigma_{22}(k)},\,
\sqrt{LJ(k)}\right\}.
\label{eq:Born-mixed-improved}
\end{equation}
The square root is real because both diagonal self-energies are
nonpositive. The $J$ bound is at least as restrictive as inserting
the largest allowed $-\Sigma_{22}$ into the first inequality.
In particular, reducing the quadratic scattering also restricts the
interference that can accompany it. These inequalities are the determinant conditions of the real
ensemble Gram matrices
\begin{equation}
\begin{pmatrix}
-\Sigma_{11} & -\Sigma_{\rm mix}/2\\
-\Sigma_{\rm mix}/2 & -\Sigma_{22}
\end{pmatrix}\succeq0,
\qquad
\begin{pmatrix}
J & -\Sigma_{\rm mix}/2\\
-\Sigma_{\rm mix}/2 & L
\end{pmatrix}\succeq0.
\label{eq:Born-moment-positivity}
\end{equation}
Here the real inner product is the real part of the ensemble-averaged
amplitude inner product. The construction therefore has the structure
of a truncated moment bootstrap: known moments and positivity
constrain unknown higher correlations~\cite{Lin2020,HanHartnollKruthoff2020}.
The constraints require no expansion in the non-Gaussian statistics;
the self-energy they bound remains a Born-order quantity.

\begin{figure}[!htbp]
\centering
\includegraphics[width=\textwidth,height=0.66\textheight,keepaspectratio]{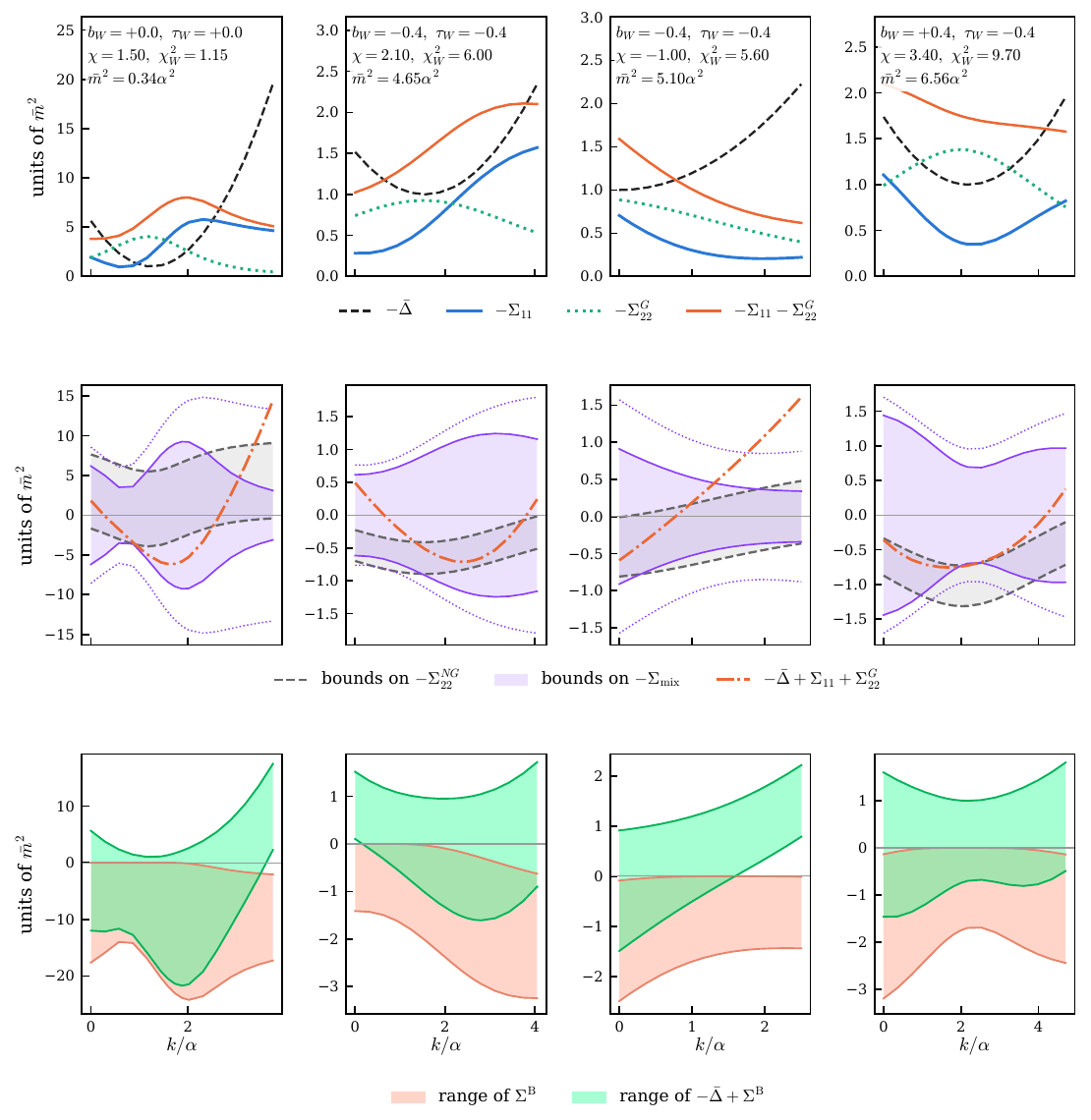}
\caption{Known contributions and outer bounds at four backgrounds
satisfying the classical moment equations with a positive Hartree
gap. All quantities are in units of $\bar m^2$.
Top: $-\bar\Delta$, $-\Sigma_{11}$, $-\Sigma_{22}^{G}$ and
$-\Sigma_{11}-\Sigma_{22}^{G}$.
Middle: bounds on $-\Sigma_{22}^{NG}$ (grey) and
$-\Sigma_{\rm mix}$ (purple), together with
$-\bar\Delta+\Sigma_{11}+\Sigma_{22}^{G}$ (dash-dotted).
The dotted purple curves show the weaker marginal
Cauchy--Schwarz bound. Positivity requires the sum of the two plotted
unknown contributions to lie below the dash-dotted curve;
their bands cannot be varied independently.
Bottom: joint outer bounds on $\Sigma^{\rm B}$ (peach) and on
$-\bar\Delta+\Sigma^{\rm B}$ (orchid), whose minimum over $k$ is
$m^2_{\rm gap}$; the two bands differ by $-\bar\Delta$.
In the top and middle rows negative self-energies are plotted as
positive scattering strengths.}
\label{fig:bounds-k}
\end{figure}

\paragraph{Excluding or certifying a Born gap.}
The quadratic and mixed contributions must be varied jointly. Their combined
statistical bounds are
\begin{equation}
    \Sigma_{\min}^{\rm B}(k)
    \leq\Sigma^{\rm B}(k)
    \leq\Sigma_{\max}^{\rm B}(k)\leq0,
    \label{eq:Born-total-band}
\end{equation}
with $\Sigma_{\rm min}^B$ and $\Sigma_{\rm max}^B$ given by
\begin{equation}
\begin{aligned}
\Sigma_{\min}^{\rm B}(k)
&=\Sigma_{11}(k)-\frac{L}{m_H^2}-2\sqrt{LJ(k)},\\[2pt]
\Sigma_{\max}^{\rm B}(k)
&=\max_{\substack{\Sigma_{22} \, \rm{allowed}}}
\left[
\Sigma_{11}(k)+\Sigma_{22}
+2\min\left\{
\sqrt{\Sigma_{11}(k)\Sigma_{22}},
\sqrt{LJ(k)}
\right\}
\right].
\end{aligned}
\label{eq:Born-total-band2}
\end{equation}
\begin{figure}[h]
\centering
\includegraphics[width=0.8\textwidth]{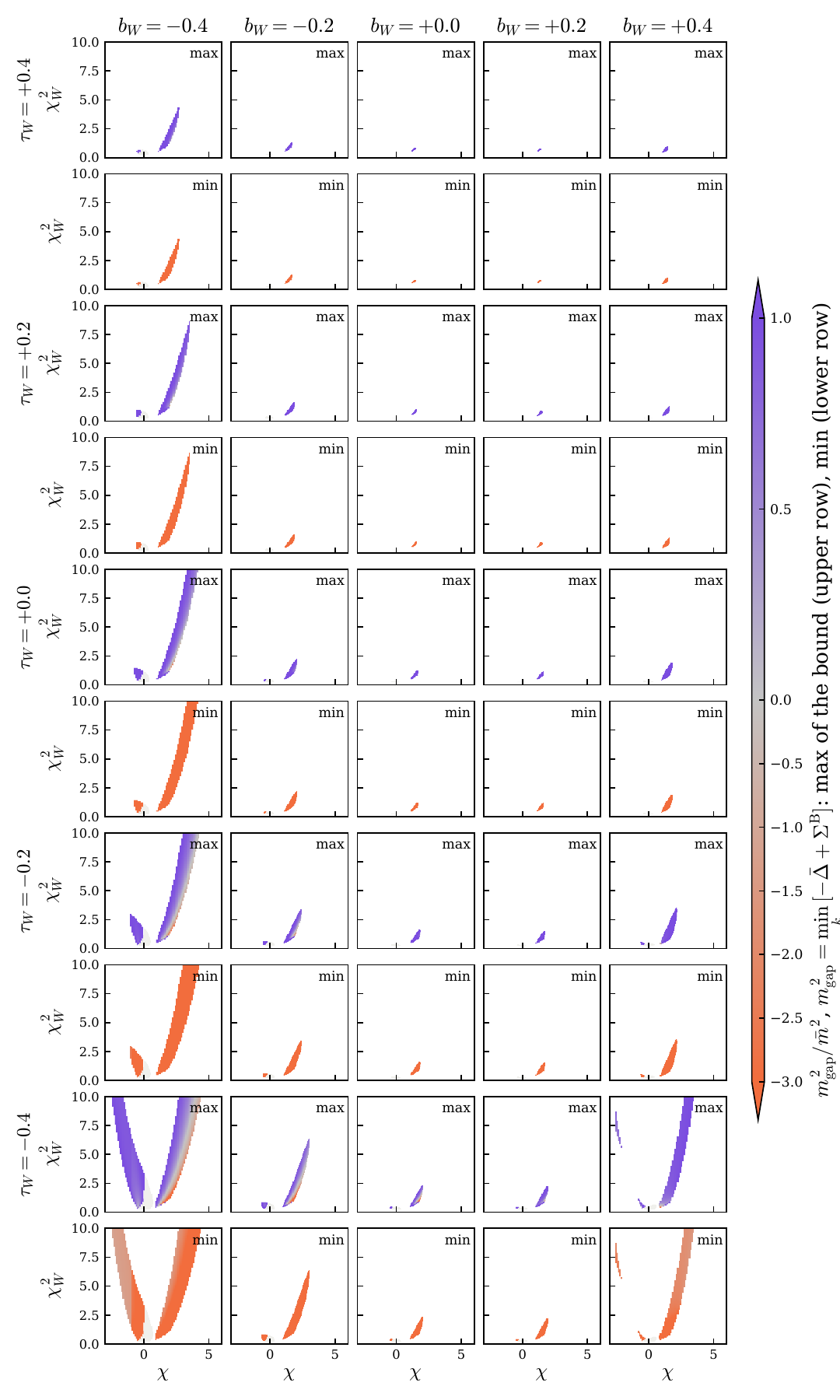}
\caption{
Outer bounds on the Born-corrected tensor mass gap, normalized by
$\bar m^2$, in the scanned region $-3<\chi<6$ and
$\chi_{\!W}^2<10$.
For each $\tau_W$, the rows marked ``max'' and ``min'' show
$m_{{\rm B},\max}^2/\bar m^2$ and
$m_{{\rm B},\min}^2/\bar m^2$, respectively.
Orange in an upper row excludes a positive Born gap, while purple in a
lower row would certify one; no certified case occurs in the scan.
}
\label{fig:gap-range}
\end{figure}

\begin{figure}[h]
\centering
\includegraphics[width=0.8\textwidth]{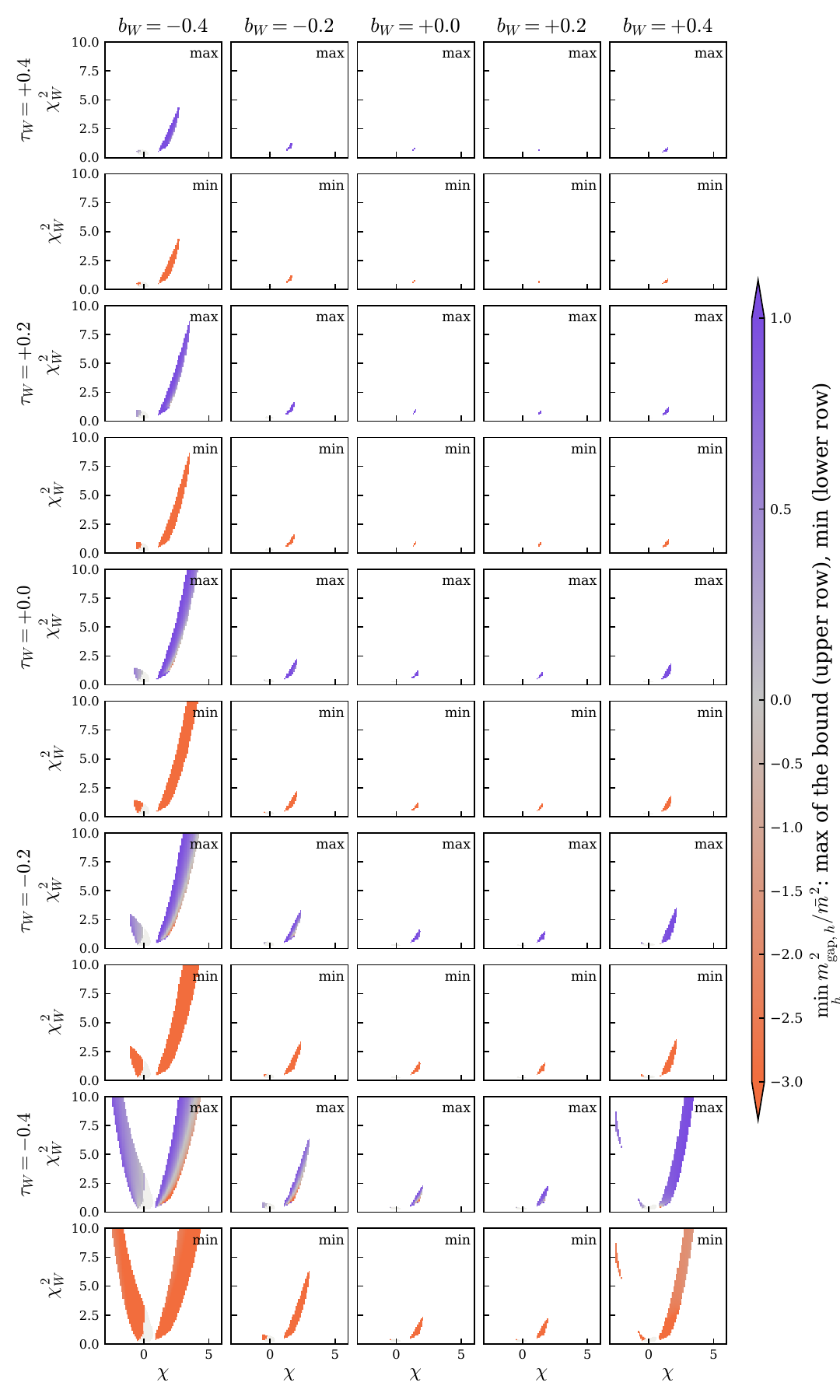}
\caption{Bounds on the Born-corrected gap across all helicities. For each
sector $h=0,\pm1,\pm2$ the bounds on the non-Gaussian scattering terms confine
$m^2_{{\rm gap},h}=\min_k\lambda_{\min}[-\bar\Delta_h+\Sigma^{\rm B}_h]$ to an
interval $\underline{m^2_{{\rm gap},h}}\le m^2_{{\rm gap},h}\le\overline{m^2_{{\rm gap},h}}$;
the upper and lower rows show $\min_h\overline{m^2_{{\rm gap},h}}$ and
$\min_h\underline{m^2_{{\rm gap},h}}$, respectively. A negative upper row
excludes a gap in some sector; a positive lower row would guarantee a gap in
all sectors. Light grey denotes backgrounds lacking a Hartree gap in at least
one sector. Units, layout, and colours follow \cref{fig:gap-range}.}
\label{fig:gap-range-all}
\end{figure}

The first line uses the bound on $J$ in \cref{eq:Born-J}, while the
second gives the least negative self-energy allowed by the joint
inequalities. The corresponding lower and upper bounds on the Born-corrected mass
gap are
\begin{equation}
m_{{\rm B},\min}^2
\equiv
\min_k\!\left[-\bar\Delta(0,k)+\Sigma_{\min}^{\rm B}(k)\right]
\leq
m_{\rm gap}^2
\leq
\min_k\!\left[-\bar\Delta(0,k)+\Sigma_{\max}^{\rm B}(k)\right]
\equiv
m_{{\rm B},\max}^2 .
\label{eq:Born-gap-bounds}
\end{equation}
Thus $m_{{\rm B},\max}^2\leq0$ excludes a positive Born gap for every
compatible ensemble, whereas $m_{{\rm B},\min}^2>0$ guarantees one.
Otherwise the gap remains unresolved. Equivalently, a nonpositive
upper bound at even one momentum is sufficient for exclusion.
Conversely, a positive upper bound does not establish a gapped
ensemble, since the required correlations at all momenta must arise
from one common probability distribution.

The same bounds show when interference is essential. If
$\Sigma_{\rm mix}=0$, the most favourable corrected spectrum obeys
\begin{equation}
-\bar\Delta(0,k)+\Sigma^{\rm B}(k)
\leq
-\bar\Delta(0,k)+\Sigma_{11}(k)
-\frac{L}{\Lambda_{\max}(k)}.
\label{eq:Born-zero-mixed}
\end{equation}
Where the right-hand side is nonpositive, positive mixed interference
is necessary to preserve a gap. In particular, if
$-\bar\Delta(0,k)+\Sigma_{11}(k)\leq0$, suppressing the quadratic
channel alone cannot restore one, since its self-energy contribution
remains nonpositive.

\paragraph{Born-stability domains.}
The scan excludes a positive Born tensor gap for $795$ of the $6757$
coarse-grid backgrounds and $373$ of the $3611$ fine-grid
backgrounds ($790$ and $373$ when the admissible $\Sigma_{22}$ is
minimised exactly rather than on a grid, as in \cref{sec:quantum-structure-Hartree}). No sampled
background has $m_{{\rm B},\min}^2>0$; the remaining $5962$ and $3238$
backgrounds, respectively, therefore remain unresolved by these inequalities.
The excluded points lie on the lower flanks of the positive-$\chi$
branches, with $\tau_W\leq0$, $b_W\leq0.2$, and $\chi>1$ in this scan
(\cref{fig:gap-range}). These exclusions rely on the sharper $J$ bound:
the weaker marginal Cauchy--Schwarz bound excludes none of them.
Physically, the linear scattering is too large for the allowed
interference to compensate, even after optimizing the quadratic
contribution. Including the scalar and vector sectors
(\cref{fig:gap-range-all}) changes this picture only marginally: the
excluded set grows by three backgrounds, all through the scalar block at
$b_W=\tau_W=-\tfrac25$, $\chi=-0.4$, and no background acquires a
guaranteed gap in all sectors, the lower edge of $\min_h m^2_{{\rm gap},h}$
being $\le-1.47\,\bar m^2$ everywhere. The four examples in
\cref{fig:bounds-k} remain unresolved. In the first,
$-\bar\Delta+\Sigma_{11}<0$ near the Hartree minimum, so positive mixed
interference is necessary. In the other three, the bounds still permit a
positive gap near the minimum through sufficient suppression of quadratic
scattering. At $\chi=2.1$ and $\chi=3.4$, the variance bound already
requires a positive $\Sigma_{22}^{NG}$ there. Establishing a Born gap
nevertheless requires compatible higher-point correlations over the full
momentum range. These results concern the static sector at Born order;
control of higher scattering orders is a separate requirement for
perturbative stability.

\FloatBarrier

\paragraph{Single shells.}
We repeated the analysis for $1085$ gapped single-shell solutions on
the boundary ${\rm Var}(x)=0$ of the broad-spectrum region. Their Born-gap classification closely
tracks the broad-spectrum results: $179$ shells are excluded, none is
certified, and the remaining shells are unresolved. At matched
backgrounds the gap bounds differ by only
$0.1$--$0.3\,\bar m^2$ with no systematic sign, indicating that the
spectral width is not a significant stability lever at Born order.
We therefore do not show the single-shell maps separately.

\paragraph{Numerical resolution of the band.}
For broad spectra, the kernels are averaged over the normalized power density
$\rho(\chi_k)$ defined in \cref{eq:broad-rho-def},
\[
\Sigma_{11}\propto
\int d\chi_q\,\rho(\chi_q)\,
\mathcal I_{11}(k;\chi_q),
\qquad
\Sigma_{22}^{G}\propto
\iint d\chi_q\,d\chi_{q'}\,
\rho(\chi_q)\rho(\chi_{q'})\,
\mathcal I_{22}(k;\chi_q,\chi_{q'}),
\]
with $\chi_q=\lambda q/\alpha$ and
$\chi_{q'}=\lambda' q'/\alpha$.
We evaluate these integrals by Gauss--Legendre quadrature with three nodes per
band segment, splitting the support at $\chi_k=0$ when it crosses the helicity
boundary (six nodes in total). The nodes are numerical quadrature points, not
physical shells. Since the single-shell kernels change only smoothly across the typical band
half-width $\sqrt{3\,{\rm Var}(\chi_k)}\simeq0.1$--$0.6$, they are well
approximated by a low-order polynomial over the integration range. A
three-point Gauss--Legendre rule, which integrates polynomials up to degree
five exactly, therefore already provides high accuracy.
Increasing to $5$, $8$, and $12$ nodes changes $\Sigma_{11}$ and
$\Sigma_{22}^{G}$ by less than $3\times10^{-5}$ and the Born gap by less than
$1.2\times10^{-4}\alpha^2$. By comparison, the one-node approximation at
$\bar\chi_k$ differs from the converged broad-spectrum result by
$0.1$--$5\%$. Angular integrations use $48$ Gauss--Legendre nodes in
$\cos\theta$ for $\Sigma_{11}$ and $12^3$ nodes in
$(\cos\theta_1,\cos\theta_2,\phi_2)$ for $\Sigma_{22}^{G}$, converged to
$10^{-14}$ and $4\times10^{-6}$, respectively.

\subsection{Gauge dependence of the tensor gap}
\label{sec:gauge-dependence}

Physical stability must be gauge independent, whereas the
Hartree approximation need not be. To clarify this distinction,
consider the background-field gauge-fixing term from \cref{eq:gf-BG2}
where $\xi=1$ corresponds to Feynman gauge, adopted throughout
our analysis.

For each classical background realization satisfying the equations
of motion, pure-gauge fluctuations are zero modes of the unfixed
quadratic operator. However, these directions depend on the
stochastic background, so the projection onto physical modes
does not generally commute with ensemble averaging. Consequently,
the Hartree tensor gap depends explicitly on the gauge choice:
\begin{equation}
\bar m^2(\xi)
=
-g^2Q^2-\frac{\alpha^2}{4}
+g^2W_2\left(
\frac{7}{10}+\frac{3}{10\xi}
\right),
\end{equation}
recovering our expression
$\bar m^2=g^2W_2-g^2Q^2-\alpha^2/4$
at $\xi=1$. Thus, a positive Hartree gap alone does not
establish gauge-invariant stability.

The Born correction must be evaluated consistently in the same
gauge. In our Lorentzian convention, the corrected static tensor
kernel is
\begin{equation}
D_{\rm B}^{(\xi)}(k)
=
-\bar\Delta^{(\xi)}(0,k)
+\Sigma_W^{B,(\xi)}(0,k),
\end{equation}
where the gauge dependence enters both the dressed propagator
and the scattering vertices. Numerically, although the Hartree
gap and individual Born contributions vary appreciably with
$\xi$, the spectrum-determined correction
$\Sigma_{11}+\Sigma_{22}^{G}$ closes the tensor gap at the
representative backgrounds for all gauge choices examined.
This indicates that the observed gap closure is not merely
an artefact of Feynman gauge. Nevertheless, the minimum of a truncated, off-shell static
kernel is not itself gauge invariant. Moreover, the mixed and
connected non-Gaussian Born contributions remain undetermined.
Our results therefore provide evidence for the robustness of
the gap closure, rather than a definitive proof of physical
instability, which requires the complete correction projected
onto physical fluctuations.

\section{Conclusion}
\label{sec:conclusion}

In this work, we investigated whether stochastic spin-$2$ dressing can
stabilize the chromomagnetic condensate of axion--Yang--Mills theory.
We considered static $SU(2)$ backgrounds of the form
$A_i^a=Q\delta_i^a+W_i^a$ (\cref{eq:BG-gauge-field})
at fixed axion shift charge (\cref{eq:B-eq-alpha}), with $W_i^a$
statistically homogeneous and isotropic. Rather than requiring each
realization to satisfy the classical field equations pointwise, we
imposed self-consistency through the coarse-grained moment equations,
which constrain the spectrum and higher moments of the dressing
(\cref{eq:tensor-dispersion-responses,eq:scalar-trace---}).
For thin-shell spectra, the scalar and tensor equations determine
the signed wavenumber and amplitude (\cref{sec:thin-shell}).
More generally, the broad-spectrum equations fix the mean and
variance of the signed wavenumber (\cref{sec:broad-spectrum-W}),
with $\rm Var(\sigma)=0$ defining the thin-shell boundary and $\rm Var(\sigma)>0$
corresponding to finite-width spectra. We also established the
realizability bounds $\tau_W\ge-2/5$ and
$6b_W^2\le7/5+\tau_W$, constraining the allowed bispectrum
and connected-trispectrum moments.

We next examined quantum fluctuations in the Hartree approximation,
replacing the full fluctuation operator by its stochastic average
$\bar\Delta$ (\cref{sec:Hartree--Delta}).
At this level, the dressing enters $\bar\Delta$ only through its
two-point intensity $\mathcal W_2$, while $b_W$ and $\tau_W$
indirectly determine which points in the $(\chi,\chi_W)$ plane
satisfy the background equations. Combining classical
self-consistency with a positive Hartree gap identifies the
candidate gapped condensates shown in
\cref{fig:param-panels,fig:param-phase}. The helicity sector setting the tightest Hartree-gap bound
varies across parameter space (\cref{fig:smallest-gap-helicity}). The size and location of this region depend on the higher moments of the stochastic dressing through the background
equations. In \cref{sec:confinement-validity}, we discuss the weak-coupling regime
and the conditions under which confinement effects and the running
of the gauge coupling can be neglected.

The Hartree approximation, however, neglects scattering of quantum
fluctuations from the stochastic background. We therefore computed
the leading Born self-energy (\cref{sec:Born-Scattering-----}).
Its power-spectrum-determined contributions, $\Sigma_{11}$ and
$\Sigma_{22}^{G}$ (\cref{eq:born-split}), are already sufficient,
when taken alone, to close the Hartree gap throughout the backgrounds
examined (\cref{fig:gbox-map}).
A positive Born gap therefore requires compensating contributions
from the genuinely non-Gaussian channels, $\Sigma_{\rm mix}$ and
$\Sigma_{22}^{\rm NG}$, governed by the momentum-dependent
bispectrum and connected trispectrum, respectively.

The classical moment equations constrain only particular projections
of these higher-point functions, leaving the non-Gaussian scattering
contributions partly undetermined. We therefore derived bounds
using exact statistical positivity inequalities
(\cref{sec:bounds-k,fig:bounds-k}).
These bounds exclude a positive Born gap in part of the
Hartree-gapped parameter space and delimit the region where
non-Gaussian correlations may preserve it (\cref{fig:gap-range}). Extending the analysis to the scalar and vector sectors
(\cref{fig:gap-range-all}) enlarges the excluded region only marginally and
leaves no background with a guaranteed gap in all sectors.
The remaining region thus represents candidate Born-gapped
configurations rather than established stable states. Demonstrating
stability requires a realizable non-Gaussian ensemble whose
momentum-dependent correlations sustain a positive gap, together
with perturbative control beyond Born order. The closure of the gap persists for all gauge parameters examined (\cref{sec:gauge-dependence}), whereas the Hartree gap itself is gauge dependent; the instability is therefore a property of the dressed condensate rather than of Feynman gauge.

Several directions follow from these results. A natural next step
is to construct explicit non-Gaussian ensembles within the
remaining region, determine their higher-point correlations,
and assess higher-order scattering corrections. Including
dynamical fermions would extend the analysis to fermion
production, chiral backreaction, anomaly-induced transport,
and infrared axion--QCD dynamics. Ultimately, determining the nonlinear fate of the color--spin--locked
condensate, and whether stochastic spin-$2$ dressings can form
and persist dynamically, requires following the real-time evolution
of the coupled axion--gauge system. Lattice simulations at fixed
conserved axion shift charge provide a natural framework for
this investigation, which we leave for future work.

\section*{Acknowledgments}
\addcontentsline{toc}{section}{Acknowledgments}
We would like to thank Simon Caron-Huot, Gia Dvali, Zohar Komargodski, Harvey Meyer, and Raju Venugopalan for helpful discussions. We are particularly grateful to Eiichiro Komatsu for his
thoughtful feedback,  and support during
the development of this work. We thank Anthropic’s Claude for assistance with the numerical scans. All scientific content, calculations, and conclusions have been independently verified by the authors. %; the numerics were independently redone by the authors, who take full responsibility for the results. 
The work of AM was supported by the Royal Society through a University Research Fellowship (Grant No. RE22432). 

\appendix

\crefalias{section}{appendix}

\section{Mathematical  Tools}\label{app:Math}

This appendix collects the mathematical identities, matrix relations, and algebraic tools used throughout the derivations in the main text. We summarize them here to keep the discussion self-contained and to streamline the presentation of the calculations.

\subsection*{The Cayley--Hamilton theorem:}

For a traceless matrix $W_\mu^a$, the Cayley--Hamilton theorem yields the identities
\be
\epsilon_{ijk}\epsilon^{abc}\big[ W^b_j(W^2)^c_k
-\frac{1}{2}
W^b_j\delta^c_k\,\tr(W^2)\big]
=
\left[
W^a_i \, \tr(W^2)-(W^3)^a_i]
\right],
\label{eq:CH-1}
\ee
as well as
\be
\frac12 \epsilon_{ijk} \epsilon^{abc} W^b_j W^c_k  = \text{det}W (W^{-1})^a_i  = (W^2)^a_i - \frac12 \, \tr[W^2] \, \delta^a_i.
\label{eq:CHI}
\ee

\subsection*{Helicity basis and polarization tensors}

For later use, we summarize the helicity basis for the spin-$2$ sector.
The polarization tensors are defined as
\begin{equation}
e^\lambda_{ij}(\hat{\bk})
=
\epsilon^\lambda_i(\hat{\bk})\,
\epsilon^\lambda_j(\hat{\bk}),
\qquad
i\hat{\bk}\times\boldsymbol\epsilon^\lambda
=
\lambda\,\boldsymbol\epsilon^\lambda,
\qquad
\lambda=\pm,
\end{equation}
so that $\lambda=\pm$ labels helicity $\pm2$. They are symmetric,
transverse, and traceless, and satisfy
\begin{equation}
i\epsilon^{ikj}\hat k_k e^\lambda_{jl}
=
\lambda e^\lambda_{il}.
\label{eq:k-e-2}
\end{equation}
Accordingly, $\epsilon_{ijk}\partial^j$ acting on the spatial index of
a helicity-$\lambda$ Fourier mode gives $\lambda |\bk|$.
We choose
\begin{equation}
e^\lambda_{ij}(-\hat{\bk})
=
\big[e^\lambda_{ij}(\hat{\bk})\big]^*,
\qquad
e^\lambda_{ij}(\hat{\bk})\,
e^{\lambda'}_{ij}(-\hat{\bk})
=
\delta_{\lambda\lambda'}.
\end{equation}

\section{Details of the Bispectrum and Trispectrum Calculations}
\label{app:bispectrum-trispectrum}

In this appendix we give the details of the bispectrum and trispectrum
calculations summarised in \cref{sec:stochastic-spin2-background}. The two correlators rest on
ingredients of rather different character. The Gaussian part of the
trispectrum follows from the disconnected Wick contractions and depends
on the background only through its helicity power spectrum
$P_\lambda(k)$; we evaluate it in \cref{app:trispectrum-gaussian}. The
remaining contributions are controlled by how the composite operators
$Z$, $U$ and $V$ respond to the background, through the cross-spectra
$\langle W^*_\lambda O_\lambda\rangle'_{\rm st}$: the responses of $Z$
and $U$ determine the bispectrum, while that of $V$ determines the
connected, non-Gaussian part of the trispectrum. These cross-spectra
involve higher moments of $W$ that the power spectrum alone does not
fix. In \cref{app:bispectrum-responses} we show that, for a background concentrated
on a single thin shell populated in one helicity, these responses collapse to
a handful of constants fixed by the moments of $W$.

\subsection{Gaussian and connected trispectrum}\label{app:trispectrum-gaussian}

For Gaussian statistics, Wick's theorem gives
\be
\begin{aligned}
\left\langle
W_\lambda^*(\bk)
W_\lambda(\bp)
W_\lambda(\bq)
W_\lambda(\bl)
\right\rangle_{\rm G}
 &=
P_\lambda(k)\,
\delta^{(3)}(\bk-\bp)\,
P_\lambda(q)\,
\delta^{(3)}(\bq+\bl)
\nonumber\\
&+
P_\lambda(k)\,
\delta^{(3)}(\bk-\bq)\,
P_\lambda(p)\,
\delta^{(3)}(\bp+\bl)
\nonumber\\
&+
P_\lambda(k)\,
\delta^{(3)}(\bk-\bl)\,
P_\lambda(p)\,
\delta^{(3)}(\bp+\bq).
\end{aligned}
\label{eq:Wick-four-app}
\ee
Using the above and \cref{eq:def-V} in $\left\langle
W_\lambda^*(\bk)V_\lambda(\bk)
\right\rangle'_{\rm G}$, the last contraction gives precisely
$\mathcal W_2P_\lambda(k)$,
and is cancelled by the subtraction $-\mathcal W_2W_\lambda$. The two remaining exchange contractions are equal and give
\be
\left\langle
W_\lambda^*(\bk)V_\lambda(\bk)
\right\rangle'_{\rm G}
=
2P_\lambda(k)
\int d^3q\,
P_\lambda(q)
\left|
e^\lambda(\hat k):e^\lambda(\hat q)
\right|^2.
\label{eq:WV-G-before-angle}
\ee
where the prime on the correlator means that the overall momentum-conserving delta function has been removed.
Without introducing an explicit polarization basis, the angular integral can be evaluated as follows
\be
\int\frac{d\Omega_{\hat q}}{4\pi}\,
\left|
e^\lambda(\hat k):e^\lambda(\hat q)
\right|^2
=
\frac15.
\label{eq:polarization-average-one-fifth}
\ee
The result is independent of whether the occupied mode is $\lambda=+$ or $\lambda=-$. The radial shell integral then gives
\be
\left\langle
W_\lambda^*(\bk)V_\lambda(\bk)
\right\rangle'_{\rm G}
=
\frac{2}{5}\,
\mathcal W_2P_\lambda(k).
\ee
The factor $2$ comes from the two exchange Wick contractions, whereas the factor $1/5$ follows from rotational averaging in the five-dimensional $j=2$ representation.

For general, non-Gaussian statistics, the four-point function
decomposes into its Gaussian Wick contractions and a connected
trispectrum. Consequently, the full cross-spectrum reads
\be
\begin{aligned}
\left\langle W_\lambda^*(\bk)V_\lambda(\bk)\right\rangle'
=
\underbrace{\frac{2}{5}\mathcal W_2P_\lambda(k)}
_{\text{Gaussian}}+
\underbrace{
\sum_{\lambda',\lambda'',\lambda'''}
\int_{\bp,\bq}
\mathcal K_{\lambda\lambda'\lambda''\lambda'''}
(\bk,\bp,\bq,\mathbf s)\,
T^c_{\lambda\lambda'\lambda''\lambda'''}
(\bk,\bp,\bq,\mathbf s)
}_{\text{Non-Gaussian}},
\end{aligned}
\label{eq:WV-G-NG}
\ee
where $\mathbf s=\bk-\bp-\bq$ and $\mathcal K_{\lambda\lambda'\lambda''\lambda'''}$
is the quartic Yang--Mills polarization kernel,
\be
\mathcal K_{\lambda\lambda'\lambda''\lambda'''}
=
\operatorname{tr}\left[
e^{\lambda'}(\hat{\bp})
e^{\lambda''}(\hat{\bq})
\right]
\operatorname{tr}\left[
e^{\lambda *}(\hat{\bk})
e^{\lambda'''}(\hat{\mathbf s})
\right],
\label{eq:app:Math-K}
\ee
and $T^c_{\lambda\lambda'\lambda''\lambda'''}$
denotes the connected tensor trispectrum,
\be
T^c_{\lambda\lambda'\lambda''\lambda'''}
(\bk,\bp,\bq,\mathbf s)
\equiv
\left\langle
W_\lambda^*(\bk)
W_{\lambda'}(\bp)
W_{\lambda''}(\bq)
W_{\lambda'''}(\mathbf s)
\right\rangle'_c.
\label{eq:app-T}
\ee
The first term contains the two surviving Gaussian Wick
contractions, while the second captures genuine connected
four-point correlations, including mixed-helicity contributions.
The latter vanishes for Gaussian statistics.

\subsection{Response relations for a single-helicity shell}
\label{app:bispectrum-responses}

In this part we derive the relations used in
\cref{sec:thin-shell}. We consider a background supported on
a single thin shell in Fourier space, populated in the single helicity
$\lambda={\rm sgn}(k_\star)$, so that the signed wavenumber
$k_\star=\lambda|k_\star|$ records both the radius of the shell and its
handedness. We write
$\mathcal W_n\equiv\langle{\rm tr}[W^n]\rangle_{\rm st}$ and
$b_W\equiv\mathcal W_3/\mathcal W_2^{3/2}$. The relations we need are
\begin{align}
&\int_{\rm shell}d^3k\,
\left\langle W^*_{\lambda}O_{\lambda}\right\rangle'_{\rm st}
=\left\langle{\rm tr}[WO]\right\rangle_{\rm st},
\qquad
{\rm tr}[WZ]={\rm tr}[W^3],
\nonumber\\
&{\rm tr}[WU]=3\,{\rm tr}\!\left[W^2(\nabla\times W)\right],
\qquad
\nabla\times W=k_\star\,W.
\label{eq:app-claims}
\end{align}
The first converts a helicity-projected cross-spectrum into a
real-space average, the second and third are pointwise algebraic
identities, and the last expresses the fact that a single-helicity
shell is an eigenspace of the curl. Together they fix the response
coefficients and the helicity of the background in terms of
$\mathcal W_2$, $b_W$ and $k_\star$ alone, as summarised in
\cref{eq:app-responses,eq:app-responses2}.

\subsection*{Helicity spectra and real-space averages}

Let $O_{ai}(\bx)$ be any statistically homogeneous matrix field; it
need not be symmetric, traceless or transverse. By
\cref{eq:W-decompose}, the background on the shell is
$W_{ai}(\bk)=e^{\lambda}_{ai}(\hat\bk)\,W_{\lambda}(\bk)$, and $W$
vanishes away from the shell. Since the helicity components of $O$ are
$O_\lambda(\bk)=[e^\lambda_{ai}(\hat\bk)]^*O_{ai}(\bk)$, contracting
with $W$ projects $O$ onto the populated helicity,
\be
W_{ai}(\bk)^*\,O_{ai}(\bk)
=W^*_{\lambda}(\bk)\,O_{\lambda}(\bk),
\ee
and all other components of $O$ drop out. As $W(\bx)$ is real,
Parseval's theorem then gives
\be
\int d^3x\,W_{ai}(\bx)\,O_{ai}(\bx)
=(2\pi)^3\int_{\rm shell}d^3k\,
W^*_{\lambda}(\bk)\,O_{\lambda}(\bk).
\ee
After averaging over the ensemble, statistical homogeneity turns the
left-hand side into $\mathrm{Vol}\,\langle W_{ai}O_{ai}\rangle_{\rm st}$,
while on the right-hand side the equal-momentum correlator supplies a
factor $(2\pi)^3\delta^{(3)}(\mathbf 0)=\mathrm{Vol}$ multiplying the
primed correlator. Because $W$ is symmetric,
$W_{ai}O_{ai}={\rm tr}[WO]$, and we obtain
\be
\int_{\rm shell}d^3k\,
\left\langle W^*_{\lambda}O_{\lambda}\right\rangle'_{\rm st}(\bk)
=\left\langle{\rm tr}[WO]\right\rangle_{\rm st},
\label{eq:lemma1}
\ee
which is the first relation in \cref{eq:app-claims}. It holds for any
shell populated in a single helicity, independently of its width.

Two consequences of \cref{eq:lemma1} are used in the main text. First,
for $O=W$ it gives $\int_{\rm shell}d^3k\,P_\lambda=\mathcal W_2$. Second, on a thin shell the
cross-spectrum of any operator with the background can be written as
$\langle W^*_{\lambda}O_{\lambda}\rangle'_{\rm st}=R_O\,P_{\lambda}(k)$,
with a response coefficient $R_O$ that is constant across the shell.
Integrating over the shell then fixes
\be
R_O=\frac{\langle{\rm tr}[WO]\rangle_{\rm st}}{\mathcal W_2}.
\label{eq:corollary}
\ee
For $O=V$ one has
$W_{ai}V_{ai}=({\rm tr}[W^2]-\mathcal W_2)\,{\rm tr}[W^2]$, so that
$\langle W_{ai}V_{ai}\rangle_{\rm st}$ is the variance of
${\rm tr}[W^2]$, and \cref{eq:corollary} reproduces
$R_V=( \frac25 +\tau_W)\mathcal W_2$.

\subsection*{The $Z$-response}

The $Z$-response requires only that $W$ be symmetric and traceless.
Pointwise,
\be
{\rm tr}[WZ]
=W_{ai}W_{ab}W_{bi}-\frac13\,W_{aa}\,{\rm tr}[W^2]
={\rm tr}[W^3],
\ee
which is the second relation in \cref{eq:app-claims}. Averaging gives
$\langle{\rm tr}[WZ]\rangle_{\rm st}=\mathcal W_3=b_W\mathcal W_2^{3/2}$,
and \cref{eq:corollary} yields $R_Z=b_W\mathcal W_2^{1/2}$.

\subsection*{The $U$-response and a cubic identity}

The $U$-response involves one spatial derivative and requires more
work. Since $W$ is symmetric and traceless, the trace part of $U$ drops
out of its contraction with $W$,
\be
W_{ai}U_{ai}
=W_{ai}\bigl(Y_{ai}+Y_{ia}\bigr)-\frac23\,W_{aa}\,{\rm tr}[Y]
=2\,W_{ai}Y_{ai}
=T_1-2T_2 ,
\ee
where the two cubic invariants
\be
T_1\equiv\epsilon_{abc}\,W_{ai}W_{bj}\,\partial_iW_{cj},
\qquad
T_2\equiv\epsilon_{abc}\,W_{ai}W_{bj}\,\partial_jW_{ci}
\ee
differ only in which index of the differentiated field is contracted
with the derivative. They are in fact equal and opposite: relabelling
$a\leftrightarrow b$ and $i\leftrightarrow j$ in $T_2$ gives
$\epsilon_{bac}W_{bj}W_{ai}\partial_iW_{cj}=-T_1$. Hence
\be
{\rm tr}[WU]=3\,T_1
\label{eq:WU-T1}
\ee
pointwise.

It remains to express $T_1$ through the curl of $W$,
$(\nabla\times W)_{bm}\equiv\epsilon_{mjk}\partial_jW_{bk}$. The two
undifferentiated fields in $T_1$ are antisymmetrised in their first
index and therefore assemble into the cofactor matrix of $W$,
$C_{ck}\equiv\frac12\,\epsilon_{abc}\,\epsilon_{ijk}\,W_{ai}W_{bj}$.
Contracting this definition with $\epsilon_{lmk}$ and using
$\epsilon_{ijk}\epsilon_{lmk}=\delta_{il}\delta_{jm}-\delta_{im}\delta_{jl}$
gives $\epsilon_{abc}W_{ai}W_{bj}=\epsilon_{ijk}C_{ck}$, so that
\be
T_1=\epsilon_{kij}\,C_{ck}\,\partial_iW_{cj}=C_{ck}\,(\nabla\times W)_{ck}.
\ee
For a $3\times3$ matrix the Cayley--Hamilton theorem expresses the
cofactor matrix as a quadratic polynomial in the matrix
itself.\footnote{In general
$C_{kc}=(W^2)_{ck}-{\rm tr}[W]\,W_{ck}+\frac12\big({\rm tr}[W]^2-{\rm tr}[W^2]\big)\delta_{ck}$,
i.e.\ $C^{T}$ is the adjugate $\det(W)\,W^{-1}$.}
For symmetric traceless $W$ it reads
\be
C_{ck}=(W^2)_{ck}-\frac12\,{\rm tr}[W^2]\,\delta_{ck}.
\ee
The trace term does not contribute, because
${\rm tr}[\nabla\times W]=\epsilon_{bjk}\partial_jW_{bk}$ vanishes for
symmetric $W$. We thus arrive at
\be
\epsilon_{abc}\,W_{ai}W_{bj}\,\partial_iW_{cj}
={\rm tr}\!\left[W^2(\nabla\times W)\right],
\label{eq:lemma2}
\ee
valid pointwise for any symmetric traceless field; transversality is
not needed. Together with \cref{eq:WU-T1}, this is the third relation
in \cref{eq:app-claims}.

\subsection*{Curl on a single-helicity shell}

A shell populated in a single helicity is an eigenspace of the curl.
With $e^\lambda_{ak}=\epsilon^\lambda_a\epsilon^\lambda_k$ and
$i\hat\bk\times\boldsymbol\epsilon^\lambda=\lambda\boldsymbol\epsilon^\lambda$,
each Fourier component satisfies
\be
\epsilon_{ijk}\,\partial_j\!\left[e^\lambda_{ak}\,e^{i\bk\cdot\bx}\right]
=\epsilon^\lambda_a\,(i\bk\times\boldsymbol\epsilon^\lambda)_i\,e^{i\bk\cdot\bx}
=\lambda|\bk|\,e^\lambda_{ai}\,e^{i\bk\cdot\bx}.
\ee
The components at $-\bk$ required by the reality of $W$ are the complex
conjugates of these, and since the curl is a real operator with a real
eigenvalue they carry the same eigenvalue; equivalently,
$e^\lambda(-\hat\bk)=[e^\lambda(\hat\bk)]^*$ is again of helicity
$\lambda$ with respect to $-\hat\bk$. On a thin shell every component
has $|\bk|=|k_\star|$, so that
\be
\nabla\times W=k_\star\,W
\ee
pointwise, which is the last relation in \cref{eq:app-claims}.

\subsection*{The $U$-response and the helicity of the background}

Combining \cref{eq:WU-T1,eq:lemma2} with the curl eigenvalue gives
\be
{\rm tr}[WU]
=3\,{\rm tr}\!\left[W^2(\nabla\times W)\right]
=3k_\star\,{\rm tr}[W^3]
=3k_\star\,{\rm tr}[WZ]
\ee
pointwise. On a single shell, $U$ and $3k_\star Z$ therefore have
identical projections onto the background, and the $U$-response is
fixed by the $Z$-response and the shell wavenumber,
$R_U=3k_\star R_Z$. The same eigenvalue determines the helicity of the
background, $\langle{\rm tr}[W(\nabla\times W)]\rangle_{\rm st}=k_\star\mathcal W_2$,
so that $H_W=-k_\star\mathcal W_2$ and $h_W=-k_\star$. Collecting
results,
\be
R_Z=b_W\,\mathcal W_2^{1/2},
\qquad
R_V=(\frac25+\tau_W)\,\mathcal W_2,
\qquad
R_U=3k_\star\,b_W\,\mathcal W_2^{1/2}=-3h_W\,b_W\,\mathcal W_2^{1/2},
\label{eq:app-responses}
\ee
and 
\be
H_W=-k_\star\,\mathcal W_2 .
\label{eq:app-responses2}
\ee

All the algebraic steps above are exact pointwise identities, and
\cref{eq:lemma1} is exact for any single-helicity shell. The only
approximation is the thin-shell limit, in which the tensor kernel, the
response coefficients and the eigenvalue relation
$\nabla\times W=k_\star W$ are constant across the shell up to
corrections of relative order $\Delta k/|k_\star|$.

\section{One-Loop Effective Action:  Renormalization and Instabilities}\label{app:1-loop-action}

In this section, we first provide a brief review of one-loop effective field theory
based on the effective action expanded around a classical background.
We then work out the explicit form of the effective action. The one-loop effective action encodes renormalization effects, such as the running of the gauge coupling, as well as genuine quantum phenomena. In particular, the appearance of an imaginary part signals an instability of the background and is directly related to particle production, as in pair creation in external fields. This formalism will be used throughout this work.

At the quantum level, the fluctuations $\mathcal{A}_\mu$ and the ghost fields $c^a$, $\bar c^a$ can be integrated out in the functional integral. At one-loop order, the action naturally separates as
\begin{align}
    \Gamma[A_\mu] 
    = S_{_{\text{YM}}}[\bar{A}_\mu] + S_{_{\text{CS}}}[\bar{A}_\mu] + \Gamma_{\text{1-loop}}[\bar{A}_\mu], 
\label{Eq:EffectiveAction_Decomposition}
\end{align}
where $S_{\text{YM}}[\bar A_\mu]$ is the classical Yang-Mills action (with gauge fixing), $S_{\text{CS}}[\bar A_\mu]$ is the classical CS term and $\Gamma_{\text{1-loop}}[\bar A_\mu]$ encodes the quantum corrections from integrating fluctuations of the gauge field
\be
e^{i \Gamma_{\text{1-loop}}[\bar A]}
=
\int \mathcal{D}\mathcal{A}\, \mathcal{D}\bar c\, \mathcal{D}c\,
\exp\!\left[
i \int d^4x\, \mathcal{L}^{(2)}
\right].
\ee
The quadratic action for the gauge and ghost fluctuations can be written in Gaussian form as
\begin{equation}
\mathcal{L}^{(2)}=
\frac12\, 
\mathcal{A}_\mu^a\,
\Delta_{ab}^{\mu\nu}\,
\mathcal{A}_\nu^b
+
\bar c^a \,
\Delta_{\rm gh}^{ab}\,
c^b,
\label{eq:Gaussian--}
\end{equation}
where the gauge-field fluctuation operator is decomposed as
\begin{equation}
\Delta=
\Delta_{\text{YM}}
+
\Delta_{\text{CS}}.
\end{equation}
Here, \(\Delta_{\rm YM}\) arises from the Yang--Mills action together with the gauge-fixing term, while \(\Delta_{\rm CS}\) denotes the axion-induced contribution. The operator \(\Delta_{\rm gh}\) is the corresponding Faddeev--Popov ghost operator. The one-loop effective action for the axion-Yang--Mills theory is expressed in terms of functional determinants of the corresponding fluctuation operators,
\be
\Gamma_{\text{1-loop}}[\bar A]
= \frac{i}{2} \ln \det (\Delta_{\textmd{YM}} + \Delta_{\textmd{CS}})
- i \ln \det \Delta_{\textmd{gh}}.
\ee
In the following, we derive these fluctuation operators explicitly.

\subsection{BRST quantization in the background-field $R_{\xi}$-gauge}

In the following, we derive the explicit forms of the ghost action and the gauge-fixing term in the background-field \(R_\xi\) gauge, demonstrating how they emerge from a BRST-invariant gauge-fixing construction. Our starting point will be the $SU(N)$ Yang--Mills gauge theory described by the action
\begin{equation}
\label{eq:YM}
S_{_{\textmd{YM}}}[\boldsymbol{A}]
= \int d^4x \,
\left[
-\frac{1}{2}\, \tr\!\left(\boldsymbol{F}_{\mu\nu}\boldsymbol{F}^{\mu\nu}\right)
\right],
\end{equation}
where the non-Abelian field strength tensor is
\begin{equation}
\boldsymbol{F}_{\mu\nu}
= \partial_\mu \boldsymbol{A}_\nu - \partial_\nu \boldsymbol{A}_\mu - i g [\boldsymbol{A}_\mu , \boldsymbol{A}_\nu] ,
\end{equation}
with $\boldsymbol{A}_\mu = A_\mu^a \bT_a$ taking values in the Lie algebra $\mathfrak{su}(N)$.
The generators $\bT_a$ satisfy
\be
[\bT_a , \bT_b] = i f_{abc} \bT_c , \qquad
\{\bT_a , \bT_b\} = \frac{1}{N}\delta_{ab}\,\boldsymbol{1}_N + d_{abc}\, \bT_c ,
\ee
where $f_{abc}$ and $d_{abc}$ denote the totally antisymmetric and symmetric structure constants of $\mathfrak{su}(N)$, respectively. Upon gauge fixing, the complete Lagrangian density becomes
\begin{equation}
\mathcal{L}_{_{\textmd{YM}}}
= -\frac{1}{2} \tr(\boldsymbol{F}_{\mu\nu} \boldsymbol{F}^{\mu\nu})
  + \bar{c}^a \, \partial^\mu D_\mu^{ab} \, c_b
  + \mathcal{L}_{\textmd{gf}},
\label{eq:Lagrangian}
\end{equation}
where $\mathcal{L}_{\textmd{gf}}$ is the gauge-fixing term and the covariant derivative  is defined as
\begin{equation}
D_\mu^{ab}  \equiv \delta^{ab} \, \nabla_\mu  -g f^{abc} \, A^c_\mu.
\end{equation}

The ghost fields $c^a$ and $\bar{c}^a$ are the Faddeev--Popov fields, introduced by the gauge-fixing procedure; they encode the Jacobian associated with the change of variables in field space and play a crucial role in preserving unitarity and gauge invariance at the quantum level. The classical action is invariant under infinitesimal gauge transformations
\begin{equation}
A_\mu^a \;\to\; {}^{\alpha}\!A_\mu^a =A_\mu^a + D_\mu^{ab}\alpha^b , \end{equation}
where $\boldsymbol{\alpha}=\alpha^a \bT_a$ is the Lie-algebra valued gauge parameter. That implies that the functional integral naively sums over infinitely many physically equivalent configurations along each gauge orbit. Gauge fixing is therefore essential to render the path integral well defined and to allow for a consistent perturbative quantization of the theory.

\textbf{Background-field formalism:}
In the presence of a background field, the gauge potential is decomposed into a classical background configuration $\bar A_\mu$ and a quantum fluctuation $\mathcal{A}_\mu$,
\be
\boldsymbol{A}_\mu =  \boldsymbol{\bar{A}}_\mu + \boldsymbol{\mathcal{A}}_\mu .
\label{eq:A-decom}
\ee
This split allows one to preserve explicit background gauge invariance while performing the perturbative quantization of the theory.  In the background-field $R_\xi$ gauge the gauge-fixing term becomes
\be
\mathcal{L}_{\rm gf}
= -\frac{1}{2\xi}\,(\bar D^\mu \mathcal{A}_\mu)^2 ,
\label{eq:gf-BG}
\ee
where $\xi$ is the gauge-fixing parameter.
Note that the gauge is fixed only for the quantum fluctuation
$\mathcal{A}_\mu$ while preserving the gauge invariance of the
background field $\bar A_\mu$.

Now we consider the Faddeev--Popov identity,
\be
1 = \Delta_{\text{FP}}[A] \int \mathcal{D}\alpha \;
\delta\!\left(G[A^\alpha]\right),
\ee
where $G[A]=0$ is the chosen gauge-fixing condition, $\alpha(x)$ denotes the gauge parameter generating the gauge transformation, and $\mathcal{D}\alpha$ is the invariant Haar measure on the gauge group. The object $\Delta_{\text{FP}}[A]$ is the Faddeev--Popov determinant. The unfixed path integral,
\be
Z = \int \mathcal{D}A_\mu \, e^{iS[A]},
\ee
includes infinitely many gauge-equivalent field configurations, which must be removed.  
Inserting the identity above yields
\be
Z = \int \mathcal{D}A_\mu \; \Delta_{\text{FP}}[A]
\int \mathcal{D}\alpha \; \delta\!\left(G[A^\alpha]\right)
e^{iS[A]} .
\ee
Factoring out the gauge-group volume, and imposing the gauge condition sharply through a delta functional then leads to the gauge-fixed expression
\be
Z = \int \mathcal{D}A_\mu \; \Delta_{\text{FP}}[A] \,
\delta\!\left(G[A]\right) \, e^{iS[A]} .
\ee
That restricts the functional integral to a single representative of each gauge orbit. Following the Faddeev--Popov procedure, 't~Hooft observed that one may instead average over nearby gauge slices with a Gaussian weight, $
\exp\left[
-\frac{i}{2\xi}\int d^4x\,\omega^a\omega^a
\right]$,
where \(\omega^a(x)\) parametrizes deviations from the gauge condition. Integrating over \(\omega^a\) replaces the delta functional by the gauge-fixing term
\[
\mathcal L_{\rm gf}
=
-\frac{1}{2\xi}\,
\mathcal{R}_{\rm Har}^a[A]\mathcal{R}_{\rm Har}^a[A].
\]
Physically, this procedure removes the infinite overcounting associated with gauge redundancy while preserving locality and Lorentz covariance, and renders the quadratic fluctuation operator invertible. In the present work, we employ the background-field \(R_\xi\)-gauge, which maintains covariance with respect to the background gauge symmetry and considerably simplifies the one-loop effective action.

The determinant can be written as a Gaussian integral over anticommuting (ghost) fields $c,\bar c$,
\be
\Delta_{\text{FP}}[A]
= \int \mathcal{D}c \, \mathcal{D}\bar{c} \;
\exp\!\left(
i \int d^{4}x \; \bar{c} \, M[A] \, c
\right),
\ee
where the Faddeev--Popov operator is defined by
\be
M[A] = \left. \frac{\delta G[A^\alpha]}{\delta \alpha} \right|_{\alpha = 0}.
\ee

Under the action of a generic small gauge transformation, we have
\begin{align}
    A_\mu \mapsto A_\mu^U = -\frac{1}{ig} \, U D_\mu U^{-1},
\end{align}
We can decompose it into the background and perturbations as
\begin{align}
  &  \bar{A}_\mu + \mathcal{A}_{\mu} \mapsto A_\mu^U = \bar{A}_\mu + \mathcal{A}^U_\mu, \nonumber\\
    &  \mathcal{A}^U_\mu = -\bar{A}_\mu -\frac{1}{ig} \, U D_\mu U^{-1}.
\end{align}

Next we consider a local gauge transformation
$U = e^{i\alpha^a T^a}$ acting only on the quantum fluctuations of the gauge field.  
To leading order in perturbation theory,
\be
\mathcal{A}_\mu^a \to \mathcal{A}_\mu^a + D_\mu^{ab}\alpha^b.
\ee
Imposing the background-field $R_\xi$ gauge condition on the fluctuations,
\be
G[A] = \bar{D}_{\mu} \mathcal{A}^{\mu} = 0,
\ee
we obtain
\be
\delta G^a(A^U) = [\bar{D}_{\mu} ( D_\mu\alpha)]^a,
\ee
from which it follows that
\be
M^a_{\;b} = \frac{\delta G^a(A^U)}{\delta \alpha^b} = [\bar{D}_{\mu}D^{\mu}]^a_{\;b}.
\ee

The corresponding ghost Lagrangian in the background-covariant $R_\xi$ gauge is therefore
\begin{equation}
\mathcal{L}_{\text{gh}} = \bar c^a\,(\bar D_\mu D^\mu)^b_a\,c_b.
\label{eq:L-gh--}
\end{equation}
In this framework the gauge-fixing term becomes
\be
\mathcal{L}_{\rm gf}
= -\frac{1}{2\xi}\,(\bar D^\mu \mathcal{A}_\mu)^2 ,
\label{eq:L-gf--}
\ee
reflecting the fact that only the quantum fluctuation $\mathcal{A}_\mu$ is gauge-fixed, while gauge invariance of the background field $\bar A_\mu$ is preserved.

\subsection{Quadratic action of the axion-Yang--Mills theory}

Expanding the gauge-fixed axion-Yang--Mills action to second order in the quantum fluctuations $\mathcal{A}_\mu$, one obtains the quadratic action as
\begin{align}
\mathcal{L}^{(2)}
= &\; -\frac{g}{2}\, f^{abc}\, \bar F_{\mu\nu}^a\,
      \mathcal{A}^{b\,\mu}\, \mathcal{A}^{c\,\nu}
 -\frac14\,
   \big(\bar D_\mu \mathcal{A}_\nu - \bar D_\nu \mathcal{A}_\mu\big)^a
   \big(\bar D^\mu \mathcal{A}^\nu - \bar D^\nu \mathcal{A}^\mu\big)^a 
   \nonumber\\
   & + \frac{1}{8}\,\frac{\kappa\, g^2}{f}\,\phi\,
\epsilon^{\mu\nu\lambda\sigma}
\Big[
 2 g\, f^{abc}\, \bar F^a_{\mu\nu}\,
   \mathcal{A}^b_\lambda \mathcal{A}^c_\sigma+ \big(\bar D_\mu \mathcal{A}_\nu - \bar D_\nu \mathcal{A}_\mu\big)^a
   \big(\bar D_\lambda \mathcal{A}_\sigma - \bar D_\sigma \mathcal{A}_\lambda\big)^a \Big] \nonumber\\
   & -\frac{1}{2\xi}\,\left( \bar D^\mu \mathcal{A}_\mu \right)^2 + \bar c^a\, (\bar D_\mu \bar D^\mu)^{ab} c^b.
\end{align}
Including the axion fluctuations $\delta\phi$ introduces additional contributions to the quadratic Lagrangian. In particular, the second-order action contains the terms
\be
\mathcal{L}^{(2)} \supset \frac{1}{2}\,\frac{\kappa\, g^2}{f}\,\delta\phi\,
\epsilon^{\mu\nu\lambda\sigma}
\,\bar F^a_{\mu\nu}\,
   \big(\bar D_\lambda \mathcal{A}_\sigma\big)^a.
\ee
After integrating by parts and discarding boundary terms, the action reduces to a Gaussian action
\begin{align}
\mathcal{L}^{(2)}
= & \frac12 \, \mathcal{A}_\nu^a \Big[ 2g\, f^{abc}\, \bar F^{c\,\mu\nu}\,
      \mathcal{A}^{b}_{\mu}  + \,
   \big(\bar{D}^2 \mathcal{A}^\nu\big)^a - (1-\frac{1}{\xi}) \big(\bar{D}^\nu\bar D_\mu \mathcal{A}^\mu\big)^a   - \frac{\kappa\, g^2}{f}\, \p_\mu\phi\,
\epsilon^{\mu\nu\lambda\sigma}
   \big(\bar D_\lambda \mathcal{A}_\sigma\big)^a \Big] 
    \nonumber\\
   & + \bar c^a\, (\bar D_\mu \bar D^\mu)^{ab} c^b,
   \label{eq:2nd-action---}
\end{align}
in which we used the following identity $
\epsilon^{\mu\nu\lambda\sigma}
   \big(\bar D_\mu \bar D_\lambda \mathcal{A}_\sigma\big)^a  =  - g\, f^{abc}\, \tilde{\bar F}^{c\, \mu\nu} \mA^b_\mu$.
In the $R_\xi$ family, we work in the 't Hooft--Feynman gauge, $\xi=1$. 

Writing the quadratic action in the Gaussian form given in \cref{eq:Gaussian--}, and using \cref{eq:L-gh--,eq:L-gf--}, we obtain
 the gauge-field fluctuation operator from YM term
\be
[\Delta_{\textmd{YM}}]_{\mu\nu}^{ab} = \Big[
\eta_{\mu\nu}\big(\bar{D}^2\big)^{ab}
 - \Big(1-\frac{1}{\xi}\Big)
   \big(\bar{D}_\mu \bar D_\nu\big)^{ab}
 - 2 g\, f^{abc}\, \bar F^{c}_{\mu\nu}
 \Big],
 \label{eq:Delta-gluon-app}
\ee
while the gauge-field fluctuation operator interacting with the axion, given by
\be
[\Delta_{\textmd{CS}}]_{ab}^{~~\mu\nu} =  \frac{\kappa \, g^2}{f} \, \epsilon^{\mu\nu\lambda\sigma} [\p_{\lambda} \phi \bar{D}_\sigma]_{ab},
\ee
and $\Delta_{\textmd{gh}}$ is the Faddeev--Popov ghost operator
\be
[\Delta_{\textmd{gh}}]^{ab}
=  (\bar D_\mu^2)^{ab}.
 \label{eq:Delta-gh-app}
\ee

For the $SU(2)$ gauge field, the expression can be simplified by returning to the quadratic action as
\begin{align}
\mathcal{L}^{(2)}
 = \frac{1}{2}\mathcal{A}_{~\nu}^a \Big[ 2g\, \epsilon_{abc}\, \bar F^{c\,\mu\nu} +  \, \eta^{\mu\nu} \, \big(\bar D^2\big)_{ab}  - \frac{\kappa \, g^2 \, \dot\phi}{f} \, \epsilon^{ijk} \, \delta^\nu_i \, \delta^\mu_k \, \big(\bar D_j\big)_{ab}\Big]\,        \mathcal{A}^{b}_{~\mu} .
   \label{eq:2nd-action-app}
\end{align}
The corresponding quadratic fluctuation operator are 
\begin{align}
\Delta^{~~00}_{ab} &= - \, \big(\bar D^2\big)_{ab}, \qquad
\Delta^{~~0i}_{ab} = 0,\\
\Delta^{~~ij}_{ab} &= - 2g\, \epsilon_{abc}\, \bar F^{c\,ij} + \, \delta^{ij} \, \big(\bar D^2\big)_{ab}  + \frac{\kappa \, g^2\,  \dot\phi}{f} \, \epsilon^{ijk} \, \big(\bar D_k\big)_{ab}.
\end{align}

\subsubsection*{Quadratic action around the spin-2-dressed chromomagnetic background}

We now consider the static chromomagnetic background ($A^a_0=0$)
\be
\bar{A}^a_i = Q \delta^a_i + W^a_i,
\ee
where the first term is the color-locked component and $W^a_i$ denotes the tensor deformation. Expanding the action to quadratic order around this background, we obtain the fluctuation operator 
\begin{align}
\mathcal{L}^{(2)}
& =  \frac12 \, \mathcal{A}_{~\nu}^a \Big[ 2g\, \epsilon_{abc}\, \bar F^{c\,\mu\nu}\,
      \mathcal{A}^{b}_{~\mu}  + \,
   \big(\bar{D}^2 \mathcal{A}^\nu\big)_a   - \frac{\kappa \, g^2}{f}\, \p_\mu\phi\,
\epsilon^{\mu\nu\lambda\sigma}
   \big(\bar D_\lambda \mathcal{A}_\sigma\big)_a \Big] \nonumber\\
   & = \frac12 \mathcal{A}_{~\nu}^a \Big[ 2g\, \epsilon_{abc}\, \bar F^{c\,\mu\nu} + \, \eta^{\mu\nu} \, \big(\bar D^2\big)_{ab}  - \frac{\kappa \, g^2 \dot\phi}{f} \, \epsilon^{ijk} \, \delta^\nu_i \, \delta^\mu_k \, \big(\bar D_j\big)_{ab}\Big]\,        \mathcal{A}^{b}_{~\mu} .
   \label{eq:2nd-action-}
\end{align}

That gives 
\begin{align}
\Delta^{~~00}_{ab} &= - \, \big(\bar D^2\big)_{ab}, \\
\Delta^{~~0i}_{ab} &= - 2g\, \epsilon_{abc}\, \bar F^{c\,0i} = 0 \quad \text{(static solution)},\\
\Delta^{~~ij}_{ab} &= - 2g\, \epsilon_{abc}\, \bar F^{c\,ij} +  \, \delta^{ij} \, \big(\bar D^2\big)_{ab}  + \frac{\kappa \, g^2 \dot\phi}{f} \, \epsilon^{ijk} \, \big(\bar D_k\big)_{ab}.
\end{align}
with 
\begin{align}
    \bar F^a_{~\mu \nu} & = \partial_\mu \bar A^a_{~\nu} - \partial_\nu \bar A^a_{~\mu} + g \epsilon^{a}_{~bc} \bar A^b_{~\mu} \bar A^c_{~\nu}, \nonumber \\
    \big(\bar D_\mu \big)_{ab} &= \partial_\mu \eta_{ab} - g \epsilon_{abc} \bar{A}^{c}_{~\mu}, \nonumber \\
    \big(\bar D^2\big)_{ab} &= \partial^\mu \partial_\mu \eta_{ab} + g^2 \epsilon_{acd} \epsilon^{de}_{~~b} \bar A^{c \, \mu} \bar A^e_{~\mu} - g \epsilon_{abc} \partial^\mu \bar{A}^c_{~\mu} - 2g \epsilon_{abc} \bar{A}^c_{~\mu} \partial^\mu, \nonumber \\
    &= \partial^\mu \partial_\mu \eta_{ab} + g^2 (\bar{A}_a^{~\mu} \bar{A}_{b \ \mu} -\bar{A}^{e \ \mu} \bar{A}_{e \ \mu} \delta_{ab}) - g \epsilon_{abc} \partial^\mu \bar{A}^c_{~\mu} - 2g \epsilon_{abc} \bar{A}^c_{~\mu} \partial^\mu.
\end{align}
So in total we find
\begin{align}
    \Delta_{ab}^{~~ij} =&- 2g \epsilon_{abc} (\partial^i \bar A^{c \, j} - \partial^j \bar A^{c \, i} + g \epsilon^{c}_{~de} \bar A^{i\,d} \bar A^{j \, e}) \nonumber \\ &+\eta^{ij} (\partial^\mu \partial_\mu \eta_{ab} + g^2 \epsilon_{acd} \epsilon^{de}_{~~b} \bar A^{c \, \mu} \bar A^e_{~\mu} - g \epsilon_{abc} \partial^\mu \bar{A}^c_{~\mu} - 2g \epsilon_{abc} \bar{A}^c_{~\mu} \partial^\mu)  \nonumber \\
    &-\alpha \epsilon^{ijk}  (\partial_k \eta_{ab} - g \epsilon_{abc} \bar{A}^{c}_{~k})
\end{align}

For the spin-2-dressed background from Eq.~\eqref{eq:BG-gauge-field}, we find 
\begin{align}
    \baF^a_{ij}={}&
    \partial_iW^a_j-\partial_jW^a_i
    +gQ^2\eps^a{}_{ij} +gQ\left(\eps^a{}_{ic}W^c_j+\eps^a{}_{cj}W^c_i\right)
    +g\eps^a{}_{bc}W^b_iW^c_j, \nonumber \\
    (\baD_k)_{ab} &= \p_k\delta_{ab}-gQ\eps_{abk}-g\eps_{abc}W^c_k, \nonumber \\
    (\baD^2)_{ab} &= \p^\mu\p_\mu\delta_{ab}
    +g^2\left[-2Q^2\delta_{ab}+2QW_{ab}+(W^2)_{ab}- \tr(W^2) \delta_{ab}\right] \nonumber \\  
    &\quad  -2gQ\eps_{abi}\p^i-2g\eps_{abc}W^c_{i}\p^i.
\end{align}
We can decompose the fluctuation operator as
\begin{equation}
  \Delta=\Delta_Q+\Delta_W+\Delta_{W^2}.
\end{equation}
\begin{equation}
\begin{aligned}
(\Delta_Q)^{ij}_{ab}={}&
\delta^{ij}\left[\left(\partial^\mu\partial_\mu-2g^2Q^2\right)\delta_{ab}
-2gQ\eps_{ab\ell}\partial_\ell\right]
\\
&-\alpha \eps^{ijk}\partial_k\delta_{ab}
+gQ\left(\alpha-2gQ\right)
\left(\delta_a^i\delta_b^j-\delta_a^j\delta_b^i\right).
\end{aligned}
\end{equation}
The part linear in $W$ is
\begin{equation}
\begin{aligned}
(\Delta_W)^{ij}_{ab}={}&
2\delta^{ij}\left[g^2QW_{ab}-g\eps_{abc}W_{c\ell}\partial_\ell\right]
-2g\eps_{abc}\left(\partial^iW^c_j-\partial^jW^c_i\right)
\\
&-2g^2Q\eps_{abc}\left(\eps^c{}_{ie}W^e_j+\eps^c{}_{dj}W^d_i\right)
+\alpha g \eps^{ijk}\eps_{abc}W^c_k.
\end{aligned}
\label{eq:Delta-W-}
\end{equation}
The quadratic part is 
\begin{equation}
    (\Delta_{W^2})^{ij}_{ab}= g^2 \delta^{ij}\left[(W^2)_{ab}- \tr(W^2)\delta_{ab}\right]
  -2 g^2\eps_{abc}\eps^c{}_{de}W^{di}W^{ej}.
  \label{eq:Delta-W2}
\end{equation}

The exact dispersion relations of the averaged fluctuation operators $\langle \Delta^{00}_{ab} \rangle$ and $\langle \Delta^{ij}_{ab} \rangle$ are given by
\begin{align}
    &\Omega_{00,h}^2
    = k^2
    +2g^2Q^2
    +\frac{2}{3}g^2\mathcal{W}_2
    -2h\,gQk, \quad h=0,\pm1,\\
    &\Omega^2_{0,n} = k^2 + \frac{10}{3}g^2Q^2 - \frac{2}{3}gQ \alpha + \frac{4}{9}g^2\mathcal{W}_2 -\frac{2\mathcal R}{3}\cos\left(\Theta-\frac{2\pi n}{3}\right), \quad  n=0,1,2,\\
    &\Omega^2_{\pm1,m} = k^2 + 2g^2 Q^2+ \frac{2}{3}g^2 \mathcal{W}_2 \mp (gQ + \frac{\alpha}{2})k \nonumber \\
    &\qquad \qquad \quad +m \sqrt{\left(gQ-\frac{\alpha}{2} \right)^2 k^2 + \left[ -\frac{1}{3}g^2 \mathcal{W}_2 + gQ\left(2gQ -\alpha\right) \right]^2}, \quad m =\pm1,\\
    &\Omega^2_{\pm2} = (k \mp (gQ+\frac{\alpha}{2}))^2 -g^2Q^2 -\frac{\alpha^2}{4}+g^2 \mathcal{W}_2,
\end{align}
with
\begin{align}
    &\mathcal R = \sqrt{ 7\left[ -\frac{1}{3}g^2\mathcal{W}_2+ gQ\left(2gQ -\alpha\right) \right]^2 + 3k^2 \left(2gQ-\alpha\right)^2}, \nonumber \\
    &\eta = \frac{\left[-\frac{1}{3}g^2\mathcal{W}_2+gQ\left(2gQ -\alpha\right)\right] \left\{ 10\left[-\frac{1}{3}g^2 \mathcal{W}_2 +gQ\left(2gQ-\alpha\right) \right]^2 - 9k^2\left(2gQ-\alpha\right)^2 \right\}}{ \left\{ 7\left[ -\frac{1}{3}g^2 \mathcal{W}_2 + gQ\left(2gQ -\alpha \right) \right]^2 + 3k^2\left(2gQ-\alpha\right)^2 \right\}^{3/2} },\nonumber \\
    &\Theta = \frac{1}{3} \arccos \eta,
\end{align}

\section{Heat-kernel derivation of the Yang--Mills running coupling}
\label{app:heat-kernel-running}

The ultraviolet structure of the one-loop effective action can be obtained particularly efficiently using the Schwinger--DeWitt, or heat-kernel, expansion. This method has the advantage that it makes gauge invariance manifest and directly identifies the local operators required as counterterms. Here we use it only to determine the renormalization of the Yang--Mills coupling. In particular, although we allow for a spacetime-dependent axion background coupled through the Chern--Simons interaction, we do not consider the renormalization of the axion coupling itself.

It is convenient first to absorb the gauge coupling into the gauge field,
\begin{equation}
    \mathbb A_\mu^a \equiv g A_\mu^a,
\end{equation}
so that
\begin{equation}
    \mathbb F_{\mu\nu}^a
    =
    \partial_\mu\mathbb A_\nu^a
    -
    \partial_\nu\mathbb A_\mu^a
    +
    f^{abc}\mathbb A_\mu^b\mathbb A_\nu^c,
    \qquad
    \mathbb F_{\mu\nu}^a=gF_{\mu\nu}^a .
\end{equation}
The Euclidean action may then be written as
\begin{equation}
    S_E
    =
    \frac{1}{4g_0^2}
    \int d^d x\,
    \mathbb F_{\mu\nu}^a\mathbb F_{\mu\nu}^a
    +
    i\frac{\kappa}{4f}
    \int d^d x\,
    \phi\,
    \mathbb F_{\mu\nu}^a
    \widetilde{\mathbb F}_{\mu\nu}^a ,
    \qquad
    d=4-2\epsilon .
    \label{eq:Euclidean-rescaled-action}
\end{equation}
The factor of $i$ in the second term is associated with the Wick rotation and will not play any role below. The important point is that the Chern--Simons interaction is independent of $g$ in terms of the rescaled field $\mathbb A_\mu$. Consequently, the running of the gauge coupling is determined by the coefficient multiplying the operator
$\mathbb F_{\mu\nu}^a\mathbb F_{\mu\nu}^a$.

The heat-kernel, or Schwinger--DeWitt, expansion provides a convenient
way of extracting the ultraviolet divergences of the one-loop effective
action; see, e.g., Refs.~\cite{Vassilevich2003,Barvinsky:1985an,Avramidi:2000bm}.
The heat kernel associated with an operator $\Delta$ is defined by
\begin{equation}
    K(s)\equiv\Tr e^{-s\Delta}.
\end{equation}
For a Laplace-type operator, its small-$s$ expansion takes the form
\begin{equation}
    \Tr e^{-s\Delta}
    \simeq
    \frac{1}{(4\pi s)^{d/2}}
    \int d^d x\,
    \tr\left[
        a_0+s\,a_2+s^2\,a_4+\cdots
    \right].
\end{equation}
In four dimensions, the logarithmic ultraviolet divergence is determined
by the coefficient $a_4$ which determines the logarithmically divergent part of
the one-loop effective action \cite{Vassilevich2003,Barvinsky:1985an,Avramidi:2000bm}. Working in dimensional regularization,
$d=4-2\epsilon$, and using the $\overline{\rm MS}$ scheme, one finds
\begin{equation}
    \Gamma^{\rm div}
    =
    -\frac{1}{(4\pi)^2}
    \frac{1}{\bar\epsilon}
    \int d^4x\,a_4(x),
    \qquad
    \frac{1}{\bar\epsilon}
    \equiv
    \frac{1}{\epsilon}-\gamma_E+\ln 4\pi .
    \label{eq:a4}
\end{equation}
The pole $1/\bar\epsilon$ is the dimensional-regularization
representation of the logarithmic ultraviolet divergence.

Using the relation between the heat-kernel coefficient $a_4$ and the logarithmic ultra-
violet divergence given in \cref{eq:a4}, and combining the gauge-field and Faddeev--Popov
ghost contributions for pure Yang--Mills theory (see \cref{eq:Delta-gluon-app,eq:Delta-gh-app}), the logarithmically divergent part of the one-loop effective action is
\begin{equation}
    \Gamma^{\rm div}
    =
    -\frac{b_0}{64\pi^2}
    \frac{1}{\bar\epsilon}
    \int d^4x\,
    \bar{\mathbb F}_{\mu\nu}^a
    \bar{\mathbb F}^{a\mu\nu},
    \label{eq:Gamma-div-YM}
\end{equation}
where
\begin{equation}
    b_0=\frac{11}{3}C_A
\end{equation}
is the one-loop coefficient of the pure Yang--Mills beta function, with
$C_A=N$ for $SU(N)$. The crucial observation is that this divergence has precisely the same
local operator structure as the classical Yang--Mills action. In terms of the rescaled gauge field, the bare Yang--Mills action is
\begin{equation}
    S_{\rm YM}
    =
    \frac{1}{4g_0^2}
    \int d^d x\,
    \bar{\mathbb F}_{\mu\nu}^a
    \bar{\mathbb F}^{a\mu\nu}.
    \label{eq:bare-YM-action}
\end{equation}
Since the divergence in \cref{eq:Gamma-div-YM} has the same
$\bar{\mathbb F}^2$ structure, it can be absorbed into the
renormalization of the gauge coupling. In dimensional regularization,
$d=4-2\epsilon$, we introduce the renormalization scale $\mu$ and define
the renormalized coupling $g(\mu)$ through
\begin{equation}
    \frac{1}{g_0^2}
    =
    \mu^{-2\epsilon}
    \left[
        \frac{1}{g^2(\mu)}
        +
        \delta\!\left(\frac{1}{g^2}\right)
    \right].
\end{equation}
In the $\overline{\rm MS}$ scheme, cancellation of the one-loop pole
requires
\begin{equation}
    \delta\!\left(\frac{1}{g^2}\right)
    =
    \frac{b_0}{16\pi^2}
    \frac{1}{\bar\epsilon}, \quad \rightarrow \quad
    \frac{1}{g_0^2}
    =
    \mu^{-2\epsilon}
    \left[
        \frac{1}{g^2(\mu)}
        +
        \frac{b_0}{16\pi^2}
        \frac{1}{\bar\epsilon}
    \right].
    \label{eq:g0-renormalized}
\end{equation}
The corresponding counterterm, $S_{\rm ct} =
   -\Gamma^{\rm div}$, precisely cancels the divergence in \cref{eq:Gamma-div-YM}. The
remaining dependence on $\mu$ is encoded in the running coupling
$g(\mu)$.

The running coupling follows from the $\mu$-independence of the bare
coupling, $\mu\,d g_0/d\mu=0$. In four dimensions this gives the
one-loop Yang--Mills beta function
\begin{equation}
    \beta(g)
    \equiv
    \mu\frac{dg}{d\mu}
    =
    -\frac{b_0}{16\pi^2}g^3
    +\mathcal O(g^5).
    \label{eq:YM-beta}
\end{equation}
Since $b_0>0$, pure Yang--Mills theory is asymptotically free. Integrating
the renormalization-group equation gives
\begin{equation}
    \frac{1}{g^2(\mu)}
    =
    \frac{1}{g^2(\mu_0)}
    +
    \frac{b_0}{8\pi^2}
    \ln\left(\frac{\mu}{\mu_0}\right).
    \label{eq:running-g-YM}
\end{equation}

It is convenient to introduce the RG-invariant strong-coupling scale
\begin{equation}
    \Lambda_{\rm conf}
    =
    \mu\,
    \exp\left[
        -\frac{8\pi^2}{b_0g^2(\mu)}
    \right],
\end{equation}
so that
\begin{equation}
    g^2(\mu)
    =
    \frac{8\pi^2}
    {b_0\ln(\mu/\Lambda_{\rm conf})}.
\end{equation}
The scale $\Lambda_{\rm conf}$ signals the onset of strong coupling and,
in pure Yang--Mills theory, is associated with the nonperturbative
confinement scale.

For the background considered here, the renormalization scale should be
chosen of order its characteristic physical scale,
\begin{equation}
    \mu_{\rm bg}
    \sim
    \max\left\{
        |\alpha|,\,
        |gQ|,\,
        g\sqrt{\mathcal W_2},\,
        k_\star
    \right\},
\end{equation}
up to factors of order unity. Perturbation theory is reliable provided
\begin{equation}
    \frac{g^2(\mu_{\rm bg})}{4\pi}\ll1,
    \qquad
    \mu_{\rm bg}\gg\Lambda_{\rm conf}.
    \label{eq:weak-coupling-condition}
\end{equation}
Thus, the relevant weak-coupling criterion is set by the running gauge
coupling evaluated at the characteristic scale of the background.
The axion--Chern--Simons interaction does not alter this pure
Yang--Mills running of $g$ at the order considered here.

\section{Details of the Stochastic Scattering Self-Energy}
\label{app:self-energy}

In this appendix, we present the details of the self-energy computation
associated with the stochastic tensor background. We keep the full Born
kernel of \cref{eq:born-selfenergy-diagrams},
\be
\Sigma^B
=
\Sigma_{11}
+
\Sigma_{22}
+
\Sigma_{12}
+
\Sigma_{21},
\label{eq:SigmaW-12-decomposition}
\ee
with
\be
\begin{aligned}
\Sigma_{11}
&\equiv
\left\langle
\Delta_W\,\bar G_0\,\Delta_W
\right\rangle_{\rm st},
&\qquad
\Sigma_{22}
&\equiv
\left\langle
\delta\Delta_{W^2}\,
\bar G_0\,
\delta\Delta_{W^2}
\right\rangle_{\rm st},
\\
\Sigma_{12}
&\equiv
\left\langle
\Delta_W\,\bar G_0\,\delta\Delta_{W^2}
\right\rangle_{\rm st},
&\qquad
\Sigma_{21}
&\equiv
\left\langle
\delta\Delta_{W^2}\,
\bar G_0\,
\Delta_W
\right\rangle_{\rm st}.
\end{aligned}
\ee
The linear--linear term involves only the power spectrum, the
quadratic--quadratic term involves the power spectrum and the connected
trispectrum, and the mixed terms are cubic in $W$ and therefore proportional
to the bispectrum. The mixed terms vanish only for a sign-symmetric
ensemble, $\mathcal P[W]=\mathcal P[-W]$; otherwise they are of the
same order as the diagonal ones and must be retained.

\subsection{Dressed propagator and its helicity structure}
\label{app:dressed-propagator}

The isotropically dressed fluctuation operator is given in
\cref{eq:Deltabar-explicit}. The corresponding isotropically dressed
propagator is $\bar G_0\equiv\bar\Delta^{-1}$. The inverse is taken in the
combined color--spatial index space. In momentum space, it is defined by
\be
\bar\Delta^{i\ell}_{ac}(\omega,\bk)
\left(\bar G_0\right)^{\ell j}_{cb}(\omega,\bk)
=
\delta^{ij}\delta_{ab}.
\ee
Let $I=(i,a)$ be a composite index and let
$u_I^{(n)}(\bk)$ denote an orthonormal eigenmode of $\bar\Delta$:
\be
\bar\Delta_{IJ}(\omega,\bk)
u_J^{(n)}(\bk)
=
\bar{\mathcal{D}}_n(\omega,k)
u_I^{(n)}(\bk).
\ee
Orthonormality and completeness require
\begin{equation}
    u_I^{(n)*}(\bk)u_I^{(m)}(\bk) =
    \delta_{nm}, \quad \sum_n u_I^{(n)}u_J^{(n)*}
    =\delta_{IJ}.
\end{equation}
The dressed propagator then has the spectral representation
\be
\left(\bar G_0\right)_{IJ}(\omega,\bk)
=
\sum_n
\frac{
1}{
\bar{\mathcal{D}}_n(\omega,k)
}\, u_I^{(n)}(\bk)
u_J^{(n)*}(\bk).
\label{eq:barG-0-}
\ee

Since $\bar\Delta(\omega,\bk)=\omega^2-\bar H(\bk)$ with $\bar H$ a
Hermitian $9\times9$ matrix, $\bar{\mathcal D}_n(\omega,k)=\omega^2-\omega_n^2(k)$
with $\omega_n^2(k)$ the eigenvalues of $\bar H(\bk)$. The eigenmodes are
organized by the total helicity about $\hat\bk$. In the basis
\be
u^{(\lambda_1\lambda_2)}_{I}(\hat\bk)
\equiv
\epsilon^{\lambda_1}_i(\hat\bk)\,
\epsilon^{\lambda_2}_a(\hat\bk),
\qquad
\lambda_{1,2}\in\{+,0,-\},
\label{eq:helicity-basis}
\ee
with $\epsilon^{\pm}(\hat\bk)$ the helicity vectors of
\cref{eq:W-decompose} and $\epsilon^0(\hat\bk)=\hat\bk$, rotations about
$\hat\bk$ conserve $J_z=\lambda_1+\lambda_2$, and $\bar H(\bk)$ is block
diagonal with blocks of dimension $1+1+2+2+3$, as shown in \cref{sec:Hartree--Delta}.

\subsection{The linear--linear contribution}
\label{app:Sigma11}

The part of the fluctuation operator linear in $W$ is presented in
\cref{eq:Delta-W}. Its momentum-space matrix element between a fluctuation
of incoming momentum $\bp$ and outgoing momentum $\bk$ is
\be
\left(\Delta_W\right)^{ij}_{ab}(\bk,\bp)
=
g\sum_{\lambda=\pm}
W_{\lambda}(\bk-\bp)
\left[
\mathcal V_{\lambda}(\bk-\bp;\bp)
\right]^{ij}_{ab},
\label{eq:DeltaW-momentum-vertex}
\ee
where the vertex carries two arguments, the momentum $\bq$ of the
background mode and the momentum $\bp$ of the fluctuation.
The effective vertex $\mathcal V_\lambda$ has the following Feynman-diagram representation
\begin{equation}
  \resizebox{0.93\linewidth}{!}{\LWVertexEquation}
  \label{eq:linear-W-vertex-decomposition}
\end{equation}
Here the large blue disk denotes the effective vertex $\mathcal V_\lambda$;
the three terms on the right represent, respectively, the cubic
Yang--Mills derivative coupling, the quartic Yang--Mills coupling with
one homogeneous-background insertion $Q\delta_i^a$, and the axion
Chern--Simons coupling.

It is obtained from \cref{eq:Delta-W} as
\be
\begin{aligned}
\left[
\mathcal V_{\lambda}(\bq;\bp)
\right]^{ij}_{ab}
&=
2\delta^{ij}
\left[
gQ\,e^{\lambda}_{ab}(\hat\bq)
-
i\eps_{abc}p_\ell e^{c\ell,{\lambda}}(\hat\bq)
\right]
-
2i\eps_{abc}
\left[
q^i e_j^{c,{\lambda}}(\hat\bq)
-
q^j e_i^{c,{\lambda}}(\hat\bq)
\right]
\\
&-
2gQ\eps_{abc}
\left[
\eps^c{}_{ie}e_j^{e,{\lambda}}(\hat\bq)
+
\eps^c{}_{dj}e_i^{d,{\lambda}}(\hat\bq)
\right]+
\alpha\eps^{ijk}\eps_{abc}
e_k^{c,\lambda}(\hat\bq).
\end{aligned}
\label{eq:Vplus-exact}
\ee   
The momentum $\bp$ in the first line is the momentum of the quantum
fluctuation on which the derivative acts, whereas $\bq$ in the second
line arises from derivatives acting on the stochastic background.
Hermiticity of $\Delta_W$, together with
$e^\lambda(-\hat\bq)=e^\lambda(\hat\bq)^*$ and $W_\lambda(-\bq)=W^*_\lambda(\bq)$,
gives
\be
\mathcal V_{\lambda}(-\bq;\bp+\bq)
=
\mathcal V_{\lambda}(\bq;\bp)^\dagger .
\label{eq:V-hermiticity}
\ee
In momentum space, the linear--linear self-energy is
\be
\begin{aligned}
\left[
\Sigma_{11}(\omega,\bk)
\right]_{IJ}
={}&
\int_{\bp}
\left\langle
\left[\Delta_W(\bk,\bp)\right]_{IK}
\left[\bar G_0(\omega,\bp)\right]_{KL}
\left[\Delta_W(\bp,\bk)\right]_{LJ}
\right\rangle_{\rm st}.
\end{aligned}
\label{eq:Sigma11-momentum-start}
\ee
Because the stochastic background is static, the frequency is
unchanged at each scattering. Moreover, the two stochastic insertions
carry opposite momenta, and \cref{eq:W-power,eq:V-hermiticity} give
\be
\begin{aligned}
\left[
\Sigma_{11}(\omega,\bk)
\right]_{IJ}
= g^2\,
\sum_{\lambda=\pm} \int_{\bp} \, P_{\lambda}(\lvert \bk-\bp\rvert) \,
\left[\mathcal{V}_{\lambda}(\bk-\bp;\bp)\right]_{IL}
\left[\bar G_0(\omega,\bp)\right]_{LK}
\left[\mathcal{V}^{\dag}_{\lambda}(\bk-\bp;\bp)\right]_{K J}
.
\end{aligned}
\label{eq:Sigma11-general}
\ee
Both background helicities contribute, each weighted by its own spectrum.

\paragraph{Projection on the external helicity.}
Statistical isotropy makes $\Sigma_{11}(\omega,\bk)$ invariant under
rotations about $\hat\bk$, so that, like $\bar H(\bk)$, it is block
diagonal in the basis \cref{eq:helicity-basis} taken about $\hat\bk$. For
an external tensor fluctuation of helicity $\sigma$, the block is
one-dimensional and the self-energy is the single number
$\Sigma_{11,\sigma}(\omega,k)\equiv e^{\sigma}_I(\hat\bk)^*
[\Sigma_{11}(\omega,\bk)]_{IJ}e^{\sigma}_J(\hat\bk)$. Writing
$\bq\equiv\bk-\bp$ for the momentum of the background mode and
$\bp=\bk-\bq$ for the intermediate momentum, we introduce the transition
amplitudes into the nine intermediate states \cref{eq:helicity-basis}
about $\hat\bp$,
\be
\mathcal M_{\sigma;(\lambda_1\lambda_2)}(\bk,\bq)
\equiv
e^{\sigma}_I(\hat\bk)^*
\left[
\mathcal V_{\lambda}(\bq;\bk-\bq)
\right]_{IJ}
u^{(\lambda_1\lambda_2)}_{J}(\widehat{\bk-\bq}),
\label{eq:M-general-def}
\ee
and collect these amplitudes into the block vectors
\be
\mathbf m_h
\equiv
\begin{pmatrix}
\mathcal M_{\sigma;(h0)} \\[2pt]
\mathcal M_{\sigma;(0h)}
\end{pmatrix},
\qquad h=\pm1,
\label{eq:m-vector-block}
\ee
and
\be
\mathbf m_0
\equiv
\begin{pmatrix}
\mathcal M_{\sigma;(+-)} \\[2pt]
\mathcal M_{\sigma;(00)} \\[2pt]
\mathcal M_{\sigma;(-+)}
\end{pmatrix}.
\label{eq:m-scalar-block}
\ee
Then we can write
\cref{eq:Sigma11-general} as
\be
\begin{aligned}
\Sigma_{11,\sigma}(\omega,k)
=
g^2\sum_{\lambda=\pm}\int_{\bq}P_\lambda(q)
\Bigg[
&\sum_{\lambda'=\pm}
\frac{\left|\mathcal M_{\sigma;(\lambda'\lambda')}\right|^2}
{\omega^2-\omega_{\lambda'}^2(p)}
+
\sum_{h=\pm}
\mathbf m_h^\dagger
\left[\omega^2\mathbb 1-\mathsf H^{(h)}(p)\right]^{-1}
\mathbf m_h
\\
&+
\mathbf m_0^\dagger
\left[\omega^2\mathbb 1-\mathsf H^{(0)}(p)\right]^{-1}
\mathbf m_0
\Bigg],
\qquad
p=|\bk-\bq| .
\end{aligned}
\label{eq:Sigma11-projected}
\ee
The first term is the tensor channel, in which both tensor helicities
propagate between the two insertions even if the background contains only
one of them; the second and third terms are the vector and scalar
channels, which are reached by the same vertex and are of the same order.
Since the amplitudes depend on the direction of $\bq$ only through
$\mu\equiv\hat\bk\cdot\hat\bq$, with
\be
p
=
\sqrt{
k^2+q^2-2kq\mu
},
\label{eq:p-mu-def-Sigma11}
\ee
the azimuthal integration is trivial and $\int_{\bq}\to2\pi\int_0^\infty dq\,q^2\int_{-1}^1d\mu$.

\paragraph{Transition amplitudes.}
For the external tensor state the amplitudes \cref{eq:M-general-def} have
closed forms. It is convenient to introduce
\be
\begin{aligned}
S_\pm&\equiv k+\sigma\lambda q\pm p,
\qquad
R\equiv k-\sigma\lambda q,
\qquad
\nu\equiv\sigma\lambda+\mu,
\\
s&\equiv\sqrt{1-\mu^2},
\qquad
C\equiv\alpha+3gQ,
\end{aligned}
\label{eq:Sigma11-kinematic-definitions}
\ee
which satisfy $S_+S_-=2kq\nu$. Given intermediate states labelled relative to the external helicity,
$(\lambda_1,\lambda_2)=(\sigma l_1,\sigma l_2)$, we find
\be
\begin{aligned}
8p^2\,\mathcal M_{\sigma;(\sigma,\sigma)}&=s^2\,S_+^2\,(S_+-\sigma C),\\
8p^2\,\mathcal M_{\sigma;(-\sigma,-\sigma)}&=s^2\,S_-^2\,(S_--\sigma C),\\
8p^2\,\mathcal M_{\sigma;(\sigma,0)}&=\sqrt2\,s\nu\,S_+\big[R(S_+-\sigma C)-\sigma gQ\,p\big],\\
8p^2\,\mathcal M_{\sigma;(0,\sigma)}&=\sqrt2\,s\nu\,S_+\big[R(S_+-\sigma C)-p(p+R-\sigma gQ)\big],\\
8p^2\,\mathcal M_{\sigma;(0,-\sigma)}&=\sqrt2\,s\nu\,S_-\big[R(S_+-\sigma C)-p(p+R+\sigma gQ)\big],\\
8p^2\,\mathcal M_{\sigma;(-\sigma,0)}&=\sqrt2\,s\nu\,S_-\big[R(S_--\sigma C)+\sigma gQ\,p\big],\\
8p^2\,\mathcal M_{\sigma;(\sigma,-\sigma)}&=\nu^2\,(p+R)\big[(p-R)(S_+-\sigma C)+2\sigma gQ\,p\big],\\
8p^2\,\mathcal M_{\sigma;(-\sigma,\sigma)}&=\nu^2\,(p-R)\big[(p+R)(S_--\sigma C)+2\sigma gQ\,p\big],\\
8p^2\,\mathcal M_{\sigma;(0,0)}&=2\nu^2\big[R^2(S_+-\sigma C)-pR(p+R)+\sigma gQ\,p^2\big].
\end{aligned}
\label{eq:M-closed-forms}
\ee
All nine amplitudes are real, and we have checked them against the full
operator \cref{eq:Delta-W} for arbitrary couplings, momenta and helicities.
In the tensor channel they reproduce
\be
\left|
\mathcal M_{\sigma;(\lambda'\lambda')}
\right|^2
=
\frac{(1-\mu^2)^2}{64p^4}\,
S^4\left(S-\sigma C\right)^2,
\qquad
S=k+\sigma\lambda q+\sigma\lambda' p ,
\label{eq:M-tensor}
\ee
for all eight assignments of $(\sigma,\lambda,\lambda')$: for
$\sigma=\lambda$ one has $S=k+q\pm p$ for $\lambda'=\pm\sigma$, and for
$\sigma=-\lambda$ one has $S=k-q\pm p$ for $\lambda'=\pm\sigma$. Three
features of \cref{eq:M-closed-forms} are general. First, the power of
$s=\sin\theta_{\bk\bq}$ is $2-n_0$ and the power of $\nu$ is $n_0$, where
$n_0$ is the number of vanishing helicity labels of the intermediate
state: a change $\Delta J_z$ of the total helicity of the fluctuation
costs $s^{|\Delta J_z|}\nu^{2-|\Delta J_z|}$. For collinear scattering,
$s=0$, only the scalar channel survives, and only when $\nu\neq0$, i.e.\
when the background helicity about $\hat\bk$ equals that of the
fluctuation, as required by angular-momentum conservation along the
common axis. Second, the couplings enter only through $C$ and $gQ$, while
the derivative couplings produce the explicit powers of $S_\pm$, $R$ and
$p$. Third, all amplitudes except the two tensor ones vanish at
$S_+=0$ or $S_-=0$.

\paragraph{Frequency dependence and static limit.}
For a generic real frequency the propagators in
\cref{eq:Sigma11-projected} are understood with the usual Feynman
prescription,
\be
\frac{1}{
\omega^2-\omega_n^2(p)
}
\;\longrightarrow\;
\frac{1}{
\omega^2-\omega_n^2(p)+i\epsilon
},
\qquad
\frac{1}{x+i\epsilon}
=
{\rm P}\frac{1}{x}
-i\pi\delta(x),
\ee
with P the Cauchy principal value, and an imaginary part arises whenever
an intermediate mode of any of the three sectors can go on shell,
$\omega^2=\omega_n^2(p)$, within the integration range. Since the
background is static the scattering is elastic, and the intermediate
energies are bounded below by the gap of $\bar H$. Our present purpose is
to determine whether the stochastic correction can destabilize the gapped
tensor mode, for which the relevant quantity is the static inverse
propagator at $\omega=0$. When $\bar H$ is positive, i.e.\ when
$\bar m^2>0$ and the vector and scalar blocks are positive, no on-shell
condition can be met at zero frequency, so
\be
{\rm Im}\,
\Sigma_{11,\sigma}(0,k)
=
0,
\qquad
\sigma=\pm ,
\label{eq:Im-Sigma11-static}
\ee
and the static self-energy is purely dispersive. At $\omega=0$ every
denominator in \cref{eq:Sigma11-projected} is negative definite while
every numerator is a non-negative quadratic form, so that, with the
present definition,
\be
\Sigma_{11,\sigma}(0,k)
\leq 0 ,
\ee
\be
\Sigma_{11,\sigma}(0,k)
=
-g^2\sum_{\lambda=\pm}\int_{\bq}P_\lambda(q)\,
\Bigg[
\sum_{\lambda'=\pm}
\frac{\left|\mathcal M_{\sigma;(\lambda'\lambda')}\right|^2}
{\omega_{\lambda'}^2(p)}
+
\sum_{h=\pm,0}
\mathbf m_h^\dagger
\big[\mathsf H^{(h)}(p)\big]^{-1}
\mathbf m_h
\Bigg].
\label{eq:Sigma11-static}
\ee
The dressed static kernel of the branch $\sigma$ is
$\omega_\sigma^2(k)+\Sigma_{11,\sigma}(0,k)$, so that the static
self-energy lowers the gap. The most dangerous branch is
$\sigma=\lambda_{\rm d}$ near $k=k_{\rm d}$, where the tree-level restoring
scale is $\bar m^2$; this is where the ratio of self-energy to tree-level
kernel is largest. It does not follow that
$|\Sigma_{11,\lambda_{\rm d}}|>|\Sigma_{11,-\lambda_{\rm d}}|$
in general, and both external helicities have to be evaluated.

\subsection{The quadratic--quadratic contribution}
\label{app:Sigma22}

We next consider the contribution quadratic in the residual
\(W^2\)-dependent operator,
\be
\Sigma_{22}
\equiv
\left\langle
\delta\Delta_{W^2}\,
\bar G_0\,
\delta\Delta_{W^2}
\right\rangle_{\rm st}.
\ee
In momentum space, the operator
\(\delta\Delta_{W^2}\) defined in \cref{eq:Delta-W2} is
\be
\begin{aligned}
\big[\delta\Delta_{W^2}(\bq)\big]^{ij}_{ab}
&=
g^2\int_{\bp}
\Bigg[
\delta^{ij}
\Bigg(
W_a^m(\bp)W_b^m(\bq-\bp)
-
\left\langle
W_a^m(\bp)W_b^m(\bq-\bp)
\right\rangle_{\rm st}
\\
&
-\delta_{ab}
\left(
W_c^m(\bp)W_c^m(\bq-\bp)
-
\left\langle
W_c^m(\bp)W_c^m(\bq-\bp)
\right\rangle_{\rm st}
\right)
\Bigg)
\\
&
-2\eps_{abc}\eps^c{}_{de}
\Bigg(
W^{di}(\bp)W^{ej}(\bq-\bp)
-
\left\langle
W^{di}(\bp)W^{ej}(\bq-\bp)
\right\rangle_{\rm st}
\Bigg)
\Bigg],
\end{aligned}
\label{eq:deltaDeltaW2-momentum}
\ee
where $\bq$ is the momentum transferred to the fluctuation. As in the
previous subsection, we denote a combined spatial--color index by
\(I=(i,a)\). The corresponding self-energy is
\be
\begin{aligned}
\left[
\Sigma_{22}(\omega,\bk)
\right]_{IJ}
&=
\int_{\bq}
\left[
\bar G_0(\omega,\bk-\bq)
\right]_{KL}
\,
\mathcal F_{IK;LJ}(\bq),
\label{eq:Sigma22-momentum}
\end{aligned}
\ee
where
\be
\mathcal F_{IK;LJ}(\bq)
\equiv
\left\langle
\big[\delta\Delta_{W^2}(\bq)\big]_{IK}
\big[\delta\Delta_{W^2}(-\bq)\big]_{LJ}
\right\rangle'_{\rm st}.
\label{eq:B-covariance}
\ee
For later convenience, define the symmetric bilinear form
\be
\begin{aligned}
\left[
\mathfrak{b}[X,Y]
\right]^{ij}_{ab}
\equiv
\frac{\delta^{ij}}{2}
\Big[
X_a^mY_b^m
+
Y_a^mX_b^m
-
2\delta_{ab}X_c^mY_c^m
\Big]
-\eps_{abc}\eps^c{}_{de}
\Big[
X^{di}Y^{ej}
+
Y^{di}X^{ej}
\Big].
\end{aligned}
\label{eq:Q-W2-bilinear}
\ee
Because the integration variable in
\cref{eq:deltaDeltaW2-momentum} can be exchanged as
\(\bp\leftrightarrow\bq-\bp\), this symmetrized form is equivalent to
the original quadratic operator, $g^2 \mathfrak{b}[W,W]=\Delta_{W^2}$, and
$\mathfrak{b}[X,Y]$ is a symmetric matrix in $(I,J)$ for any $X$, $Y$. For
two helicity modes, with
$\mathcal U_{\lambda\lambda'}(\hat\bp,\hat\bp')\equiv \mathfrak{b}[e^{\lambda}(\hat\bp),e^{\lambda'}(\hat\bp')]$,
the form collapses to
\be
\mathcal U_{\lambda\lambda'}(\hat\bp,\hat\bp')
=
\tfrac12\,c_{\lambda\lambda'}\,
\delta^{ij}
\Big[
\epsilon^\lambda_a\epsilon^{\lambda'}_b
+
\epsilon^{\lambda'}_a\epsilon^{\lambda}_b
-
2c_{\lambda\lambda'}\delta_{ab}
\Big]
-
\Big[
\epsilon^\lambda_a\epsilon^{\lambda'}_b
-
\epsilon^{\lambda'}_a\epsilon^{\lambda}_b
\Big]
\Big[
\epsilon^\lambda_i\epsilon^{\lambda'}_j
-
\epsilon^{\lambda'}_i\epsilon^{\lambda}_j
\Big],
\label{eq:U22-definition}
\ee
in which $c_{\lambda\lambda'}\equiv\boldsymbol\epsilon^\lambda(\hat\bp)\cdot\boldsymbol\epsilon^{\lambda'}(\hat\bp')$. It carries no
derivatives and is bounded, $\|\mathcal U_{\lambda\lambda'}\|\le\tfrac32$.
Its ensemble average is
$\langle\mathfrak{b}[W,W]\rangle_{\rm st}=-\tfrac{\mathcal W_2}{3}(2\mathbb 1- \mathsf L)$,
which is the subtraction in $\delta\Delta_{W^2}$.

\paragraph{Gaussian part.}
The four-point function in \cref{eq:B-covariance} decomposes into its
Gaussian part, given by Wick's theorem,
\be
\begin{aligned}
&
\left\langle
\left(
W_1W_2-\langle W_1W_2\rangle_{\rm st}
\right)
\left(
W_3W_4-\langle W_3W_4\rangle_{\rm st}
\right)
\right\rangle_{\rm st}^{\rm G}
 =
\left\langle W_1W_3\right\rangle_{\rm st}
\left\langle W_2W_4\right\rangle_{\rm st}
+
\left\langle W_1W_4\right\rangle_{\rm st}
\left\langle W_2W_3\right\rangle_{\rm st},
\end{aligned}
\label{eq:Wick-W2-connected}
\ee
and the connected part discussed below. The subtraction in
\(\delta\Delta_{W^2}\) removes the disconnected contraction
\(\langle W_1W_2\rangle
 \langle W_3W_4\rangle\),
while the two cross-contractions remain; after the overall
momentum-conserving delta function has been stripped off, no additional
factor proportional to \(\delta^{(3)}(\bq)\) remains in \(\mathcal F(\bq)\).
Because $\mathfrak{b}$ is symmetric the two cross-contractions are equal,
and with both helicities populated the Gaussian covariance is
\be
\begin{aligned}
\mathcal F^{\rm G}_{IK;LJ}(\bq)
=
2g^4
\sum_{\lambda,\lambda'=\pm}
\int_{\bp}
P_{\lambda}(p)
P_{\lambda'}(|\bq-\bp|)
\,
\left[
\mathcal U_{\lambda\lambda'}
(\hat\bp,\widehat{\bq-\bp})
\right]_{IK}
\left[
\mathcal U_{\lambda\lambda'}^{\dagger}
(\hat\bp,\widehat{\bq-\bp})
\right]_{LJ}.
\end{aligned}
\label{eq:C22-Wick}
\ee
Substituting this expression into
\cref{eq:Sigma22-momentum} yields
\begin{align}
 \left[
\Sigma_{22}^{\rm G}(\omega,\bk)
\right]_{IJ}  =&  2g^4
\sum_{\lambda,\lambda'=\pm}
\int_{\bp,\bp'}
P_{\lambda}(p)
P_{\lambda'}(p')
\left[
\mathcal U_{\lambda\lambda'}
(\hat\bp,\hat{\bp}')
\right]_{IK}
\left[
\bar G_0(\omega,\bk-\bp-\bp')
\right]_{KL} \nonumber\\
\times &
\left[
\mathcal U_{\lambda\lambda'}^{\dagger}
(\hat\bp,\hat{\bp}')
\right]_{LJ}.
\label{eq:Sigma22-exact-convolution}
\end{align}
The mixed pairs $\lambda\neq\lambda'$ are present whenever both
helicities carry power. For an external tensor state of helicity
\(\sigma\), define the projected quadratic vertex into the nine
intermediate states about $\widehat{\bk-\bp-\bp'}$,
\be
\mathcal N_{\sigma;(\lambda_1\lambda_2)}
(\bk;\bp,\bp')
\equiv
e_I^{\sigma *}(\hat\bk)
\left[
\mathcal U_{\lambda\lambda'}
(\hat\bp,\hat\bp')
\right]_{IK}
u^{(\lambda_1\lambda_2)}_K
(\widehat{\bk-\bp-\bp'}),
\label{eq:N22-project}
\ee
with block vectors $\mathbf n_h$, $h=\pm1,0$, built as the
$\mathbf m_h$ of \cref{app:Sigma11}, i.e.
\be
\mathbf n_h
\equiv
\begin{pmatrix}
\mathcal N_{\sigma;(h0)} \\[2pt]
\mathcal N_{\sigma;(0h)}
\end{pmatrix},
\qquad h=\pm1,
\label{eq:m-vector-block2}
\ee
and
\be
\mathbf n_0
\equiv
\begin{pmatrix}
\mathcal N_{\sigma;(+-)} \\[2pt]
\mathcal N_{\sigma;(00)} \\[2pt]
\mathcal N_{\sigma;(-+)}
\end{pmatrix}.
\label{eq:m-scalar-block2}
\ee
That gives
\be
\begin{aligned}
\Sigma_{22,\sigma}^{\rm G}
(\omega,k)
&=
2g^4
\sum_{\lambda,\lambda'=\pm}
\int_{\bp,\bp'}
P_{\lambda}(p)
P_{\lambda'}(p')
\Bigg[
\sum_{\lambda''=\pm}
\frac{
\left|
\mathcal N_{\sigma;(\lambda''\lambda'')}
\right|^2
}{
\omega^2
-
\omega^2_{\lambda''}(|\bk-\bp-\bp'|)
}
\\
&+
\sum_{h=\pm,0}
\mathbf n_h^\dagger
\left[\omega^2\mathbb 1-\mathsf H^{(h)}(|\bk-\bp-\bp'|)\right]^{-1}
\mathbf n_h
\Bigg].
\end{aligned}
\label{eq:Sigma22-projected}
\ee

\begin{figure}[t]
    \centering
    \includegraphics[width=0.45\textwidth]{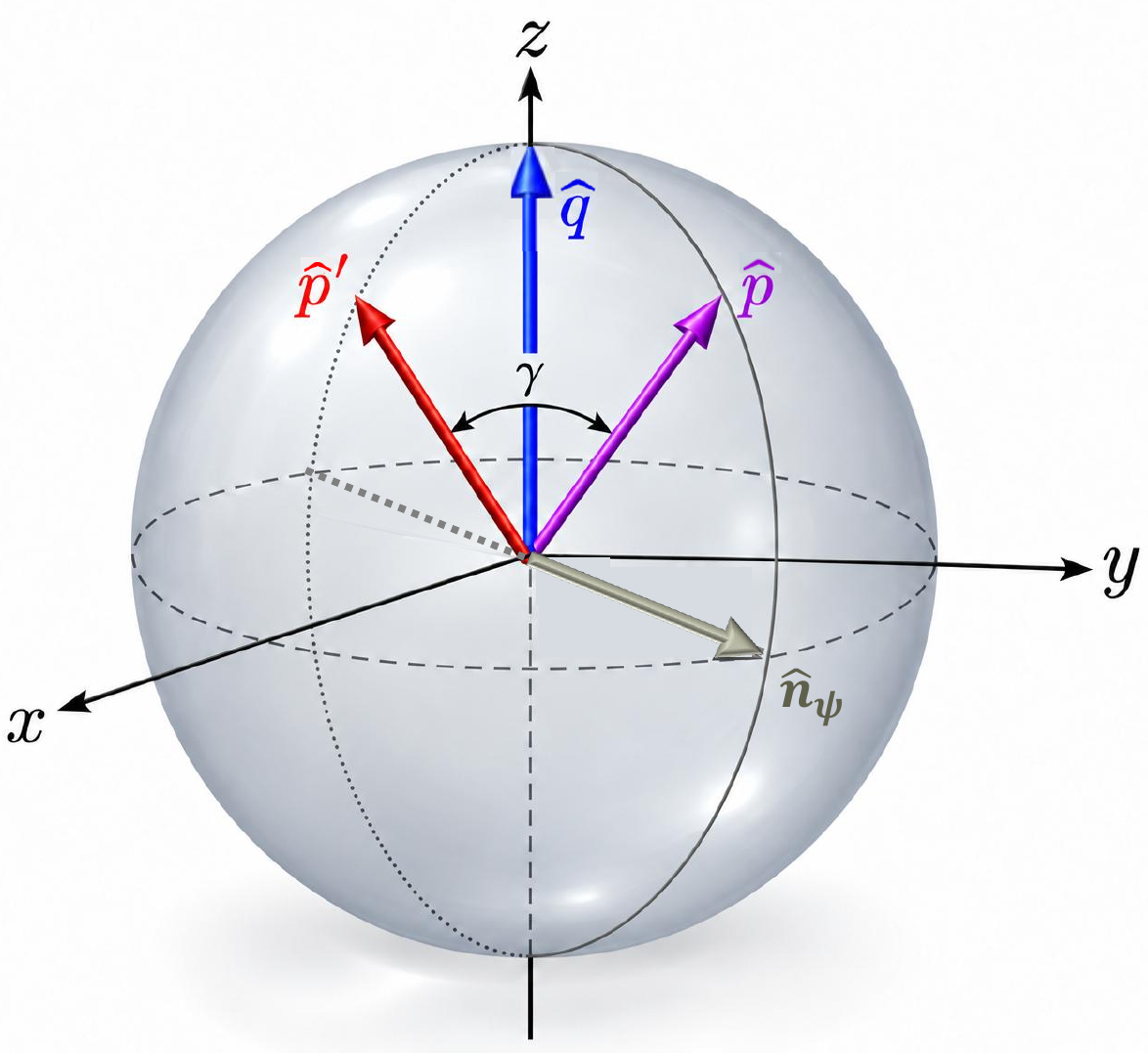}
    \caption{Geometry of the two background momenta
    $\bp$ and $\bp'$, with total momentum
    $\bq=\bp+\bp'$.
    The vectors subtend an angle $\gamma$, while
    $\bn_{\psi}$ denotes the transverse direction orthogonal to
    $\bq$ that parametrizes their relative orientation.}
    \label{fig:3d-sphere}
\end{figure}

To reduce the angular integrations, we introduce the total
momentum carried by the two stochastic modes (see \cref{fig:3d-sphere}),
\be
\bq
\equiv
\bp+\bp',
\qquad
|p-p'|\leq q\leq p+p' .
\label{eq:q-Sigma22}
\ee
If \(\gamma\) denotes the angle between
\(\bp\) and \(\bp'\), then
\be
q^2
=
p^2+p'^2+2pp'\cos\gamma,
\qquad
d\cos\gamma=\frac{q\,dq}{pp'} ,
\ee
and the two angular measures may be written as
\be
d\Omega_{\hat\bp}\,
d\Omega_{\hat\bp'}
=
\frac{q\,dq}{pp'}\,
d\Omega_{\hat\bq}\,
d\psi ,
\label{eq:two-shell-measure-Sigma22}
\ee
where \(\psi\) specifies the orientation of the pair
\((\bp,\bp')\) around \(\bq\). With
$d\equiv(p^2-p'^2)/(2q)$ and $r\equiv\sqrt{p^2-(q/2+d)^2}$, a convenient
parametrization is
\be
\begin{aligned}
\bp
=
\Big(\frac{q}{2}+d\Big)\hat\bq
+
r\,
\bn_{\psi}, \quad
\bp'
=
\Big(\frac{q}{2}-d\Big)\hat\bq
-
r\,
\bn_{\psi},
\qquad
(\bn_{\psi}\cdot\hat\bq=0),
\end{aligned}
\label{eq:p-pprime-param-Sigma22}
\ee
which for $p=p'$ reduces to equal components along $\hat\bq$ and opposite
transverse components. Choosing \(\bk\) as the polar axis, we define
\be
\mu
\equiv
\hat\bk\cdot\hat\bq,
\qquad
\bl
\equiv
\bk-\bq,
\qquad
\ell
\equiv
|\bl|
=
\sqrt{
k^2+q^2-2kq\mu
},
\label{eq:ell-Sigma22}
\ee
and the azimuthal integration about $\hat\bk$ is again trivial. The
Gaussian self-energy therefore reduces to
\be
\begin{aligned}
\Sigma_{22,\sigma}^{\rm G}
(\omega,k)
={}&
4\pi g^4
\sum_{\lambda,\lambda'=\pm}
\int dp\,dp'\,dq\;
p\,p'\,q\;
P_{\lambda}(p)P_{\lambda'}(p')
\int_{-1}^{1}
d\mu
\int_0^{2\pi}
d\psi
\\
&\times
\Bigg[
\sum_{\lambda''=\pm}
\frac{
\left|
\mathcal N_{\sigma;(\lambda''\lambda'')}
(k;p,p',q,\mu,\psi)
\right|^2
}{
\omega^2
-
\omega^2_{\lambda''}(\ell)
+i\epsilon
}
+
\sum_{h=\pm,0}
\mathbf n_h^\dagger
\left[\omega^2\mathbb 1-\mathsf H^{(h)}(\ell)\right]^{-1}
\mathbf n_h
\Bigg],
\end{aligned}
\label{eq:Sigma22-final-angular}
\ee
over the region \cref{eq:q-Sigma22}. For a single thin shell of radius
$k_{\rm s}$, the prefactor and the radial integrations collapse to
$g^4\mathcal W_2^2/(4\pi k_{\rm s}^2)\int_0^{2k_{\rm s}}q\,dq$.

As discussed for the linear--linear contribution in
\cref{app:Sigma11}, our primary interest here is the static limit,
\(\omega=0\). Since the intermediate energies are bounded below by the
gap of $\bar H$, no intermediate mode can go on shell at zero frequency,
\be
{\rm Im}\,
\Sigma_{22,\sigma}^{\rm G}(0,k)
=
0,
\qquad
\sigma=\pm ,
\label{eq:Im-Sigma22-static}
\ee
and the static quadratic--quadratic self-energy is purely real. At
\(\omega=0\) the denominators are negative definite and the numerators
are non-negative quadratic forms, so that
\be
\Sigma_{22,\sigma}^{\rm G}(0,k)
\leq 0
\ee
with the present definition of the self-energy. Two background modes can
always combine to a small total momentum, $q\to0$, for which the
propagator is evaluated at the external momentum, $\ell\to k$; at the
minimum of the dangerous branch this gives the same enhancement by
$1/\bar m^2$ as in the linear--linear term. Since $\mathcal U$ has no
derivatives, the integrand falls off like $1/q^2$ at large total
momentum, and the quadratic--quadratic term is dominated by pairs with
$q\lesssim k$.

\subsubsection*{The non-Gaussian contribution}

The Gaussian result obtained above is not, in general, the complete
quadratic--quadratic self-energy. The connected tensor trispectrum
\cref{eq:T--},
$T^c_{\lambda\lambda'\lambda''\lambda'''}(\bk,\bp,\bq,\mathbf s)$
with $\bk=\bp+\bq+\mathbf s$ (see also \cref{eq:se-legend}), generates an additional contribution,
\be
\Sigma_{22,\sigma}
=
\Sigma_{22,\sigma}^{\rm G}
+
\Sigma_{22,\sigma}^{\rm NG}.
\label{eq:Sigma22-G-NG-decomposition}
\ee
In \cref{eq:Sigma22-momentum} the first insertion of $\delta\Delta_{W^2}$
absorbs the modes $W_{\lambda'}(\bp)$ and $W_{\lambda''}(\bq)$, taking the
fluctuation from $\bk$ to $\bk+\bp+\bq$, and the second absorbs
$W_{\lambda'''}(\mathbf s)$ and $W_{\lambda}(-\bp-\bq-\mathbf s)=W^*_\lambda(\bp+\bq+\mathbf s)$,
so that
\be
\begin{aligned}
\left[
\Sigma_{22}^{\rm NG}(\omega,\bk)
\right]_{IJ}
=
g^4
\sum_{\lambda\lambda'\lambda''\lambda'''}
\int_{\bp,\bq,\mathbf s}
&
T^c_{\lambda\lambda'\lambda''\lambda'''}(\bp+\bq+\mathbf s,\bp,\bq,\mathbf s)
\left[
\mathfrak{b}\big[e^{\lambda}(\widehat{\bp+\bq+\mathbf s})^*,e^{\lambda'''}(\hat{\mathbf s})\big]
\right]_{IK}
\\
\times &
\left[
\bar G_0(\omega,\bk+\bp+\bq)
\right]_{KL}
\left[
\mathcal U_{\lambda'\lambda''}(\hat\bp,\hat\bq)
\right]_{LJ}.
\end{aligned}
\label{eq:Sigma22-NG}
\ee
As discussed around \cref{eq:response-V--} (see \cref{app:trispectrum-gaussian} for details), the connected four-point function
also contributes to the nonlinear response
$\langle W^*_\lambda V_\lambda\rangle'_{\rm st}$ in \cref{eq:response-V--}. The two quantities,
however, correspond to different momentum and polarization projections
of the trispectrum, \cref{eq:Sigma22-NG} contracting the four
polarization tensors through two factors of $\mathfrak{b}$ rather than
through the overlap $\mathcal K$ of \cref{eq:math K}, and
are therefore independent in general. It is consequently useful to
introduce a separate dimensionless parameter characterizing the
non-Gaussian correction to the quadratic--quadratic self-energy. Since
our stability analysis is concerned with the static limit at the
minimum of the dangerous tensor branch, we define
\be
\mathcal S_{\sigma}^{\rm NG}(k)
\equiv
\frac{
\Sigma_{22,\sigma}^{{\rm NG}}(0,k)
}{
\Sigma_{22,\sigma}^{{\rm G}}(0,k)
},
\label{eq:S22-NG-definition}
\ee
so that the full static contribution can be written as
\be
\Sigma_{22,\sigma}(0,k)
=
\left(
1+\mathcal S_{\sigma}^{\rm NG}(k)
\right)
\Sigma_{22,\sigma}^{{\rm G}}(0,k).
\label{eq:Sigma22-with-SNG}
\ee
For a Gaussian stochastic background,
$\mathcal S_{\sigma}^{\rm NG}=0$. In a non-Gaussian background,
$\mathcal S_{\sigma}^{\rm NG}$ encodes the particular projection
of the connected trispectrum relevant for stochastic scattering. It is
in general independent of
$\mathcal R_{\lambda}^{(4)}$, which probes a different
projection of the same four-point function, and it is bounded from below by
the positivity of the full Born kernel 
\be
1+\mathcal S^{\rm NG}_\sigma(k)\ \ge\ \frac{\Sigma_{\rm mix}(k)^2}{4\,\Sigma_{11}(k)\,\Sigma^{G}_{22}(k)}.
\label{eq:SNG-lower--bound} 
\ee

\subsection{The mixed contributions}
\label{app:Sigma12}

The mixed terms are new relative to a Gaussian or sign-symmetric
background. In \(\Sigma_{12}\), the operator \(\delta\Delta_{W^2}\)
acts first, absorbs two background modes $W_{\lambda'}(\bp)$ and
$W_{\lambda''}(\bq)$ and takes the fluctuation from \(\bk\) to
\(\bk+\bp+\bq\); the operator \(\Delta_W\) then absorbs the mode
$W_\lambda(-\bp-\bq)=W^*_\lambda(\bp+\bq)$ and returns it to \(\bk\).
Momentum conservation is supplied by the tensor bispectrum
\cref{eq:b-W----},
$B_{\lambda\lambda'\lambda''}(\bk',\bp,\bq)=\langle W^*_\lambda(\bk')W_{\lambda'}(\bp)W_{\lambda''}(\bq)\rangle'_{\rm st}$
with $\bk'=\bp+\bq$, whose first argument is the total momentum of the
pair. Using \cref{eq:DeltaW-momentum-vertex,eq:deltaDeltaW2-momentum},
\be
\begin{aligned}
\left[
\Sigma_{12}(\omega,\bk)
\right]_{IJ}
={}&
g^3
\sum_{\lambda\lambda'\lambda''}
\int_{\bp,\bq}
B_{\lambda\lambda'\lambda''}(\bp+\bq,\bp,\bq)\,
\left[
\mathcal V_{\lambda}(-\bp-\bq;\bk+\bp+\bq)
\right]_{IK}
\\
&\times
\left[
\bar G_0(\omega,\bk+\bp+\bq)
\right]_{KL}
\left[
\mathcal U_{\lambda'\lambda''}(\hat\bp,\hat\bq)
\right]_{LJ},
\end{aligned}
\label{eq:Sigma12-general}
\ee
where the linear vertex is evaluated for a background mode of momentum
$-(\bp+\bq)$, whose polarization tensor is
$e^\lambda(-\widehat{\bp+\bq})=e^\lambda(\widehat{\bp+\bq})^*$. For the
external tensor state of helicity $\sigma$ this becomes
\be
\begin{aligned}
\Sigma_{12,\sigma}(\omega,k)
&=
g^3
\sum_{\lambda\lambda'\lambda''}
\int_{\bp,\bq}
B_{\lambda\lambda'\lambda''}(\bp+\bq,\bp,\bq)
\Bigg[
\sum_{\lambda'''=\pm}
\frac{
\mathcal M_{\sigma;(\lambda'''\lambda''')}(\bk,-\bp-\bq)\;
\mathcal N^{(3)}_{(\lambda'''\lambda''');\sigma}(\bk;\bp,\bq)
}{
\omega^2-\omega^2_{\lambda'''}(|\bk+\bp+\bq|)
}
\\
&+
\sum_{h=\pm,0}
\mathbf m_h^{\!\top}
\left[\omega^2\mathbb 1-\mathsf H^{(h)}(|\bk+\bp+\bq|)\right]^{-1}
\mathbf n^{(3)}_h
\Bigg],
\end{aligned}
\label{eq:Sigma12-projected}
\ee
with $\mathcal M_{\sigma;(\lambda_1\lambda_2)}(\bk,-\bp-\bq)$ the
amplitudes \cref{eq:M-closed-forms} of the linear vertex evaluated for
the background momentum $-(\bp+\bq)$, i.e.\ with
$q=|\bp+\bq|$ and $\mu=-\hat\bk\cdot\widehat{(\bp+\bq)}$ in
\cref{eq:Sigma11-kinematic-definitions}, and
\be
\mathcal N^{(3)}_{(\lambda_1\lambda_2);\sigma}(\bk;\bp,\bq)
\equiv
u^{(\lambda_1\lambda_2)}_{K}(\widehat{\bk+\bp+\bq})^*
\left[
\mathcal U_{\lambda'\lambda''}(\hat\bp,\hat\bq)
\right]_{KJ}
e^{\sigma}_J(\hat\bk),
\label{eq:N3-project}
\ee
collected in the block vectors $\mathbf n^{(3)}_h$.

The quadratic vertex is symmetric in its
spatial--color indices and, by the reality
condition on the background helicity modes,
satisfies
\be
\mathcal U_{\lambda'\lambda''}
(\hat\bp,\hat\bq)^\dagger
=
\mathcal U_{\lambda'\lambda''}
(-\hat\bp,-\hat\bq).
\label{eq:U-hermiticity}
\ee
Consequently, the reverse transition amplitude
is simply the complex conjugate of the projected
quadratic vertex defined in \cref{eq:N22-project},
evaluated at reversed background momenta:
\be
\mathcal N^{(3)}_{(\lambda_1\lambda_2);\sigma}
(\bk;\bp,\bq)
=
\mathcal N_{\sigma;(\lambda_1\lambda_2)}^*
(\bk;-\bp,-\bq).
\label{eq:N3-N22-relation}
\ee
Thus, introducing the helicity-block vectors
$\mathbf m_h=\mathbf m_h(\bk,-\bp-\bq)$ and
$\mathbf n_h=\mathbf n_h(\bk;-\bp,-\bq)$,
with their background helicity labels implicit,
we obtain
\be
\begin{aligned}
\Sigma_{12,\sigma}(\omega,k)
&=
g^3
\sum_{\lambda,\lambda',\lambda''}
\int_{\bp,\bq}
B_{\lambda\lambda'\lambda''}
(\bp+\bq,\bp,\bq)
\Bigg[
\sum_{\lambda_1=\pm}
\frac{
\mathcal N_{\sigma;(\lambda_1\lambda_1)}^*
\mathcal M_{\sigma;(\lambda_1\lambda_1)}
}{
\omega^2-\omega_{\lambda_1}^2(\ell)
}
\\
&+
\sum_{h=\pm,0}
\mathbf n_h^\dagger
\left[\omega^2\mathbb 1-\mathsf H^{(h)}(\ell)\right]^{-1}
\mathbf m_h
\Bigg].
\end{aligned}
\label{eq:Sigma12-projected-simplified}
\ee
All transition amplitudes in the last expression
are evaluated at the momenta specified above.
The same Hartree helicity blocks enter both
linear and quadratic scattering, while the
distinct vertex structures are encoded in
$\mathbf m_h$ and $\mathbf n_h$, respectively.

 Hermiticity of $\Delta_W$ and
$\delta\Delta_{W^2}$ gives
\be
\Sigma_{21,\sigma}^{(2)}(\omega,k)=\left(\Sigma_{12,\sigma}^{(2)}(\omega,k)\right)^*,
\label{eq:Sigma21-conj}
\ee
so the mixed contribution to the Born kernel is $2\,{\rm Re}\,\Sigma_{12,\sigma}^{(2)}$,
which is real at $\omega=0$. The orientation of the triangle
$(\bp+\bq,\bp,\bq)$ relative to $\bk$ is fixed by
$\mu=-\hat\bk\cdot\widehat{(\bp+\bq)}$ and by the rotation angle $\psi$
of the triangle about $\bp+\bq$, and with $k'\equiv|\bp+\bq|$ the measure
is $\int_{\bp,\bq}=2\pi\int dk'\,dp\,dq\;k'pq\int_{-1}^1d\mu\int_0^{2\pi}d\psi$
over the triangles allowed by the triangle inequality. Only the momentum
$\bk+\bp+\bq$ enters the propagator, as in \cref{app:Sigma11}, but
$\mathcal N^{(3)}$ depends on the full triangle. For a single thin shell
all closed triangles are equilateral, the bispectrum is the one number
fixed by $b_W$, and $\Sigma_{12}^{(2)}+\Sigma_{21}^{(2)}$ reduces to
$b_W\mathcal W_2^{3/2}$ times a universal function of $k$, the shell radius
and the couplings. We have verified \cref{eq:Sigma12-projected,eq:Sigma21-conj}
against explicit phase-locked triad and cuboctahedral fields, averaged
exactly over translations and by Monte Carlo over rotations.

\bibliography{ref.bib}

\end{document}

%% file: born_selfenergy_diagrams_qa.tex
\definecolor{SEVertexOne}{HTML}{3CB4FF}
\definecolor{SEVertexTwo}{HTML}{FFBD31}
\definecolor{SEBlob}{HTML}{E6F1EF}
\tikzset{
 se/vone/.style={circle,fill=SEVertexOne,draw=SEVertexOne,
   inner sep=0pt,minimum size=8.7pt},
 se/vtwo/.style={circle,fill=SEVertexTwo,draw=SEVertexTwo,
   inner sep=0pt,minimum size=8.7pt},
 se/blob/.style={rounded corners=1.7pt,draw=QAW,fill=SEBlob,
   line width=.65pt,inner xsep=3pt,inner ysep=2.5pt,
   font=\scriptsize,text=QAInk},
 se/corr/.style={font=\scriptsize,text=QAW,inner sep=1pt},
 se/mom/.style={font=\scriptsize,text=QAInk,inner sep=.4pt},
 se/arrow/.style={-{Stealth[length=1.8mm,width=1.25mm]},
   line width=.65pt,draw=QAInk}
}

\newcommand{\SEVertexGammaOne}{%
  \QABeginVertex
  \QAIncoming\QAOutgoing{\mathbf k-\mathbf p}\QAUpperW
  \node[se/vone] at (v) {};
  \QAVertexLabel{\Gamma_1}
  \QAEndVertex
}
\newcommand{\SEVertexGammaTwo}{%
  \QABeginVertex
  \QAIncoming\QAOutgoing{\mathbf k-\mathbf p-\mathbf q}\QAUpperW
  \draw[qa/tensor] (v)--(3.13,0);
  \draw[qa/flow] (2.33,.22)--(2.96,.22);
  \node[qa/momentum,anchor=south] at (2.65,.26) {$\mathbf q$};
  \node[qa/momentum,text=QAW,anchor=west] at (3.30,0) {$W$};
  \node[se/vtwo] at (v) {};
  \QAVertexLabel{\Gamma_2}
  \QAEndVertex
}

\newcommand{\SEDiagramWWMean}{%
  \QABeginVertex
  \QAIncoming\QAOutgoing{\mathbf k}
  \draw[qa/tensor] (v) .. controls ($(v)+(-.85,1.45)$)
                            and ($(v)+(.85,1.45)$) .. (v);
  \node[se/corr,anchor=south] at ($(v)+(0,1.22)$)
     {$\langle W^*W\rangle_{\rm st}$};
  \node[qa/quarticvertex] at (v) {};
  \QAVertexLabel{g^2}
  \QAEndVertex
}

\newcommand{\SEBegin}[2]{%
 \begin{tikzpicture}[x=.84cm,y=.84cm,scale=\QADiagramScale,
   transform shape,baseline=-.50ex]
  \path[use as bounding box] (-.08,#1) rectangle (6.70,#2);
  \coordinate (va) at (1.42,0);
  \coordinate (vb) at (5.20,0);
}
\newcommand{\SEEnd}{\end{tikzpicture}}
\newcommand{\SELine}[1]{%
  \draw[qa/quantum] (0,0)--(va);
  \draw[qa/flow] (.30,.23)--(.98,.23);
  \node[se/mom,anchor=south] at (.64,.31) {$\mathbf k$};
  \draw[qa/quantum] (va)--(vb);
  \draw[qa/flow] (2.88,-.21)--(3.58,-.21);
  \node[qa/weight,anchor=north] at (3.23,-.37)
    {$\bar G_0(\omega,#1)$};
  \draw[qa/quantum] (vb)--(6.62,0);
  \draw[qa/flow] (5.63,.23)--(6.31,.23);
  \node[se/mom,anchor=south] at (5.97,.31) {$\mathbf k$};
}
\newcommand{\SEVertices}[2]{%
  \node[#1] at (va) {};
  \node[#2] at (vb) {};
}
\newcommand{\SECorrTop}[2]{% momentum, helicity
  \draw[se/arrow] (2.88,1.43)--(3.58,1.43);
  \node[se/mom,anchor=south] at (3.23,1.52) {$#1$};
  \node[se/corr,anchor=south] at (3.23,2.02)
    {$\langle W_{#2}^{*}W_{#2}\rangle_{\rm st}$};
}
\newcommand{\SECorrBottom}[2]{% momentum, helicity
  \draw[se/arrow] (2.88,-1.83)--(3.58,-1.83);
  \node[se/mom,anchor=north] at (3.23,-1.91) {$#1$};
  \node[se/corr,anchor=north] at (3.23,-2.32)
    {$\langle W_{#2}^{*}W_{#2}\rangle_{\rm st}$};
}

\newcommand{\SEDiagramEleven}{%
 \SEBegin{-1.22}{2.35}
   \SELine{\mathbf k-\mathbf p}
   \draw[qa/tensor] (va) .. controls (2.19,1.52) and (4.43,1.52) .. (vb);
   \SECorrTop{\mathbf p}{\lambda}
   \SEVertices{se/vone}{se/vone}
 \SEEnd
}

\newcommand{\SEDiagramTwelve}{%
 \SEBegin{-1.22}{2.45}
   \SELine{\mathbf k-\mathbf p}
   \node[se/blob] (B) at (3.23,1.62) {$B_W$};
   \draw[qa/tensor] (va) to[out=80,in=200] (B.west);
   \draw[qa/tensor] (vb) to[out=100,in=-10] (B.east);
   \draw[qa/tensor] (vb) to[out=66,in=25,looseness=1.05] (B.north east);
   \SEVertices{se/vone}{se/vtwo}
 \SEEnd
}

\newcommand{\SEDiagramTwentyOne}{%
 \SEBegin{-1.22}{2.45}
   \SELine{\mathbf k-\mathbf p-\mathbf q}
   \node[se/blob] (B) at (3.23,1.62) {$B_W$};
   \draw[qa/tensor] (va) to[out=80,in=190] (B.west);
   \draw[qa/tensor] (va) to[out=114,in=155,looseness=1.05] (B.north west);
   \draw[qa/tensor] (vb) to[out=100,in=-20] (B.east);
   \SEVertices{se/vtwo}{se/vone}
 \SEEnd
}

\newcommand{\SEDiagramTwentyTwoG}{%
 \SEBegin{-2.74}{2.42}
   \SELine{\mathbf k-\mathbf p-\mathbf q}
   \draw[qa/tensor] (va) .. controls (2.19,1.50) and (4.43,1.50) .. (vb);
   \draw[qa/tensor] (va) .. controls (1.81,-2.02) and (4.81,-2.02) .. (vb);
   \SECorrTop{\mathbf p}{\lambda}
   \SECorrBottom{\mathbf q}{\lambda'}
   \SEVertices{se/vtwo}{se/vtwo}
 \SEEnd
}

\newcommand{\SEDiagramTwentyTwoT}{%
 \SEBegin{-1.22}{2.55}
   \SELine{\mathbf k-\mathbf p-\mathbf q}
   \node[se/blob] (T) at (3.23,1.73) {$T_W$};
   \draw[qa/tensor] (va) to[out=90,in=186] (T.west);
   \draw[qa/tensor] (va) to[out=50,in=252] (T.south west);
   \draw[qa/tensor] (vb) to[out=90,in=-6] (T.east);
   \draw[qa/tensor] (vb) to[out=130,in=-72] (T.south east);
   \SEVertices{se/vtwo}{se/vtwo}
 \SEEnd
}

\newcommand{\SELegendBlob}[1]{%
 \tikz[baseline=-.45ex]{\node[se/blob] at (0,0) {$#1$};}}